\documentclass[fleqn,usenatbib]{mnras}
\usepackage{amssymb}
\usepackage{pifont} 
\usepackage{newtxtext,newtxmath}
\usepackage{subcaption}
\usepackage[version=4]{mhchem}
\usepackage{tabularray}
\usepackage{graphicx} 
\usepackage{xspace}
\usepackage[T1]{fontenc}
\usepackage{siunitx}
\usepackage{gensymb}
\usepackage{fix-cm}
\usepackage{booktabs}
\usepackage{xcolor}
\DeclareRobustCommand{\VAN}[3]{#2}
\let\VANthebibliography\thebibliography
\def\thebibliography{\DeclareRobustCommand{\VAN}[3]{##3}\VANthebibliography}

\usepackage{graphicx}
\usepackage{amsmath}

\newcommand{\one}{\mbox{3DOneExpl}\xspace}
\newcommand{\onescenario}{\mbox{OneExpl}\xspace}
\newcommand{\two}{\mbox{3DTwoExpl}\xspace}
\newcommand{\twoscenario}{\mbox{TwoExpl}\xspace}

\newcommand{\oneO}{\mbox{1DOneExpl}\xspace}
\newcommand{\twoO}{\mbox{1DTwoExpl}\xspace}

\newcommand{\mass}{\mbox{M$_{\odot}$}}

\newcommand{\submch}{sub-$M_\mathrm{Ch}\xspace$}

\newcommand{\microns}{\ensuremath{\text{\textmu}\mathrm{m}}\xspace}
\newcommand{\dsix}{D$^6$\xspace}

\title[Multi-D radiative transfer for nebular SNe~Ia]{
  Multidimensional Nebular-Phase Calculations of Dynamically-Driven Double-Degenerate Double-Detonation Models for Type Ia Supernovae}

\author[J. M. Pollin et al.]{J. M. Pollin,$^{1}$\thanks{E-mail: jpollin02@qub.ac.uk}
S. A. Sim,$^{1,2,3}$
L. J. Shingles,$^{1,4}$
R. Pakmor,$^{5}$
F. P. Callan,$^{1}$
C. E. Collins,$^{6}$
F. K. Röpke $^{7,8,9}$
\newauthor
L. A. Kwok$^{10}$
A. Holas$^{7}$
and S. Srivastav$^{11}$
\\
$^{1}$Astrophysics Research Center, School of Mathematics and Physics, Queen’s University Belfast, Belfast BT7 1NN, Northern Ireland, UK\\
$^{2}$Cosmic Dawn Center (DAWN), Denmark\\
$^{3}$Niels Bohr Institute, University of Copenhagen, Jagtvej 155A, DK-2200, Copenhagen N, Denmark\\
$^{4}$GSI Helmholtzzentrum für Schwerionenforschung, Planckstraße 1, 64291 Darmstadt, Germany\\
$^{5}$Max-Planck-Institut für Astrophysik, Karl-Schwarzschild-Str. 1, D-85748, Garching, Germany\\
$^{6}$School of Physics, Trinity College Dublin, The University of Dublin, Dublin 2, Ireland\\
$^{7}$Heidelberger Institut für Theoretische Studien, Schloss-Wolfsbrunnenweg 35, 69118 Heidelberg, Germany\\
$^{8}$Zentrum für Astronomie der Universität Heidelberg, Institut für Theoretische Astrophysik, Philosophenweg 12, 69120 Heidelberg, Germany\\
$^{9}$Zentrum für Astronomie der Universität Heidelberg, Astronomisches Rechen-Institut, M{\"o}nchhofstr.\ 12--14, 69120 Heidelberg, Germany\\
$^{10}$Center for Interdisciplinary Exploration and Research in Astrophysics (CIERA), 1800 Sherman Ave., Evanston, IL 60201, USA\\
$^{11}$Department of Physics, University of Oxford, Denys Wilkinson Building, Keble Road, Oxford OX1 3RH, UK\\
}

\date{Accepted XXX. Received YYY; in original form ZZZ}

\pubyear{2025}

\begin{document}
\label{firstpage}
\pagerange{\pageref{firstpage}--\pageref{lastpage}}
\sloppy
\maketitle

\begin{abstract}
The dynamically-driven double-degenerate double-detonation model has emerged as a promising progenitor candidate for Type Ia supernovae. 
In this scenario, the primary white dwarf ignites due to dynamical interaction with a companion white dwarf, which may also undergo a detonation. Consequently, two scenarios exist: one in which the secondary survives and another in which both white dwarfs detonate. In either case, substantial departures from spherical symmetry are imprinted on the ejecta.
Here, we compute full non-local thermodynamic equilibrium nebular-phase spectra in 1D and 3D to probe the innermost asymmetries. Our simulations reveal that the multidimensional structures significantly alter the overall ionisation balance, width and velocity of features, especially when the secondary detonates. 
In this scenario, some element distributions may produce orientation–dependent line profiles that can be centrally peaked from some viewing-angles and somewhat flat-topped from others.
Comparison to observations reveals that both scenarios produce most observed features from the optical to mid-infrared. However, the current model realisations do not consistently reproduce all line shapes or relative strengths, and yield prominent optical \ion{Ar}{III} emission which is inconsistent with the data.
When the secondary detonates, including 3D effects improves the average agreement with observations, however when compared to observations, particularly weak optical \ion{Co}{III} emission and the presence of optical \ion{O}{I} and near-infrared \ion{S}{I} challenge its viability for normal Type Ia supernovae. Thus, overall, our comparisons with normal Type Ia's tentatively favour detonation of only the primary white dwarf but we stress that more model realisations and mid-infrared observations are needed.

\end{abstract}

\begin{keywords}
Radiative transfer – Transients: supernovae – Methods: numerical - Stars: binaries - white dwarfs
\end{keywords}
\section{Introduction}
\label{sec:Introduction}

There is general agreement that Type Ia supernovae (SNe~Ia) originate from the explosion of a carbon-oxygen (CO) white dwarf (WD) \citep{Hoyle1960} in a close binary with a companion. It is still unclear if a WD near the Chandrasekhar mass ($M_\mathrm{Ch}$) or a sub-Chandrasekhar mass (sub-$M_\mathrm{Ch}$) progenitor should be favoured for the majority of SNe~Ia (see e.g.\,\citealt{Liu2023,Ruiter2025} for a review). Additionally, the nature of the WD's companion -- whether it is a non-degenerate companion in the single-degenerate scenario \citep{Whelan1973} or another WD in the double-degenerate scenario \citep{Iben1984} -- has remained an open question. There is, however, a growing body of evidence favouring the double-degenerate scenario. The scenario naturally explains several key characteristics: the absence of hydrogen \citep{Leonard2007} and the minimal contribution of He in the observed spectra \citep{Jiang2017,Noebauer2017,De2019} is easily explained, the rates of the systems are consistent with observed SNe~Ia rates \citep{Ruiter2009}, and they can account for the delay-time distribution of SNe~Ia \citep{Maoz2012}.

Investigations of \submch~CO WD detonations have demonstrated reasonable agreement with observations of SNe~Ia \citep{sim2010,Blondin2017,Shen2018,Shen2021a}. These models have reproduced several key observational features, such as the width-luminosity relation \citep{philips}. However, the unknown nature of the companion introduces a significant amount of uncertainty regarding the detonation mechanism. Among the possible mechanisms, the double-detonation scenario has garnered significant interest; a detonation is ignited in the helium surface layer on the WD, which sends shockwaves inward, triggering the detonation of the underlying CO core \citep{Taam1980,Livne1990,Nomoto1980,Hoeflich1996,Nugent1997}.
Moreover, high-resolution observations of a supernova remnant provide strong support for this mechanism \citep{Das2025}; however, see \citealt{Soker2025} for an alternative perspective.

It is particularly noteworthy that modern double-detonation models \citep{Bildsten2007,Shen2009,kromer2010,fink2010,Shen2010,Townsley2019,Polin2019,Gronow2020,Gronow2021,Boos2021,Shen2024} have demonstrated a continued ability to achieve CO detonations with lower-mass helium shells compared to those used in earlier models \citep{Livne1990,Livne_Glasner1990,Livne1995}. The synthetic observables of double-detonation models have also come into better agreement with observations, due to these smaller helium shells, and, as such, the double-detonation mechanism remains a subject of active investigation.

Given the aforementioned motivations for favouring the double-degenerate channel and the success of the double-detonation mechanism, it has been of particular interest to apply the mechanism to a dynamically-driven double-degenerate merger scenario.
In this scenario, the primary WD undergoes dynamical helium accretion from the secondary WD before the merger. The instabilities in the helium accretion stream result in a thermonuclear runaway in the helium shell, leading to the double-detonation of the primary \citep{Guillochon2010,Pakmor2013,Kashyap2015,Tanikawa2018,pakmor_2021,Fraile2024}. An essential difference between the double-degenerate double-detonation scenario and the classic violent merger scenario \citep{Pakmor2012b} or the helium-ignited violent merger model \citep{Pakmor2013,Tanikawa2015} lies in the timing of the detonation. In the violent merger scenario, the secondary is disrupted during the merging of the CO cores. Whereas in the dynamically-driven double-degenerate double-detonation (\dsix) scenario a detonation occurs at the earliest possible stage of the merger during the phase of rapid accretion of helium-rich material from the outer layer of the secondary before it is disrupted.

There currently exists only a handful of \dsix models. As such, significant uncertainties persist regarding the fate of the secondary WD, which is intact at the time of the primary WD's detonation \citep{Tanikawa2019,pakmor_2021,Boos2024}. There are two key open questions surrounding the system's dynamics: firstly,  the role of the accretion stream and how the helium shell detonation affects the temperature and density of the primary WD;  and secondly, the conditions of the secondary WD at the point of shock convergence.
Notably, \cite{pakmor_2021} performed two 3D hydrodynamic simulations, tracking the initial inspiral, mass transfer, helium shell detonation, and subsequent WD core detonation. In one simulation, only the primary detonates, and the secondary survives. in the other simulation, both the primary and secondary WDs ignite and explode.

It was demonstrated that the two simulations produced remarkably similar spectra (using 1D radiative transfer simulations) in the photospheric phase \citep{pakmor_2021}. This overall similarity was confirmed by subsequent 3D simulations for these models \citep{Pollin2024a}, and is supported by the 2D models of \cite{Boos2024}. In the 3D calculation, the primary difference between the two models was more considerable variation with viewing angle of the synthetic observables, such as luminosity, spectral features, and colour evolution when the secondary WD detonated. Similar to idealised double-detonation model calculations \citep{Collins2022, Holas2025}, it was found that although the spread in these observables is significant compared to observational data, it remains insufficient to rule out this explosion pathway. Moreover, this spread may be reduced with lower helium shell masses and full non-local thermodynamic equilibrium (NLTE) calculations \citep{Collins2025}.

A striking consequence of WD merger models where the secondary WD is also disrupted is the formation of an `inner bubble' structure within the ejecta (see Figure~\ref{Fig: Model abundances plot})\footnote{This `inner bubble' feature is also present in the model by \cite{Pakmor2012b}, suggesting it may be representative of WD mergers more broadly.}.
As this `inner bubble' and other core asymmetries are located in the innermost ejecta, they significantly influence synthetic observables once the ejecta has expanded sufficiently to allow spectra from the central ejecta to emerge in the nebular-phase. Thus, to understand the impact of this inner structure, it is essential to investigate the late-phase ($\sim$ one year after the explosion). At this epoch, excitation by both radiative and collisional processes becomes slow, and only the lowest energy states possess significant populations. Hence, forbidden emission, particularly from Fe, Co, and Ni, dominate \citep{Spyromilio1992}.
 
Late-phase observations have revealed that SNe~Ia features show some degree of blueshift or redshift from their rest wavelengths \citep{Motohara2006,Gerardy2007,Maeda2010,Maguire2018,Flores2020}. It has been suggested that these shifts could be a result of off-centre ignition \citep{Maeda2010}. A 3D off-centre detonation will inherently lead to an asymmetric distribution of the ejecta \citep{Holas2025}. Moreover, the distribution of ejecta can not only lead to different velocities but also different widths and line profiles of features. For a comprehensive review of the effect of different ejecta configurations on line profiles, see \cite{Jerkstrand2017}. The optical region is complicated by blending of multiple species, and many ground-based observations are hindered by telluric absorption. However, the launch of the James Webb Space Telescope (JWST) has provided an unprecedented opportunity to explore the mid-infrared (MIR) region, where features are more isolated. Of particular interest is SN~2021aefx \citep{Kwok2023}, which was observed at $\sim$270 days post-explosion using the South African Large Telescope (SALT) and, notably, JWST. 
As such, observations of SN~2021aefx at this epoch span from the optical, near-infrared (NIR) and MIR (0.35--14\microns)
This ample wavelength coverage enables a more reliable investigation of ejecta stratification, which can be used to assess different progenitor scenarios and explosion mechanisms.

Since the one-zone models of \cite{Axelrod1980}, there have been many theoretical studies exploring multi-zone models of nebular SNe~Ia \citep[e.g.][]{Ruiz1992,Liu1997,Mazzali2001,Hoflich2004,Kozma2005,Maurer2011,Li2012,Diamond2015,Fransson2015,Botyanszki2017,Shingles2020,Mazzali2020,Polin2021,Shingles2022a,Blondin2023}. These investigations have enhanced our understanding of the geometry and composition of SNe~Ia. While many of these have relied on assumptions of spherical symmetry or optically thin emission, which may not fully capture the complexity of the ejecta conditions (e.g the line shifts and morphologies), they have nonetheless provided valuable insights. Several investigations have strived to overcome these limitations by employing various methods, such as using simplified geometries (e.g., \citealt{Maeda2010}) or superposition of 1D models (e.g., \citealt{Mazzali2018}). While these approaches have provided valuable insights, they remain constrained by their simplifications. As noted by \cite{Pakmor2024}, there is an ever-growing need to perform 3D nebular-phase calculations, since modern explosion models are fundamentally multidimensional.  Given the large departures from spherical symmetry in \dsix models developed, nebular-phase spectra are expected to show significant asymmetries. Hence, 3D NLTE nebular calculations are necessary to accurately model this scenario and reconcile synthetic observables with observed properties of SNe~Ia.

In this work, we build upon the photospheric-phase calculations of \cite{pakmor_2021} and \cite{Pollin2024a}, extending our investigation to the deepest regions of the ejecta, hundreds of days after the explosion, in 3D. Section~\ref{sec:Method} outlines the models and the radiative transfer configuration. Section~\ref{sec:angleaveraged} presents the angle-averaged spectra and compares them with the corresponding spherically averaged spectra. Sections~\ref{sec:3DOneExpl_Modle_Orientation_Effects} and \ref{sec:3DTwoExpl_Modle_Orientation_Effects} explore the orientation effects for each explosion model. Throughout this work, we compare the \dsix models across the optical, NIR and MIR to the spectroscopically normal SN~2021aefx \citep{Kwok2023}. Finally, we discuss our findings and present conclusions in Section~\ref{sec:Discussion_and_Conclusions}.

\section{Methods}
\label{sec:Method}

\subsection{Hydrodynamical Ejecta Models}
\label{sec:Models}

\begin{figure*}
\centering

\begin{subfigure}[b]{1\textwidth}
\hfill
   \includegraphics[width=0.96\linewidth]{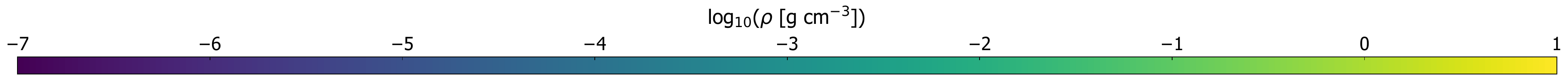}  
\end{subfigure}
\begin{subfigure}[b]{1\textwidth}
   \includegraphics[width=1\linewidth]{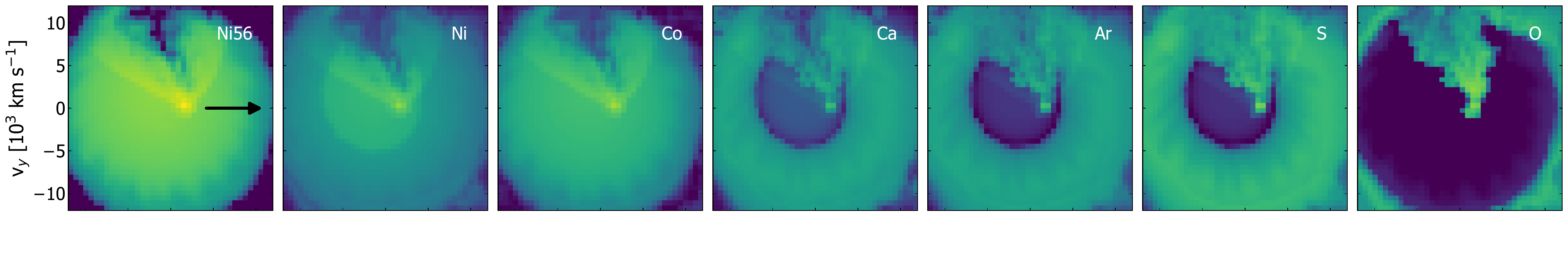}  
\end{subfigure}

\begin{subfigure}[b]{1\textwidth}
\vspace{-0.6cm}
   \includegraphics[width=1\linewidth]{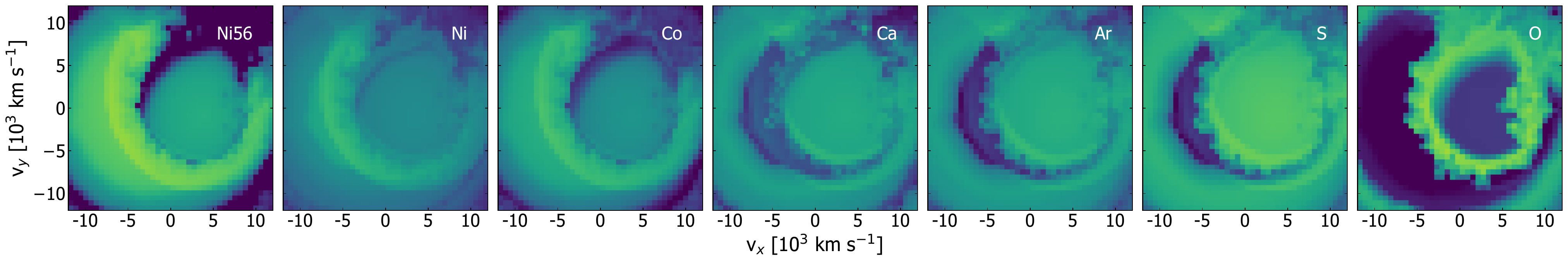}
\end{subfigure}
\caption{Density of key species in the \one (top) and \two (bottom) models, in the X-Y plane at $\cos(\theta)=0$ where the arrow in the top left panel indicate the direction of $\phi=0\degree$. The colour scale indicates the logarithmic density of each species. Note this snapshot is at 270 days after explosion, apart from\xspace $^{56}$Ni, which is shown at 0.002 days.}
\label{Fig: Model abundances plot}
\end{figure*}

The two hydrodynamical explosion models investigated in this work are the 3D models from \cite{pakmor_2021}. The first is where only the primary WD explodes and the secondary survives (\one) and the second is where both WDs explode (\two). These models originate from a binary system consisting of a primary CO WD with a mass of 1.05\mass\space and a secondary CO WD with a mass of 0.7\mass. In both scenarios, the WDs have a thin helium shell of 0.03\mass, which is dynamically transferred from the secondary WD to the primary WD \citep[for detailed nucleosynthesis, see ][]{pakmor_2021}.
The abundance structures of key species relevant to the nebular-phase for both models are presented in Figure \ref{Fig: Model abundances plot}.
To highlight the importance of multidimensional radiative transfer effects in the nebular-phase, we also performed corresponding 1D calculations of both scenarios. To impose spherical symmetry, these 1D models were constructed by averaging the 3D hydrodynamical simulations across 100 spherical shells. The resulting 1D explosion models are referred to as the \oneO and \twoO models. For clarity, when referring to the \one and \oneO models or the \two and \twoO models collectively, we describe them as the \onescenario scenario and \twoscenario scenario, respectively.

\subsection{Radiative Transfer}
\label{sec:Radiative Transfer}

We perform our radiative transfer calculations using the 3D Monte Carlo radiative transfer code \textsc{artis} \citep{sim_2007,Kromer2009}. The methods used by \textsc{artis} are based on \cite{Lucy2002,lucy2003,Lucy2005}, which divides the radiation field into indivisible Monte Carlo packet quanta. This work utilises the full NLTE approach developed by \cite{Shingles2020}, which includes an NLTE population and ionisation solver and treatment of non-thermal leptons. To follow the population of leptons with non-thermally distributed energies, \textsc{artis} solves the Spencer-Fano equation (as formulated by \citealt{Kozma1992a}). Treatment of all levels in NLTE for every ion has a significant memory and computational cost. As such, we restrict the number of NLTE levels. For most ions, we treat the first 80 levels in NLTE but increase this to 197 NLTE levels for \ion{Fe}{II} to ensure all metastable levels are treated in NLTE.  We place other levels into an additional NLTE level that can vary in population, known as a `superlevel' \citep{Anderson1989a}; the absolute population of the superlevel is determined by the NLTE solver, while the relative populations of the stages within the superlevel are set by a Boltzmann distribution at the electron temperature\footnote{For the calculations presented in Section~\ref{sec:Results}, we find that the superlevel populations are typically many orders of magnitude lower than the corresponding ground-state populations, for both iron-group and intermediate-mass elements. We therefore reaffirm that the superlevel approximation remains a valid and appropriate approach.}. We also include the heating, ionisation and excitation due to Auger electrons from ionisations of inner shells. 

The atomic data used in our calculations is based on the compilation of \textsc{CMFGEN} \citep{hillier1990a,hillier1998a} and is the same as that used by \cite{Shingles2022a}. Employing a detailed treatment \citep{lucy2003} for all photoionisation processes is memory-intensive in 3D. As such, in this work we adopt a new hybrid scheme for calculating photoionisation rate estimators, where the detailed treatment is used for bound–free transitions whose lower levels are included in the NLTE solution and the integral over the binned radiation–field model \citep{Shingles2020} for all others\footnote{See Appendix~\ref{apen:Photoionisation} for an overview of this hybrid scheme for calculating photoionisation rate estimators.}.

We employ the ejecta profiles once homologous expansion has been established, which typically occurs $\sim$100~s after the explosion. In \cite{pakmor_2021}, the simulations were evolved for more than 100~s past ignition to ensure that this condition is satisfied. We then map this homologously expanding ejecta to 230~days past explosion.
Our simulation utilises 60 logarithmically spaced time steps from 230 to 305 days post-explosion.
We simulate photons produced within the ejecta during this epoch. Given the $\sim$25-day light-crossing time for the line-forming region of the ejecta, our calculations accurately represent the observable emission between 250 and 279 days.
The simulations are initialised in LTE for the first 8 time steps, after which the full NLTE treatment is activated. We find that the plasma state converges rapidly once the calculation departs from LTE, with convergence achieved by time step 12. After time step 12, the plasma properties evolve only gradually between successive steps.
These calculations use $3.07$x$10^{10}$ Monte Carlo packets, and transport is performed on a 3D Cartesian grid that co-expands with the ejecta. The calculations carried out here maintain the same grid resolution as those by \cite{Pollin2024a}, but we exclude cells with absolute Cartesian velocities which exceed $12,000$$\,\mathrm{km}\,\mathrm{s}^{-1}$. This adjustment preserves the inner ejecta while removing the outermost cells, which have minimal impact on the nebular spectra because the fast, diffuse outer ejecta have low densities and receive little to no energy deposition at these epochs\footnote{See Appendix~\ref{apen:Resolution} for a discussion on the effects of removing the outer cells in a $25^3$ model.}.

To investigate angle-averaged synthetic observables we use all emergent packets to make the angle-averaged spectra (see section~\ref{sec:angleaveraged}). 
To examine specific lines of sight (see section~\ref{sec:3DOneExpl_Modle_Orientation_Effects} and \ref{sec:3DTwoExpl_Modle_Orientation_Effects}) we use the virtual packet scheme developed by \cite{Bulla2015}. This enhances signal-to-noise for selected viewing angles. We stress that the computational overhead for the virtual packets in the nebular-phase is negligible compared to the NLTE solver cost. The virtual packets are enabled for the entire calculation and are active between 0.35--30\microns. We have enabled virtual packets for 30 orientations, corresponding to $\cos{\theta} = 0.0,\xspace 0.4,\xspace 0.8$, with 10 equally spaced $\phi$ angles ranging from 0--360\degree, where $\theta$ is the angle from the positive z-axis, and $\phi$ is the degree of rotation of the projection in the xy-plane. Hence, rotation of $\phi$ occurs anti-clockwise from $\phi=0\degree$ and is indicated by the directional arrow in Figure~\ref{Fig: Model abundances plot}.
In this work, we focus on the merger plane ($\cos{\theta} = 0.0$), where both the \one and \two models exhibit the most significant variations in density and temperature. These differences lead to the most significant variations in synthetic observables; consequently, we exclude the $\cos{\theta} = 0.4$ and $0.8$ angles from our detailed analysis
\footnote{See Appendix~\ref{apen:additional_viewing_angles} for representative cases for a selected $\phi$ at $\cos{\theta} = 0$, $0.4$, and $0.8$ }.

\section{Results}
\label{sec:Results}

In Section~\ref{sec:angleaveraged}, we examine the spherically averaged and angle-averaged spectra and in Sections~\ref{sec:3DOneExpl_Modle_Orientation_Effects} and \ref{sec:3DTwoExpl_Modle_Orientation_Effects}, we explore the orientation effects for the \one and \two models, respectively. To facilitate detailed comparisons, we have divided the spectra into three key wavelength regions, following the same terminology of \citet{Blondin2023}: the optical (0.35--1\microns), the NIR (1--5~\microns), and the MIR (5--30~\microns).

\subsection{Angle-averaged spectra} 
\label{sec:angleaveraged}

\begin{figure*}
\centering
\begin{subfigure}{\textwidth}
    \centering
    \includegraphics[width=0.99\textwidth,height=6.5cm]{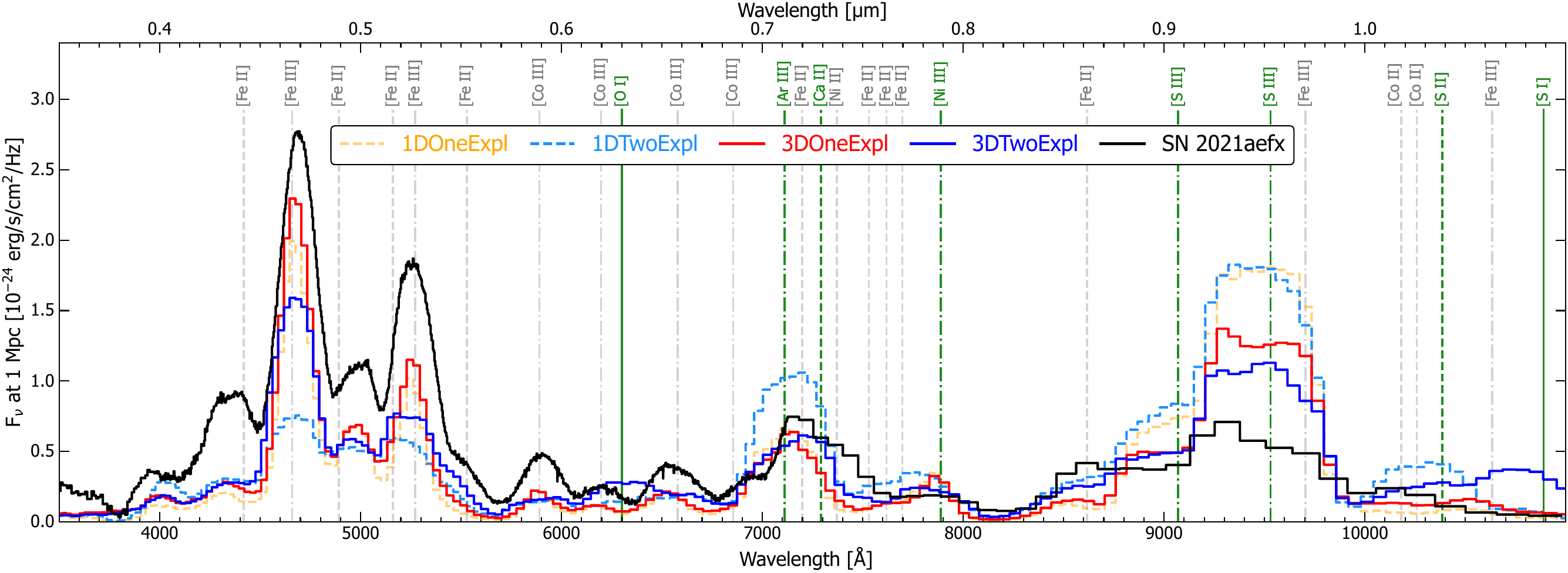}
\end{subfigure}

\vspace{1em}

\begin{subfigure}{\textwidth}
    \centering
    \includegraphics[width=0.99\textwidth,height=6.5cm]{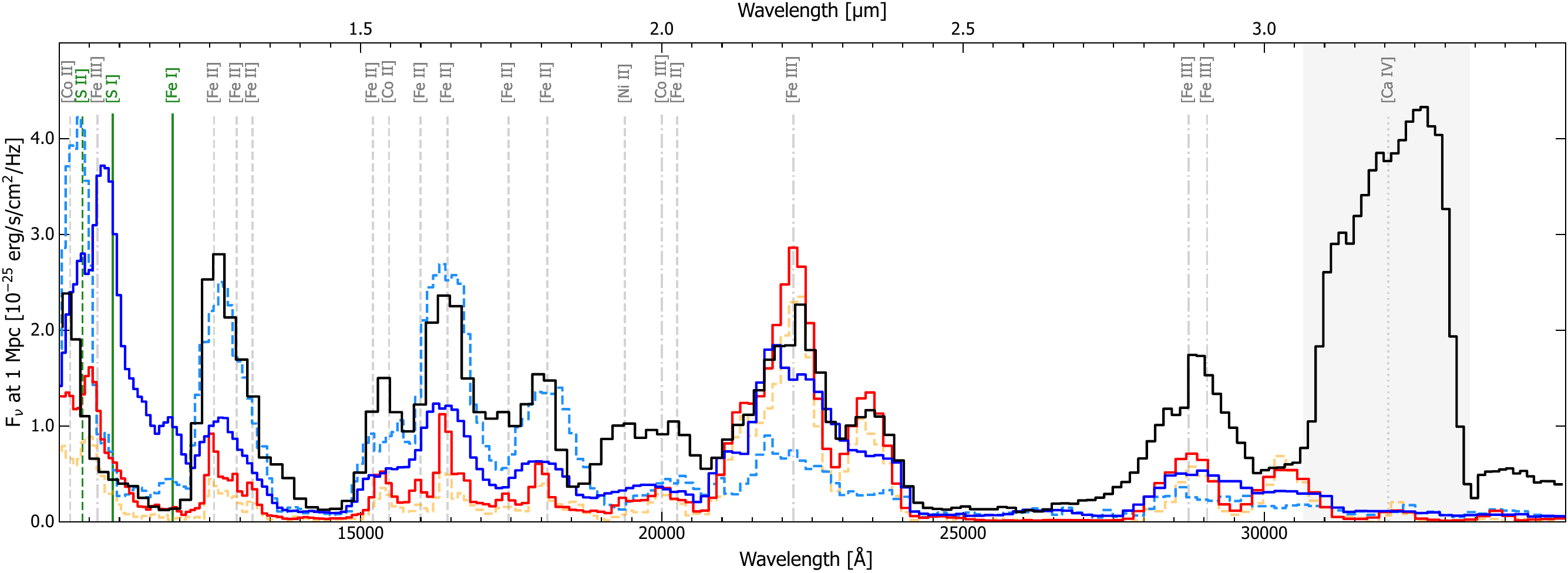}     
\end{subfigure}

\vspace{1em}

\begin{subfigure}{\textwidth}
    \centering
    \includegraphics[width=0.99\textwidth,height=6.5cm]{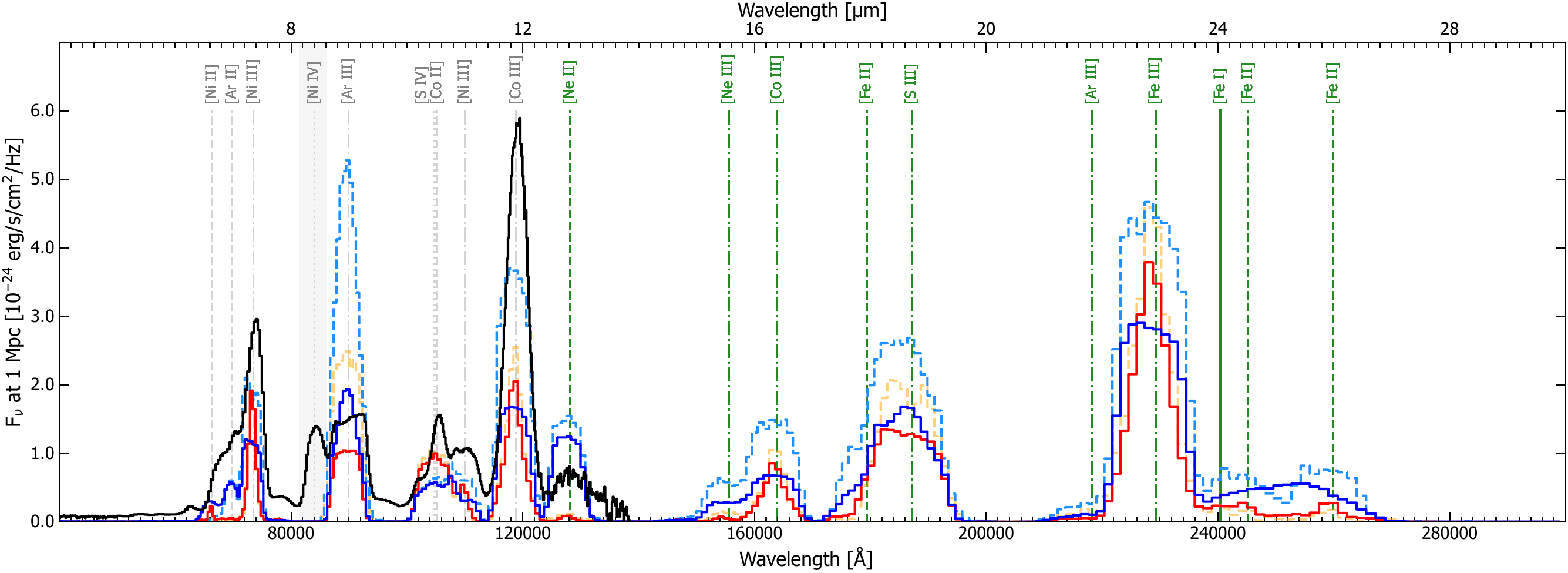}
    \label{fig:fir}
\end{subfigure}
        \caption{1D and angle-averaged 3D optical (top; $\sim$0.35–1\microns), NIR (middle; $\sim$1–4\microns), and MIR (bottom;$\sim$4–30\microns) spectra for the \oneO, \twoO, \one and \two models at 270 days post-explosion, compared to SN~2021aefx. Observed spectra are corrected for redshift and extinction \protect\citep{Hosseinzadeh2022}, and all spectra are scaled to a distance of 1 Mpc. Vertical grey lines indicate the rest wavelengths of prominent features identified by \protect\cite{Flores2020} and \protect\cite{Kwok2023}. In contrast, green lines highlight significant model features that diverge from observations and lie outside the spectral range of SN~2021aefx.
        The linestyles of the vertical lines indicate ionisation stages: solid for neutral species, dashed for singly ionised, dash-dotted for doubly ionised, and dotted for triply ionised species. Rest wavelengths identified for SN~2021aefx by \protect\cite{Kwok2023} are listed in Table~\ref{table:line_identifcation}, where we assess the presence (or absence) of species and the degree to which they are blended. Note that the shaded grey regions highlight prominent features that we do not reproduce due to their absence in our atomic data.}
    \label{fig:combined_spectra}
\end{figure*}

We begin our analysis of the \dsix scenario by examining the signatures from the 1D models and the angle-averaged signatures from 3D models, as shown in Figure~\ref{fig:combined_spectra}. Following the approach of \cite{Kwok2023} and \cite{Blondin2023}, we present the spectra in flux per unit frequency rather than per unit wavelength. This improves visual clarity and facilitates more direct comparisons between models and observations across the entire wavelength range (see Table~\ref{tab:flux_ratios} for a detailed breakdown of the flux distribution across different regions). We also summarise some of the prominent spectral lines observed in SNe~Ia in Table~\ref{table:line_identifcation}
\footnote{In this work, we refer to nebular features using the gf-weighted mean wavelength, while for plotting purposes we mark the strongest transition.}.

\begin{figure*}
\centering
\begin{subfigure}{\textwidth}
    \centering
    \includegraphics[width=0.99\textwidth]{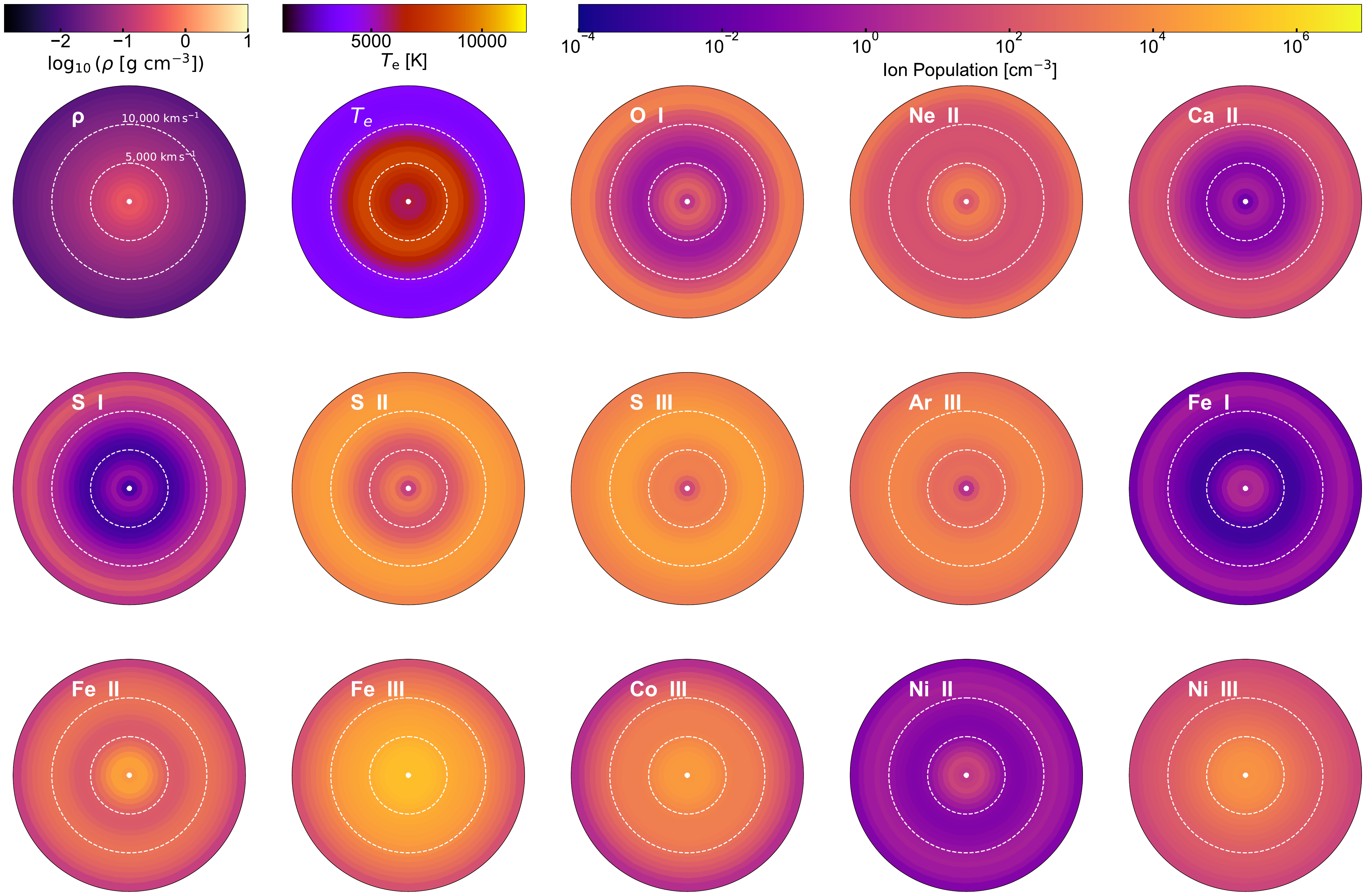}
\end{subfigure}

\caption{Ejecta properties for the \oneO model at 270 days. Each panel shows the 2D projection of the 1D property as a function of radial velocity: mass density ($\rho$), electron temperature ($T_{e}$), and ion populations of key species (\protect\ion{O}{I}, \protect\ion{Ne}{II}, \protect\ion{Ca}{II}, \protect\ion{S}{I--III}, \protect\ion{Ar}{III}, \protect\ion{Fe}{I--III} and \protect\ion{Ni}{II--III}). All panels share a common radial velocity scale, with inner and outer dashed circles marking velocities of 5,000$\,\mathrm{km}\,\mathrm{s}^{-1}$ and 10,000$\,\mathrm{km}\,\mathrm{s}^{-1}$ respectively. Colour bars are consistent with those in Figures~\ref{fig:3DOneExplionplot}, \ref{fig:1DTwoExplionplot} and \ref{fig:3DTwoExplionplot}, enabling direct comparison of ejecta properties between the 1D and 3D models. Note that a small fraction of cells ($\sim$1\%) possess populations below $10^{-4}$ and are clipped to this value to improve the overall clarity and allow for clearer comparisons.}

    \label{fig:1DOneExplionplot}
\end{figure*}

\begin{figure*}
\centering
\begin{subfigure}{\textwidth}
    \centering
    \includegraphics[width=0.99\textwidth]{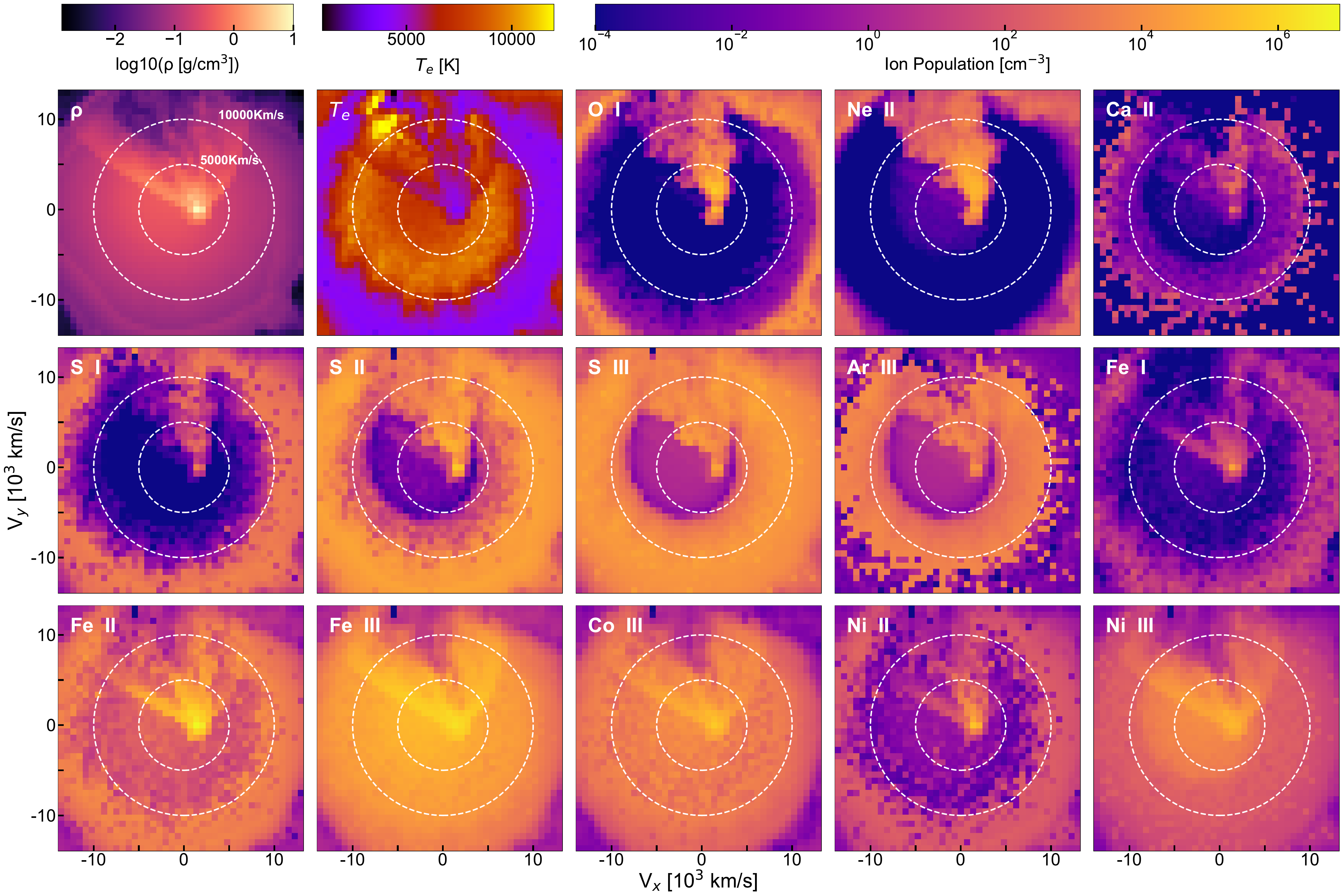}
\end{subfigure}

    \caption{Ejecta properties for a slice ($\cos(\theta)=0$; i.e., the merger plane) through the \one model at 270 days. Each panel shows a 2D slice of the 3D ejecta mapped into velocity space, displaying: mass density ($\rho$), electron temperature ($T_e$), and ion populations of key species (\protect\ion{O}{I}, \protect\ion{Ne}{II}, \protect\ion{Ca}{II}, \protect\ion{S}{I--III}, \protect\ion{Ar}{III}, \protect\ion{Fe}{I--III}, and \protect\ion{Ni}{II--III}). Dashed circles mark radial velocities of 5,000$\,\mathrm{km}\,\mathrm{s}^{-1}$ and 10,000$\,\mathrm{km}\,\mathrm{s}^{-1}$. Colour bars match those in Figures~\ref{fig:1DOneExplionplot}, \ref{fig:1DTwoExplionplot} and \ref{fig:3DTwoExplionplot}, enabling consistent comparison between 1D and 3D models. We note a small fraction of cells ($\sim$1\%) have ion populations below $10^{-4}$. These ion populations are clipped at this threshold, as such low populations have a negligible impact on the spectra. Additionally, some ($\sim$0.3\%) outer (\textgreater10,000$\,\mathrm{km}\,\mathrm{s}^{-1}$) cells in low-density regions possess temperatures above 12,000 K. As such, we also clip these cell temperatures to this value to improve the overall clarity and allow for clearer comparisons.}
    \label{fig:3DOneExplionplot}
\end{figure*}

\begin{figure*}
\centering
\begin{subfigure}{\textwidth}
    \centering
    \includegraphics[width=0.99\textwidth]{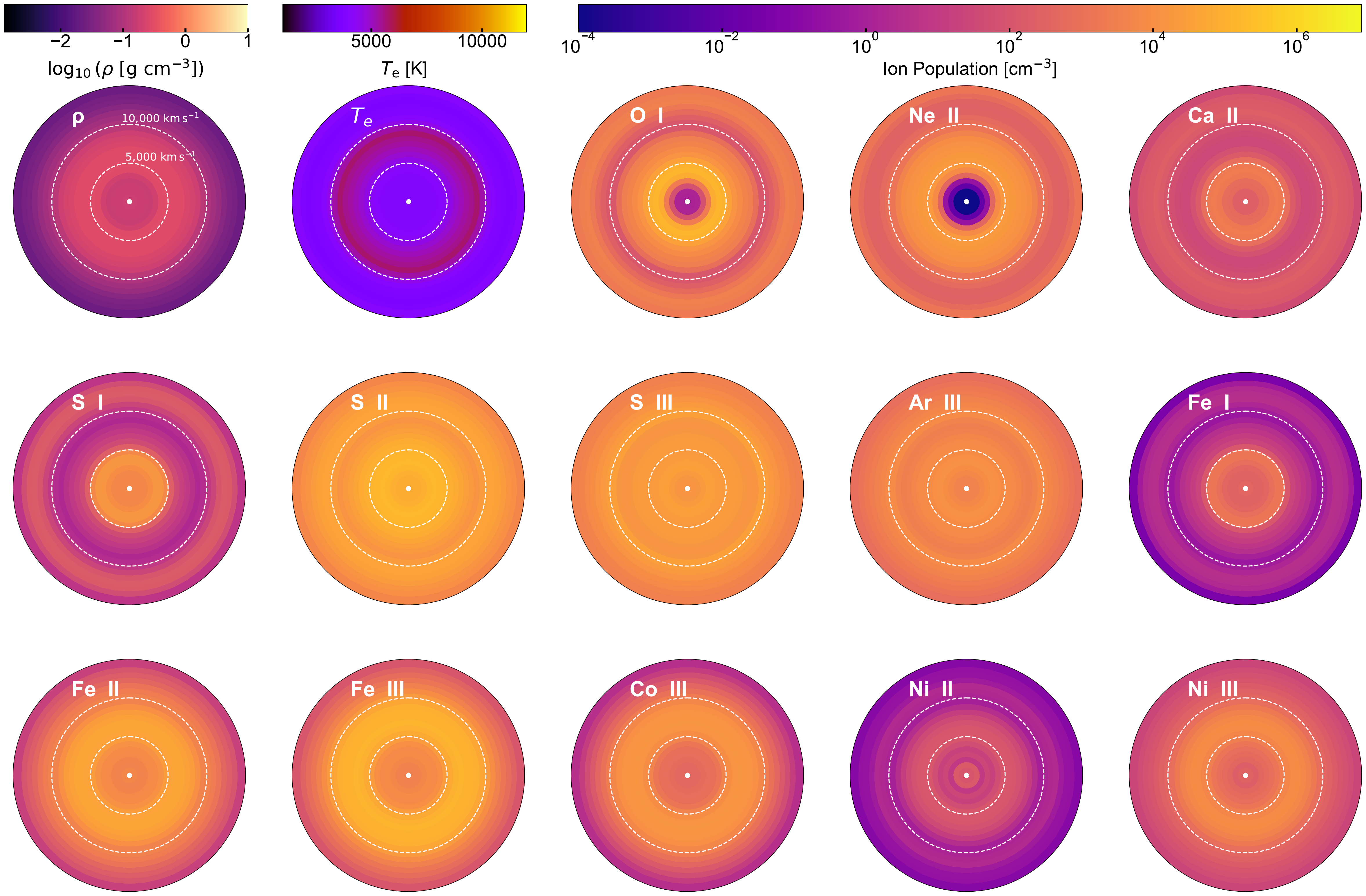}    
\end{subfigure}

    \caption{Same as Figure~\ref{fig:1DOneExplionplot} but for the \twoO model}
    \label{fig:1DTwoExplionplot}
\end{figure*}

\begin{figure*}
\centering
\begin{subfigure}{\textwidth}
    \centering
    \includegraphics[width=0.99\textwidth]{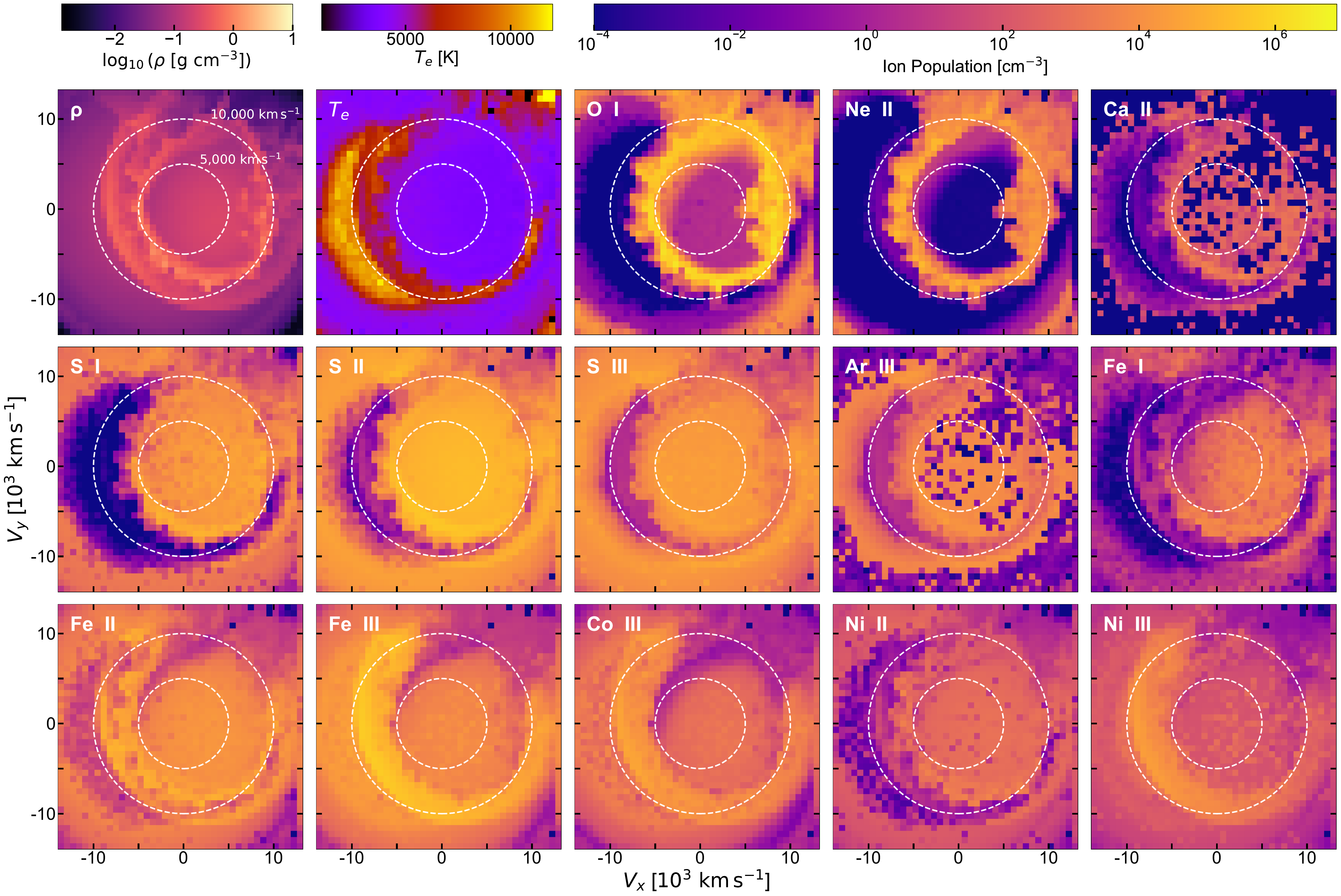}
\end{subfigure}

    \caption{Same as Figure~\ref{fig:3DOneExplionplot} but for the \two model}
    \label{fig:3DTwoExplionplot}
\end{figure*}

\begin{table*}
\caption{
Flux distribution ratios across the optical, NIR, and Lower MIR bands for each model and SN~2021aefx \protect\citep{Kwok2023}. The ratios are calculated by integrating $F_{\lambda} d\lambda$ over the full observed spectrum and then normalising to the total contribution from all bands.
}
\centering
\begin{tabular}{lccccc}
\hline
Range & \oneO & \one & \twoO & \two & SN~2021aefx \\ \hline
Optical (0--1 $\mu$m)      & 0.90 & 0.91 & 0.80 & 0.84 & 0.86 \\
NIR (1--5 $\mu$m)         & 0.03 & 0.04 & 0.07 & 0.07 & 0.06 \\
Lower MIR ($>$5--14 $\mu$m) & 0.07 & 0.05 & 0.13 & 0.09 & 0.08 \\ \hline
\end{tabular}
\label{tab:flux_ratios}
\end{table*}

\renewcommand{\arraystretch}{1.2}
\begin{table*}
\color{black}
\caption{Prominent emission lines of SN~2021aefx across the optical, NIR, and MIR wavelength ranges (see \protect\citealt{Kwok2023}). Following the notation of \protect\cite{Kwok2023}, a `?' denotes tentatively identified transitions in the observations. To evaluate the synthetic spectra for the \oneO, \one, \twoO, and \two models, we use Monte Carlo packet information to determine whether a given transition is present. For each line, we identify the observer-frame wavelength interval over which packets are tagged as having their last interaction with that transition. This interval defines the Full Width at Zero Intensity, representing the wavelength range where the feature merges into the background and provides a measure of overall blending.
We then compute the total transition specific integrated flux across this wavelength interval and report it at a distance of 1~Mpc. To assess the relative strength of the feature, we then express the contribution of the transition specific integrated flux as a percentage of the total integrated flux within the same interval at the same distance, in order to examine a feature's overall purity. 
A $\times$ denotes the absence of the species in the synthetic spectra, while a $\circ$ indicates that the feature is missing owing to its absence from our atomic dataset; in such cases, no conclusion regarding its physical presence or absence should be drawn.}
\begin{tabular}{ll *{4}{cc}}
\toprule
& & \multicolumn{2}{c}{\oneO} & \multicolumn{2}{c}{\one} & \multicolumn{2}{c}{\twoO} & \multicolumn{2}{c}{\two} \\
\cmidrule(lr){3-4} \cmidrule(lr){5-6} \cmidrule(lr){7-8} \cmidrule(lr){9-10}
$\lambda_{\rm rest}$ & Species
  & Flux [erg\,s$^{-1}$\,cm$^{-2}$]& \%
  & Flux [erg\,s$^{-1}$\,cm$^{-2}$]& \%
  & Flux [erg\,s$^{-1}$\,cm$^{-2}$]& \%
  & Flux [erg\,s$^{-1}$\,cm$^{-2}$]& \% \\ \hline
\multicolumn{10}{c}{Optical + NIR} \\ \hline
0.589  & [Co III]  & $2.4\times10^{-23}$ & 48.7 & $2.3\times10^{-23}$ & 29.4 & $2.7\times10^{-23}$ & 33.9 & $2.4\times10^{-23}$ & 31.2 \\
0.716  & [Fe II]   & $0.9\times10^{-23}$ &  3.6 & $1.6\times10^{-23}$ &  6.9 & $4.5\times10^{-23}$ &  9.0 & $2.1\times10^{-23}$ &  6.4 \\
0.738  & [Ni II]   & $0.4\times10^{-24}$ &  0.4 & $4.0\times10^{-24}$ &  2.8 & $1.6\times10^{-24}$ &  0.9 & $9.7\times10^{-24}$ &  4.4 \\
1.257  & [Fe II]   & $1.0\times10^{-23}$ & 57.2 & $1.9\times10^{-23}$ & 46.2 & $9.6\times10^{-23}$ & 43.0 & $3.8\times10^{-23}$ & 48.1 \\
1.547  & [Co II]   & $0.9\times10^{-24}$ & 11.0 & $2.9\times10^{-24}$ & 10.3 & $5.7\times10^{-24}$ &  6.6 & $4.2\times10^{-24}$ &  6.7 \\
1.644  & [Fe II]   & $1.7\times10^{-23}$ & 56.9 & $3.1\times10^{-23}$ & 49.4 & $1.5\times10^{-22}$ & 57.3 & $6.1\times10^{-23}$ & 50.7 \\
1.939  & [Ni II]   & $0.5\times10^{-24}$ & 27.6 & $5.5\times10^{-24}$ & 21.9 & $0.2\times10^{-24}$ & 17.1 & $1.0\times10^{-23}$ & 15.7 \\
2.219  & [Fe III]  & $9.3\times10^{-23}$ & 56.1 & $1.3\times10^{-22}$ & 52.9 & $4.8\times10^{-23}$ & 41.6 & $1.0\times10^{-22}$ & 39.9 \\
2.874  & [Fe III]  & $3.3\times10^{-24}$ & 31.7 & $6.2\times10^{-24}$ & 18.6 & $1.1\times10^{-24}$ & 14.3 & $7.5\times10^{-24}$ & 13.4 \\
2.905  & [Fe III]  & $2.0\times10^{-23}$ & 20.3 & $1.8\times10^{-23}$ & 26.2 & $2.1\times10^{-23}$ & 72.7 & $3.0\times10^{-23}$ & 42.1 \\ \hline

\multicolumn{10}{c}{MIR} \\ \hline
6.636  & [Ni II]   & $0.6\times10^{-23}$ & 100 & $3.1\times10^{-22}$ & 87.5 & $1.7\times10^{-22}$ & 57.8 & $8.6\times10^{-22}$ & 60.7 \\
6.985  & [Ar II]   & $1.1\times10^{-22}$ & 45.5 & $1.3\times10^{-22}$ & 32.8 & $1.5\times10^{-21}$ & 40.8 & $1.5\times10^{-21}$ & 32.7 \\
7.349  & [Ni III]  & $3.1\times10^{-21}$ & 99.0 & $3.0\times10^{-21}$ & 96.7 & $6.0\times10^{-21}$ & 92.8 & $4.0\times10^{-21}$ & 89.3 \\
8.405  & [Ni IV]   & $\circ$ & $\circ$ & $\circ$ & $\circ$ & $\circ$ & $\circ$ & $\circ$ & $\circ$ \\
8.991  & [Ar III]  & $1.3\times10^{-20}$ & 100 & $0.6\times10^{-20}$ & 99.8 & $2.3\times10^{-20}$ & 99.9 & $0.9\times10^{-20}$ & 99.7 \\
10.510 & [S IV]    & $5.6\times10^{-21}$ & 85.4 & $5.2\times10^{-21}$ & 69.1 & $3.1\times10^{-21}$ & 57.6 & $2.9\times10^{-21}$ & 57.8 \\
10.521 & [Co II]   & $0.3\times10^{-22}$ & 7.9 & $2.6\times10^{-22}$ & 5.0 & $6.1\times10^{-22}$ & 15.3 & $4.8\times10^{-22}$ & 12.9 \\
11.002 & [Ni III]  & $0.8\times10^{-21}$ & 45.7 & $1.1\times10^{-21}$ & 68.6 & $2.6\times10^{-21}$ & 17.9 & $1.5\times10^{-21}$ & 20.7 \\
11.888 & [Co III]  & $0.8\times10^{-20}$ & 97.1 & $0.7\times10^{-20}$ & 96.4 & $2.0\times10^{-20}$ & 79.0 & $1.0\times10^{-20}$ & 89.5 \\ \hline

\multicolumn{10}{c}{Tentative} \\ \hline
2.911  & [Ni II]?  & $1.1\times10^{-24}$ & 10.1 & $0.9\times10^{-24}$ & 1.5 & $\times$ & $\times$ & $0.8\times10^{-24}$ & 1.7 \\
3.044  & [Fe III]? & $5.4\times10^{-23}$ & 94.6 & $5.4\times10^{-23}$ & 70.3 & $1.4\times10^{-23}$ & 60.5 & $3.5\times10^{-23}$ & 51.7 \\
6.214  & [Co II]?  & $\circ$ & $\circ$ & $\circ$ & $\circ$ & $\circ$ & $\circ$ & $\circ$ & $\circ$ \\
6.920  & [Ni II]?  & $\times$ & $\times$ & $\times$ & $\times$ & $\times$ & $\times$ & $7.7\times10^{-24}$ & 0.7 \\
7.791  & [Fe III]? & $5.6\times10^{-23}$ & 87.3 & $6.5\times10^{-23}$ & 81.4 & $3.7\times10^{-23}$ & 76.2 & $6.9\times10^{-23}$ & 59.9 \\
10.682 & [Ni II]?  & $\times$ & $\times$ & $4.8\times10^{-23}$ & 1.0 & $3.0\times10^{-23}$ & 1.0 & $8.9\times10^{-23}$ & 2.5 \\
11.167 & [Co II]?  & $1.6\times10^{-23}$ & 49.0 & $0.4\times10^{-23}$ & 1.5 & $1.7\times10^{-23}$ & 3.6 & $1.5\times10^{-23}$ & 1.1 \\
12.729 & [Ni II]?  & $\times$ & $\times$ & $3.7\times10^{-23}$ & 14.2 & $2.3\times10^{-23}$ & 2.7 & $8.3\times10^{-23}$ & 1.5 \\ \hline
\end{tabular}
\label{table:line_identifcation}
\end{table*}

To enable a quantitative comparison between the models, we use the Monte Carlo packet data to identify the emitting regions associated with the transitions listed in Table~\ref{table:line_identifcation}. For each transition, we determine the Full Width at Zero Intensity, defining the wavelength interval over which the feature merges into the background, and integrate both the transition-specific flux and the total flux within this interval, and compute the contribution of the transition to the overall emission in that wavelength range. This provides a direct measure of the degree to which features are blended. We emphasise that this metric is not normalised to SN~2021aefx, as a normalisation would not provide a consistent basis for comparison across different SNe~Ia. Consequently, while this approach quantifies the relative contribution of a transition within a given wavelength range, it does not by itself determine whether a feature would be detectable in a particular observed spectrum. We therefore also report the total integrated flux of the specific transition under consideration in order to aid assessment of whether the transition would be detectable in an observed spectrum. Note that the synthetic spectrum should also be considered when assessing the detectability of a particular feature and the possible contribution of other transitions within the same wavelength interval\footnote{Note that the synthetic spectrum may also display emission from the same ionisation stage; however, the values presented in Table~\ref{table:line_identifcation}  specifically evaluate the presence of the specified transitions identified by \cite{Kwok2023}.}.
Overall, the explosion models recover many of the transitions reported by \cite{Kwok2023}; however, substantial blending remains evident across several wavelength intervals, particularly in the optical and NIR, while blending generally decreases towards longer wavelengths.

To better understand the differences between the 1D and 3D models we have extracted key physical properties from the models and radiative transfer calculations in Figures~\ref{fig:1DOneExplionplot}, \ref{fig:3DOneExplionplot}, \ref{fig:1DTwoExplionplot} and \ref{fig:3DTwoExplionplot}. These figures display the following physical quantities: density ($\rho$), the electron temperature ($T_{e}$) and ion populations (\ion{O}{I}, \ion{Ne}{II}, \ion{Ca}{II}, \ion{S}{I--III}, \ion{Ar}{III}, \ion{Fe}{I--III}, \ion{Co}{III}, and \ion{Ni}{II--III}). For the 1D models, which represent spherical symmetry, we project the extracted properties into 2D, for ease of visual comparison. For the 3D calculations, we extract a 2D slice through the merger plane. 
The multidimensional treatments display significant deviations from spherical symmetry in their respective ion distributions, with the \two model exhibiting more pronounced asymmetries than the \one model.

Despite the secondary WD detonation significantly altering the explosion geometry, both scenarios yield broadly similar angle-averaged synthetic spectra, although neither reproduces all observed features of SN~2021aefx at the correct strengths. The different profiles arise due to different geometries and chemical stratifications. As such, line profiles can appear broad and flat-topped (e.g., the \ion{Ar}{III} 8.991\microns feature in the \one model), narrow and centrally peaked (e.g., the \ion{Co}{III} 11.888\microns feature in the \one model), or a blend of both (e.g., the \ion{Fe}{III} 22.925\microns feature in the \two model). We emphasise, however, that the angle-averaged spectrum represents a superposition of multiple viewing angles and does not correspond to any one line-of-sight from the 3D calculation. Consequently, the morphology of a feature in the averaged 3D spectrum must be interpreted with caution. For example, a broad or flat-topped profile does not uniquely imply emission from a geometrically thin shell, but may instead arise from intrinsically narrower features whose line centres shift with observer orientation. This effect is particularly important for iron-group elements (IGEs), such as \ion{Co}{III}, in the \two model, where significant velocity offsets between viewing angles can artificially impact the averaged profile but are also important for intermediate-mass elements (IMEs), such as \ion{Ar}{III}. For this reason, direct comparisons of line shapes between 1D models and angle-averaged 3D models can be misleading. Instead, the integrated line flux provides a more robust diagnostic, as in optically thin ejecta, it should be independent of viewing angle, except in wavelength regions where residual optical depth effects remain important.

\begin{figure} 
    \centering
    \includegraphics[height=6cm,width=0.49\textwidth]{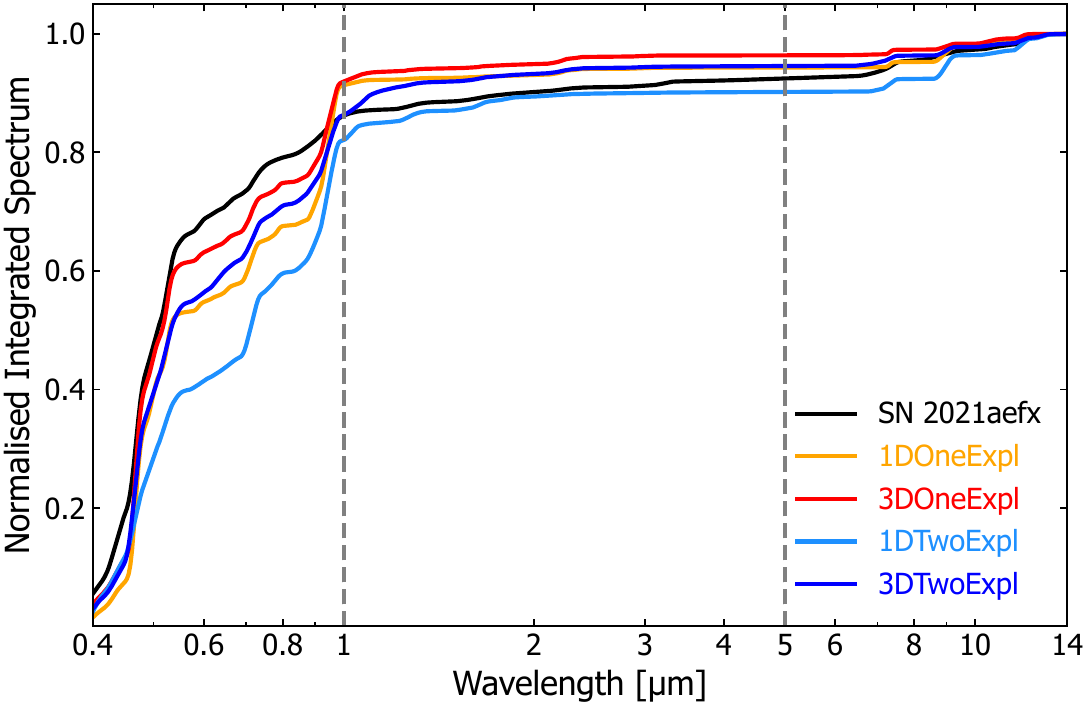}    
    \includegraphics[height=6cm,width=0.49\textwidth]{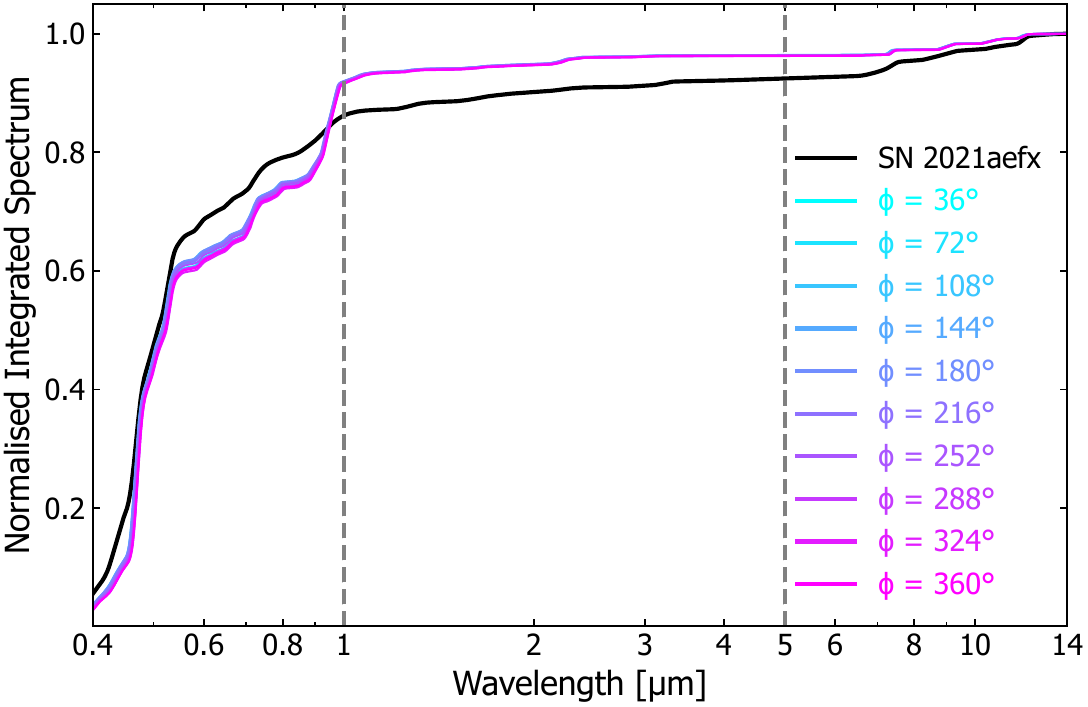}
    \includegraphics[height=6cm,width=0.49\textwidth]{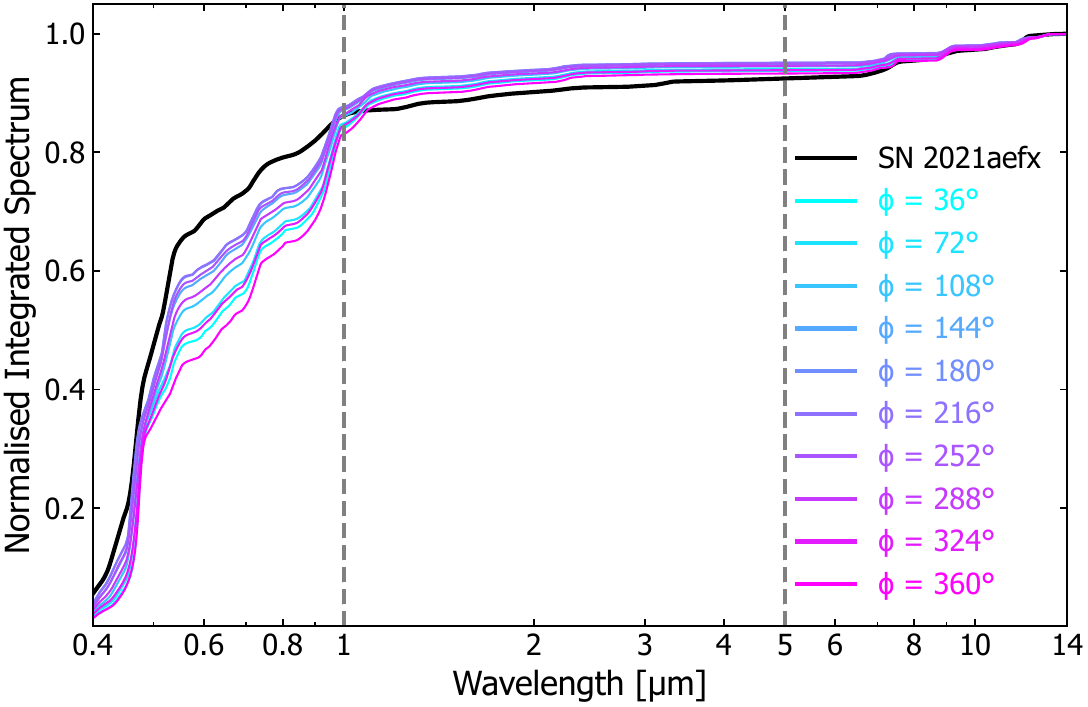}
    
    \caption{Integral of the normalised cumulative flux per unit wavelength over the range 0.4–14\microns for both explosion scenarios and SN~2021aefx. Dashed vertical lines indicate the boundaries between spectral regions as defined in Section~\ref{sec:Results}. The top panel shows the normalised cumulative flux for the angle-averaged and spherically averaged cases. The middle and bottom panels show the normalised cumulative flux for different observer orientations in the merger plane for the \one and \two models, respectively.}
    \label{fig:integrated_spectra}
\end{figure}

\begin{figure*} 
    \centering    \includegraphics[width=0.99\textwidth,height=5cm]{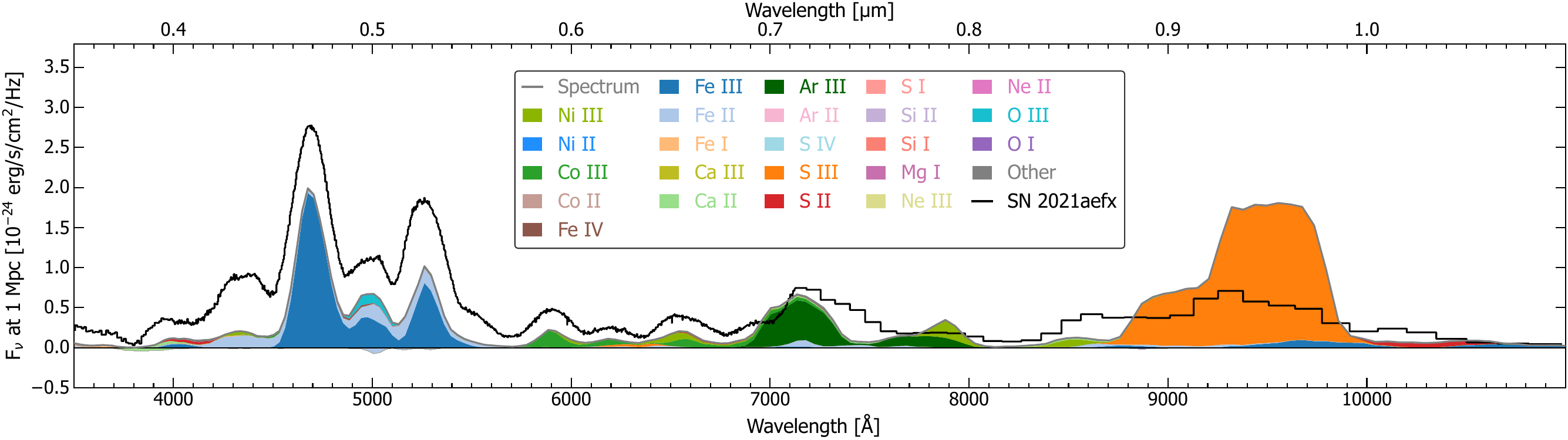}
    \vspace{0.5em}
    
    \includegraphics[width=0.99\textwidth,height=5cm]{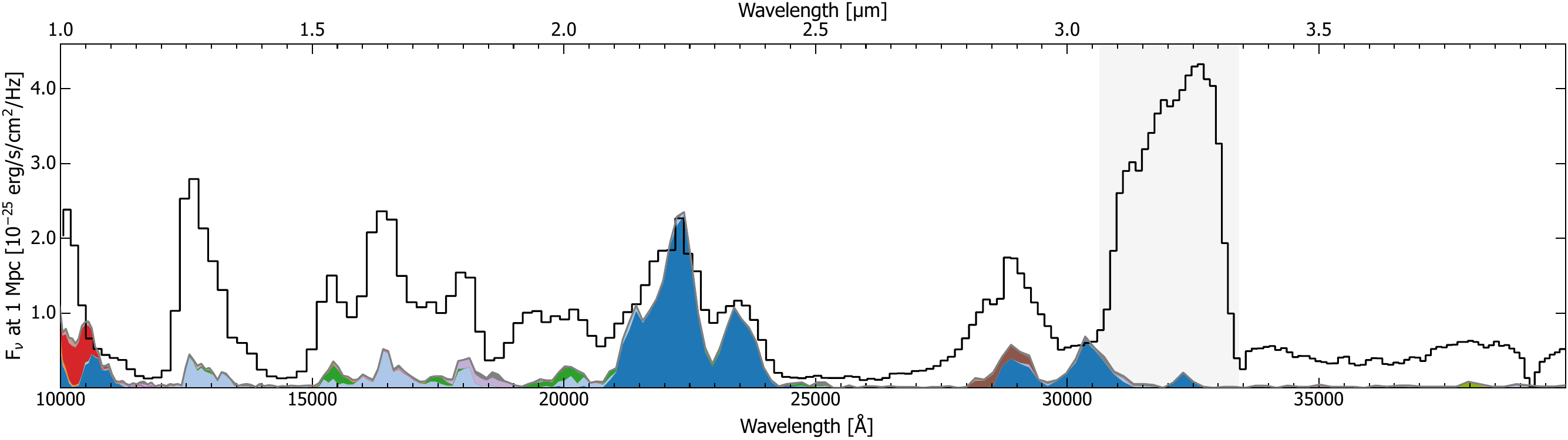}     
    \vspace{0.5em}
    
    \includegraphics[width=0.99\textwidth,height=5cm]{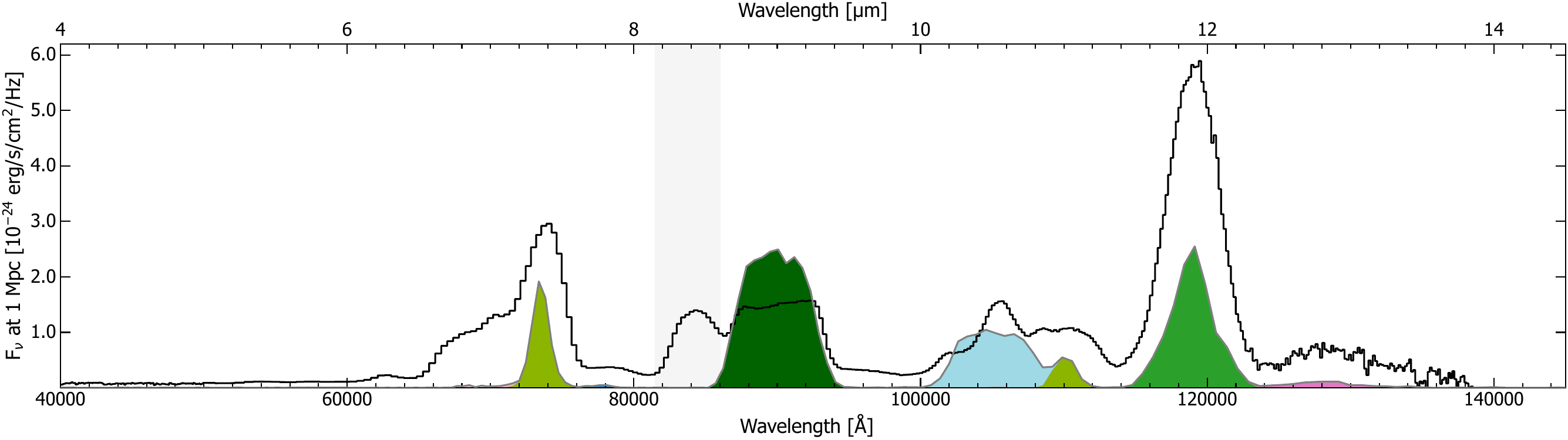}
    \vspace{0.5em}
    
    \includegraphics[width=0.99\textwidth,height=5cm]{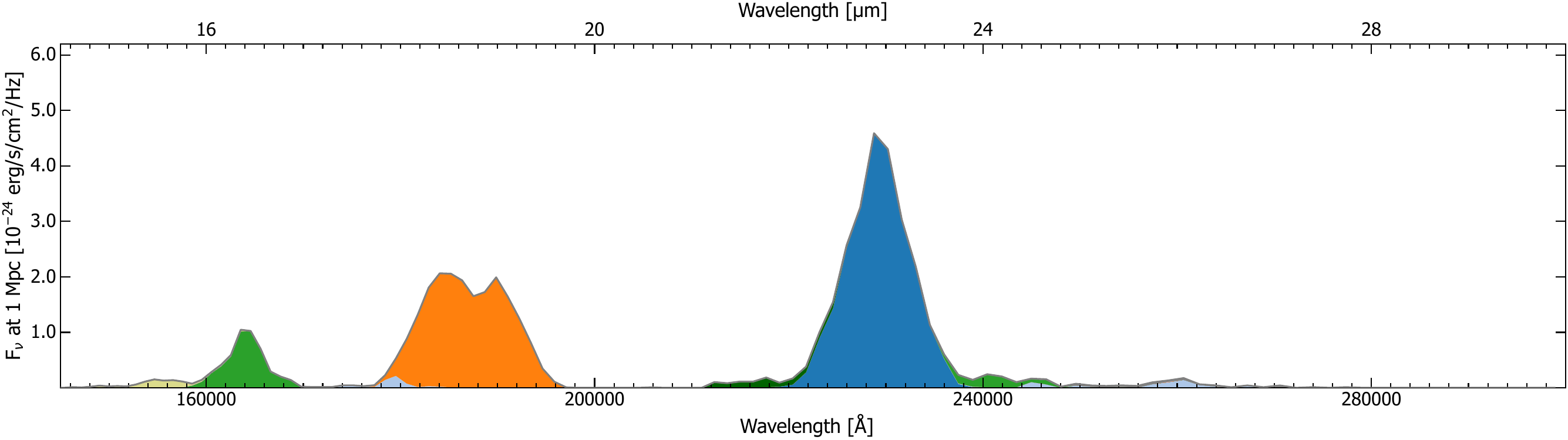}
    
    \caption{
    Late-time emission and absorption spectra for the \oneO model at 270 days across all wavelength ranges, from top to bottom: optical, NIR, lower MIR ($\sim$4–15\microns), and upper MIR ($\sim$15–30~\microns).
    The positive axis is colour-coded to indicate the emitting ions, based on each Monte Carlo packet’s last interaction. The negative axis shows the corresponding absorption contributions from each ion, which only appear in the optical region.    
    The total spectrum is overlaid as a grey curve, with the shaded regions indicating the contribution of individual ions. Observations of SN~2021aefx \protect\cite{Kwok2023} are included for comparison. Note that the shaded grey regions highlight prominent features that we do not reproduce due to their absence in our atomic data.}
    \label{fig:Kromer_1DOneExpl}
\end{figure*}

\begin{figure*} 
    \centering
    \includegraphics[width=0.99\textwidth,height=5cm]{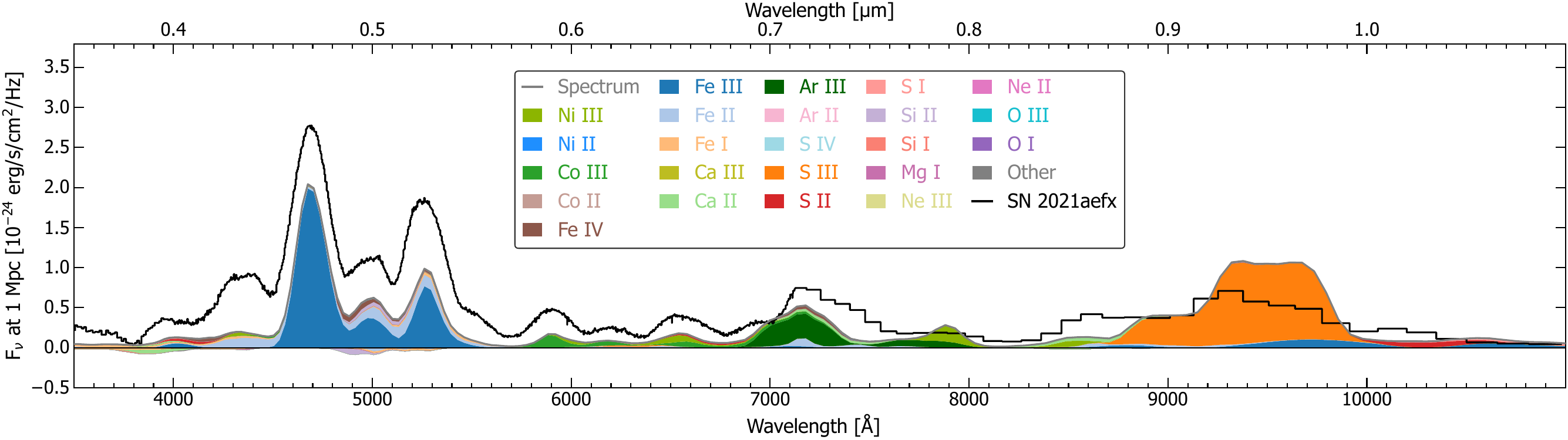}
    \vspace{0.5em}
    
    \includegraphics[width=0.99\textwidth,height=5cm]{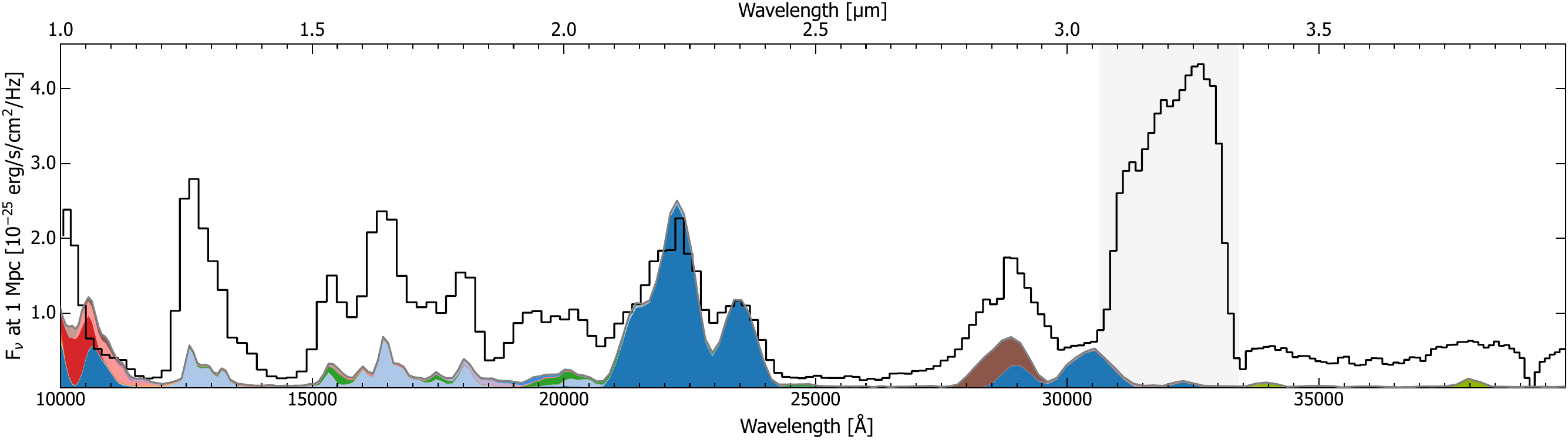}
    \vspace{0.5em}
    
    \includegraphics[width=0.99\textwidth,height=5cm]{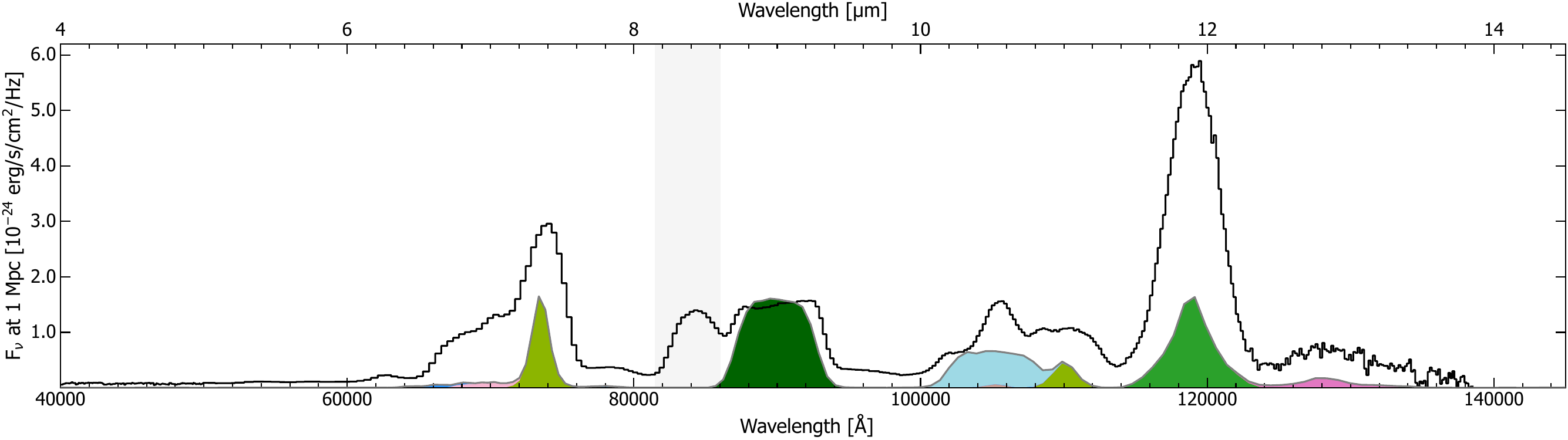}
    \vspace{0.5em}
    
    \includegraphics[width=0.99\textwidth,height=5cm]{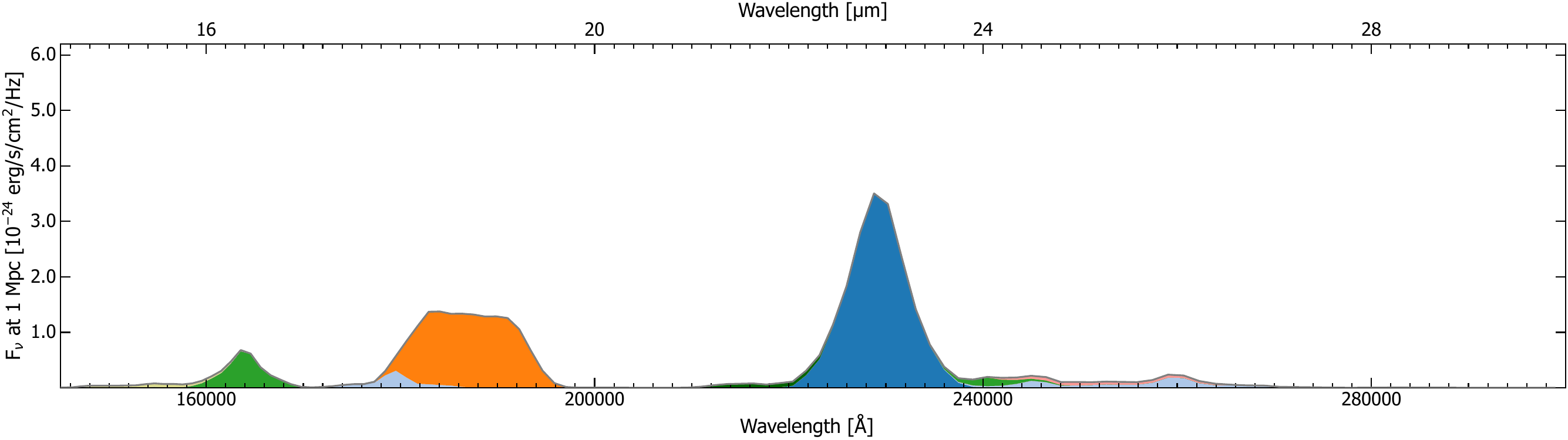}
    
    \caption{Same as Figure~\ref{fig:Kromer_1DOneExpl} but for the \one model}
    \label{fig:Kromer_3DOneExpl}
\end{figure*}

\begin{figure*} 
    \centering
    \includegraphics[width=0.99\textwidth,height=5cm]{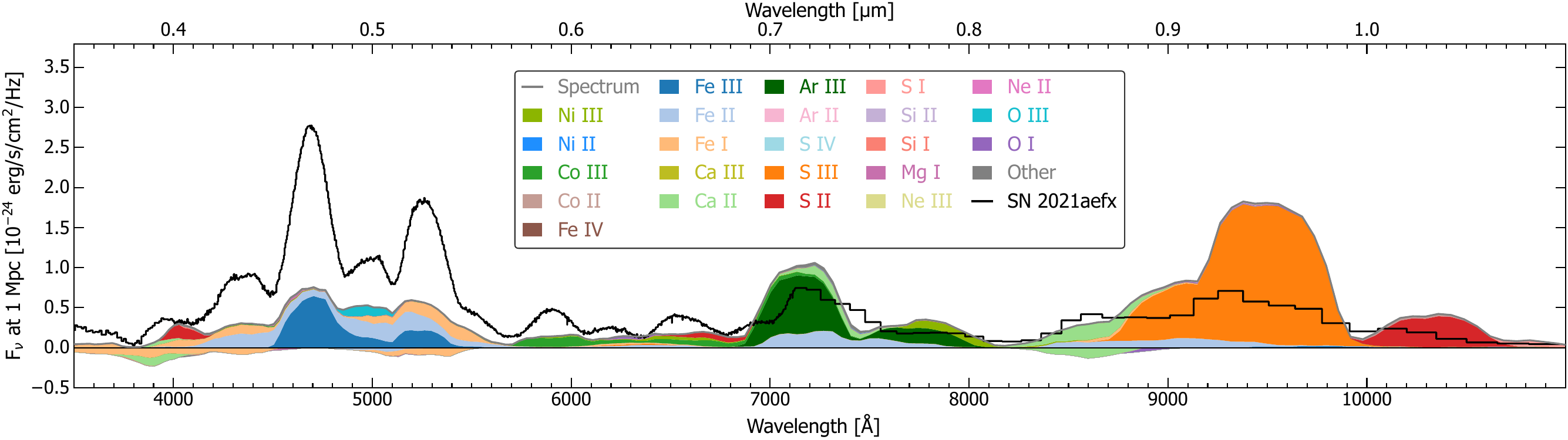}
    \vspace{0.5em}
    
    \includegraphics[width=0.99\textwidth,height=5cm]{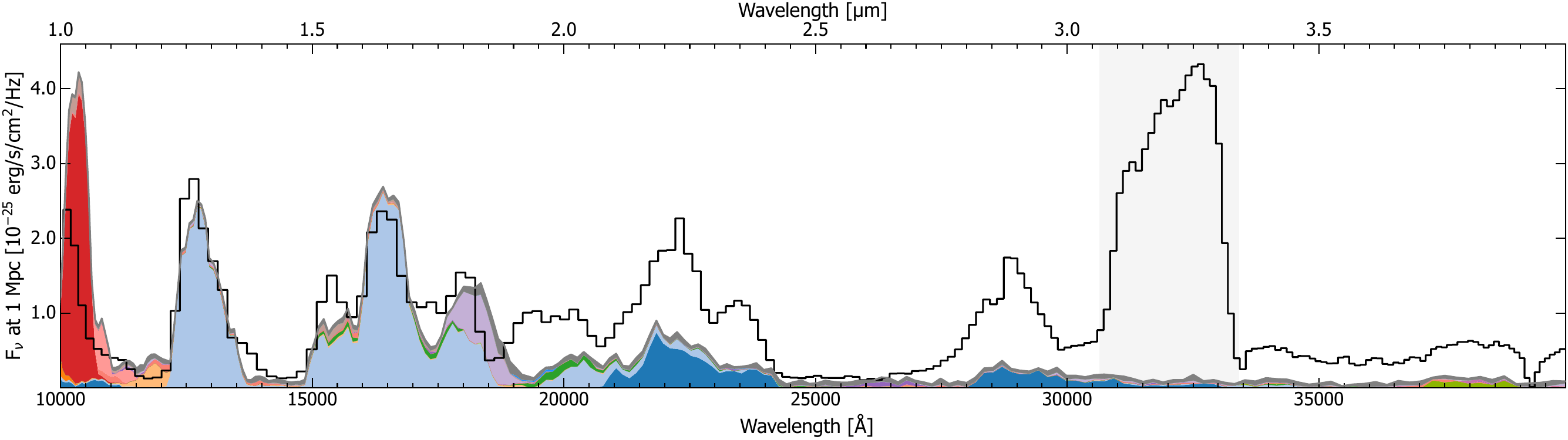}  
    \vspace{0.5em}
    
    \includegraphics[width=0.99\textwidth,height=5cm]{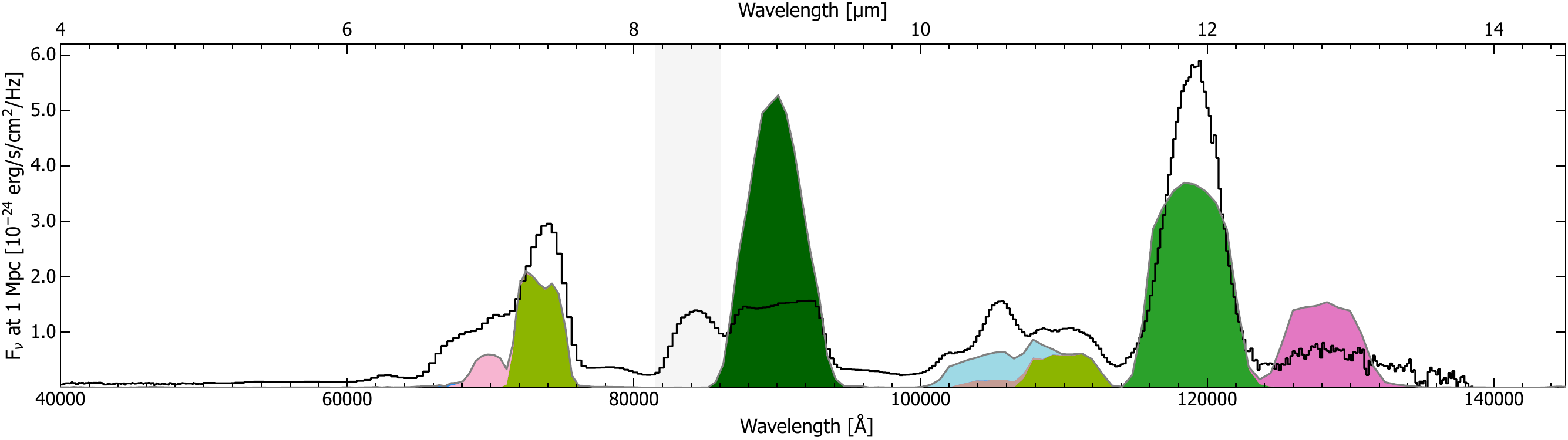}
    \vspace{0.5em}
    
    \includegraphics[width=0.99\textwidth,height=5cm]{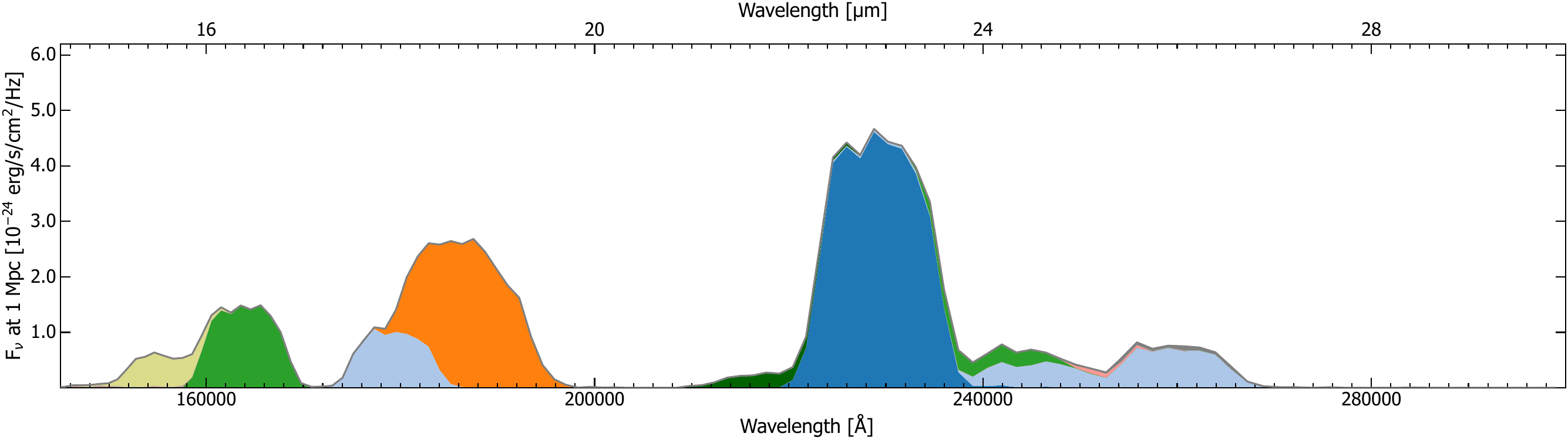}
    
    \caption{Same as Figure~\ref{fig:Kromer_1DOneExpl} but for the \twoO model}
    \label{fig:Kromer_1DTwoExpl}
\end{figure*}

\begin{figure*} 
    \centering
    \includegraphics[width=0.99\textwidth,height=5cm]{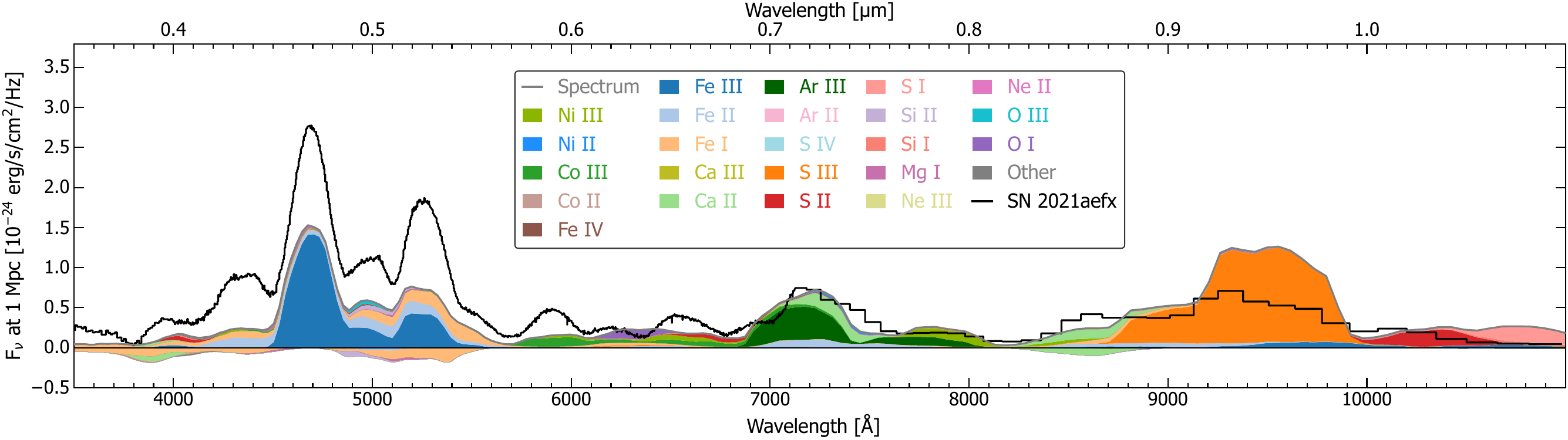}
    
    \includegraphics[width=0.99\textwidth,height=5cm]{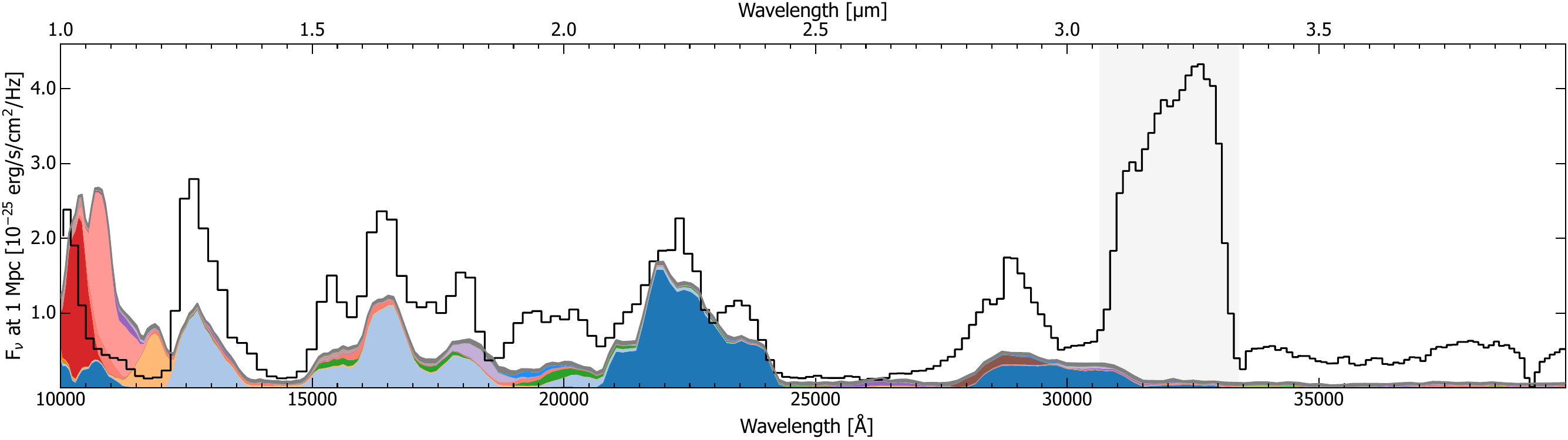}   
    
    \includegraphics[width=0.99\textwidth,height=5cm]{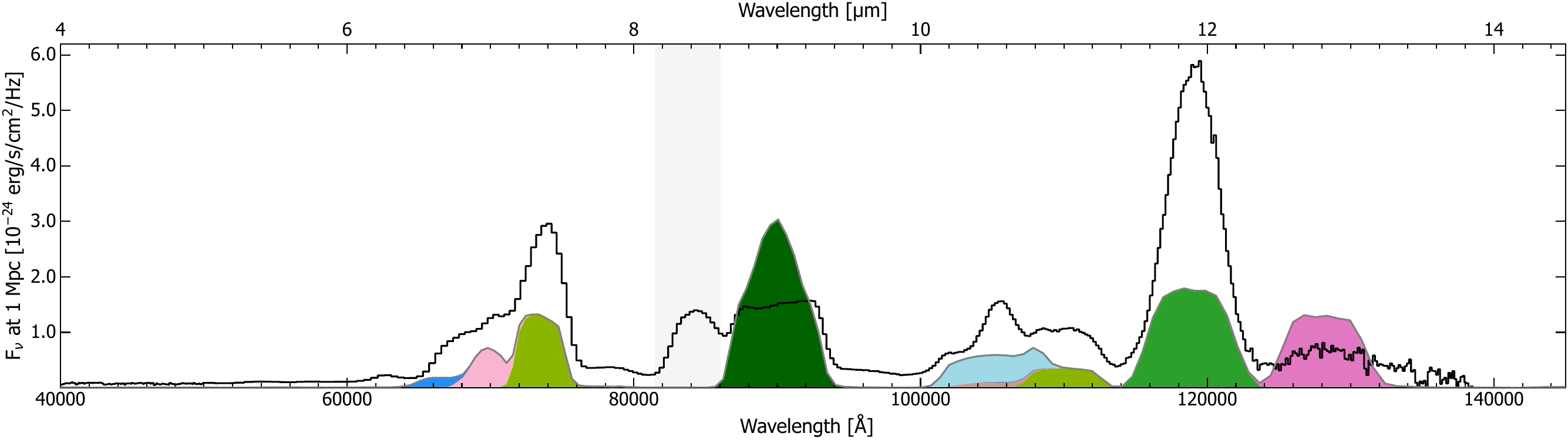}
    
    \includegraphics[width=0.99\textwidth,height=5cm]{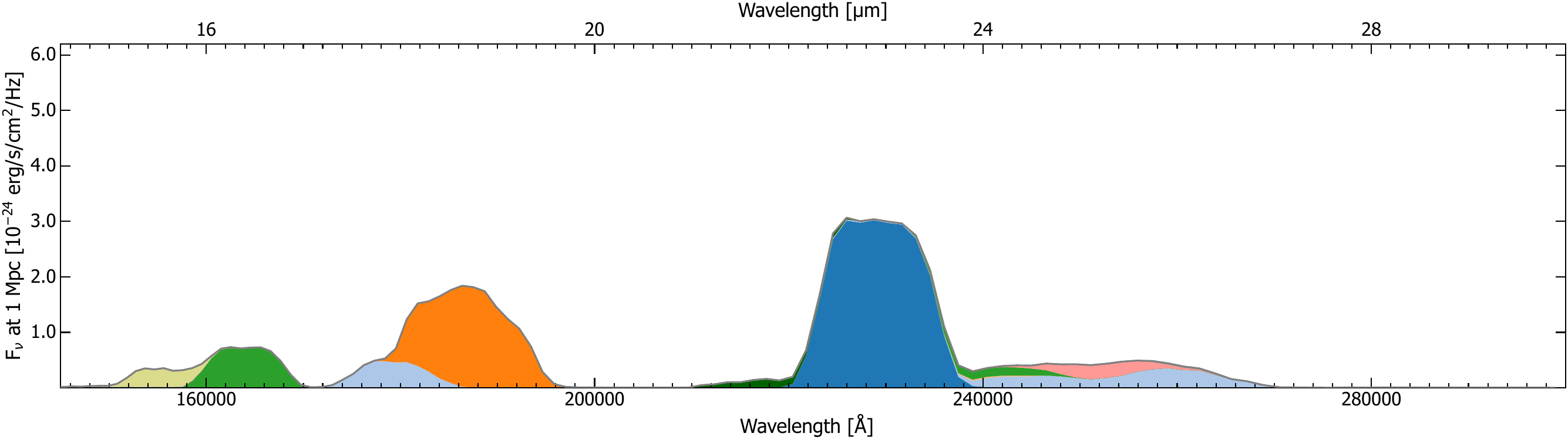}
    
    \caption{Same as Figure~\ref{fig:Kromer_1DOneExpl} but for the \two model}
    \label{fig:Kromer_3DTwoExpl}
\end{figure*}

To place our results in context, we compare to the normal SN~2021aefx \citep{Kwok2023} at 270 days post explosion during its nebular phase. The observational data have been corrected for a redshift of $z = 0.005017$ and for reddening due to host galaxy extinction, $E(B-V)_{\text{host}} = 0.097$ mag \citep{Hosseinzadeh2022}, as well as Milky Way extinction, $E(B-V)_{\text{MW}} = 0.008$ mag \citep{Schlafly2011}. 
We note that the reported peak $B$-band magnitude of SN~2021aefx ranges from $-19.28$ \citep{Ashall2022} to $-19.62$ \citep{Hosseinzadeh2022}. For further details on the data calibration and collection of the nebular spectrum, see \cite{Kwok2023}. As noted by \cite{Blondin2023}, the distance uncertainty of $\sim$2\,Mpc leads to inferred $^{56}$Ni masses ranging from $\lesssim 0.6$ to $>0.8\,\mathrm{M_\odot}$. Both explosion scenarios considered here produce nearly identical $^{56}$Ni yields, as this is primarily set by the mass of the primary white dwarf. The \onescenario and \twoscenario scenarios possess 0.45\,M$_\odot$ and 0.46\,M$_\odot$ of $^{56}$Ni, respectively. After removing the line-blanketing effects of the helium shell detonation, the peak $B$-band magnitudes of the \one and \two models are $-19.04$ and $-19.26$, respectively \citep{Pollin2024a}. However, these correspond to only a few lines of sight; on average, our explosion models are expected to be slightly too faint to match SN~2021aefx. We note that \cite{Blondin2023} considered models with $^{56}$Ni masses in the range 0.5--0.8\,M$_\odot$ as suitable progenitor candidates for SN~2021aefx. Therefore, our models are only marginally fainter than those considered in previous investigations.

\subsubsection{Optical Comparison}
\label{sec:average_spectra_optical}

The optical spectra of both explosion models are dominated by doubly ionised species, including \ion{Fe}{III}, \ion{Ar}{III}, \ion{S}{III}, \ion{Co}{III}, and \ion{Ni}{III}, with additional contributions from singly ionised elements such as \ion{Fe}{II} and \ion{Co}{II}. All models produce distinct features broadly resembling those observed, and in most cases reproduce the relative strengths of the dominant transitions. The spectra of the \one and \oneO models shown in Figure~\ref{fig:combined_spectra} are broadly similar, although quantitative differences are evident. For example, the integrated line flux of the \ion{Fe}{III} 0.470\microns feature in the 3D calculation is approximately $17\%$ larger than in the corresponding 1D model. Moreover, IME features also differ between the 1D and 3D treatments; the integrated flux of the \ion{S}{III} feature between 0.9--1\microns is around $21\%$ lower in 3D than in 1D. In contrast, the \twoscenario scenario exhibits a stronger sensitivity to multidimensional effects. In the \twoO calculation, the integrated flux of the 0.470\microns \ion{Fe}{III} feature is substantially lower than in the \two model, with the 3D treatment increasing the integrated line flux by approximately $90\%$. We do however note that there is still a small amount of opacity in this region in both the 1D and 3D calculation. Other features, such as the \ion{Co}{III} emission at 0.589\microns, remain largely unchanged between 1D and 3D despite Fe and Co being approximately co-spatial. Furthermore, the \two model shows a reduction of approximately $40\%$ in the integrated flux between 0.69--0.74\microns when compared to the corresponding 1D calculation, which brings the synthetic spectrum into better agreement with the observations.

While the models predict comparable flux levels across the optical (differing by only $\sim$10\%) and produce broadly similar spectral features, they also exhibit the persistent shortcomings of previous investigations \citep[e.g.,][]{Shingles2020,Shingles2022a,Blondin2023}. The most prominent of these is the overionisation of Fe, which is most apparent in the region around 0.73\microns, where both models fail to reproduce the observed blend of \ion{Fe}{II} and \ion{Ni}{II} at 0.720\microns and 0.735\microns. Instead, both predict only minor amounts of these singly ionised species (contributing less than 10\% to the synthetic spectra in this region; see Table~\ref{table:line_identifcation}), and are instead dominated by a strong \ion{Ar}{III} 0.714\microns feature, consistent with other double-detonation model investigations \citep[e.g.,][]{Shingles2020,Blondin2023}. 

Emission features near 0.73\microns are commonly associated with \ion{Ca}{II} in theoretical explosion models \citep[e.g.,][]{Mazzali2015,Blondin2018,Polin2021,Blondin2023}, and our calculations likewise predict the \ion{Ca}{II} lines at 0.729\microns and 0.732\microns in some of the explosion models. Although \ion{Ca}{II} cannot be observationally ruled out for normal SNe~Ia in this region \citep{Maguire2018,Flores2020,Kwok2023}, its presence has little impact on the inferred Ni/Fe mass ratio \citep{Flores2020}. The \two model exhibits stronger \ion{Ca}{II} emission than the \one model; however, in both cases the emission in this region is dominated by \ion{Ar}{III}, rather than the \ion{Fe}{II} and \ion{Ni}{II} features typically associated with these wavelengths. Prominent \ion{Ar}{III} emission is also commonly predicted in nebular-phase explosion models \citep[e.g.,][]{Shingles2020,Shingles2022a,Blondin2023}. Although the inclusion of \ion{Ca}{II} brings the models into somewhat better agreement with observations, the predicted features are too blue and too narrow to fully explain the discrepancy between observations and simulations on their own. While \ion{Ca}{II} emission is generally disfavoured in normal SNe~Ia, it may be more consistent with lower-mass explosions \citep[e.g., SN~1999by;][]{Silverman2013}, including sub-$M_\mathrm{Ch}$\xspace~CO WD double-detonation scenarios \citep[see][and references therein]{Polin2021}. However, the \one model, which most closely resembles a standard double-detonation model, shows essentially no \ion{Ca}{II} contribution in this region. We stress that our models synthesise too much $^{56}$Ni to provide a plausible match to subluminous 91bg-like SNe~Ia, and that exploring this regime further would require lower-mass progenitor systems.

All explosion models except the \twoO model, consistently overproduce \ion{Fe}{III} relative to \ion{Fe}{II}, which is clearly seen in Figures~\ref{fig:Kromer_1DOneExpl}--\ref{fig:Kromer_3DTwoExpl}. 
We utilise each Monte Carlo packet's last interaction to generate these figures as some packets undergo scattering or fluorescence before escaping which is especially evident in the 3D calculations, where scattering from \ion{Fe}{I} is present, and is notably weak in the 1D models.
It can be seen that the 3D models generally show more complex line blending in the Fe-dominated features around 0.5\microns, with increased absorption and a more balanced contribution from \ion{Fe}{I--III}.
We note that there is a substantial contribution from \ion{Fe}{I} in this region and larger \ion{Fe}{I} populations in the \twoscenario scenario, when compared to the \onescenario scenario.

\subsubsection{NIR Comparison}
\label{sec:average_spectra_NIR}

The top panel of Figure~\ref{fig:integrated_spectra} shows the integral of the normalised cumulative flux. In relative terms, significantly more flux emerges in the NIR for the \twoscenario scenario compared to the \onescenario scenario, which exhibits approximately 50\% less flux in this region (see Table~\ref{tab:flux_ratios} for a detailed breakdown). This difference in flux can be seen across the NIR spectra in Figure~\ref{fig:combined_spectra}.

The NIR spectra of all models are dominated by IGEs, primarily Fe, with the most substantial contributions arising from \ion{Fe}{III} and \ion{Fe}{II}. In most models, the dominant IGE feature in terms of peak flux is is the \ion{Fe}{III} emission between 2--2.5\microns. However, the \twoO model diverges from this trend, with its strongest feature being \ion{Fe}{II} emission at 1.5–1.7\microns. Ionisation effects play a critical role in shaping these differences: the \onescenario scenario is generally more over-ionised, yielding strong \ion{Fe}{III} emission with only minor \ion{Fe}{II} contributions (see Figures~\ref{fig:Kromer_1DOneExpl}--\ref{fig:Kromer_3DOneExpl}). In contrast, the \twoO model shows improved \ion{Fe}{II} and suppressed \ion{Fe}{III} emission, producing a weaker 2--2.5\microns \ion{Fe}{III} feature (see Figure~\ref{fig:Kromer_1DTwoExpl}). 
The \two model possesses a more balanced ionisation structure between \ion{Fe}{II} and \ion{Fe}{III}, though it remains over-ionised (see Figure~\ref{fig:Kromer_3DTwoExpl}). 
Both scenarios also show distinct \ion{Fe}{I--III} distributions: in the \oneO and \one models \ion{Fe}{I--III} populations are partially co-spatial, with \ion{Fe}{I} most centrally concentrated, \ion{Fe}{III} the most extended, and \ion{Fe}{II} being in the intermediate regions (see Figures~\ref{fig:1DOneExplionplot}--\ref{fig:3DOneExplionplot}). While the \twoO model (see Figure~\ref{fig:1DTwoExplionplot}) mirrors \oneO model's gradual radial increase in \ion{Fe}{I--III} populations, the \two model (see Figure~\ref{fig:3DTwoExplionplot}) contrasts sharply with these distributions: \ion{Fe}{I} dominates lower temperature and density regions, while \ion{Fe}{II} and \ion{Fe}{III} trace higher temperature and density regions, overlapping only marginally with \ion{Fe}{I}. Although the exact ion populations do vary from 1D to 3D treatments, we find that the \one model's distributions of \ion{Fe}{I--III} are well approximated by spherically averaged ejecta. Critically, however, 3D effects significantly alter the \twoscenario scenarios' distribution of \ion{Fe}{I--III} populations throughout the ejecta, substantially modifying ionisation states and, consequently, the NIR spectra which is most clearly seen by the emergence of an \ion{Fe}{I} feature around 1\microns.

It can be seen in Figures~\ref{fig:1DOneExplionplot}--\ref{fig:3DTwoExplionplot} that the models possess distinctly different \ion{Ni}{II}-rich regions within the ejecta: the \oneO and \one models have centrally concentrated populations, while the \twoO and \two models show more extended distributions at higher velocities. However, none of the models reproduce the strength of the \ion{Ni}{II} 1.939\microns feature seen in SN~2021aefx, with the \two model producing the largest contribution despite the low stable Ni yields in the \dsix scenario (0.016\mass\xspace and 0.018\mass\xspace for the \onescenario and \twoscenario scenarios, respectively). As discussed in the context of Fe, the models are over-ionised, which explains the lack of a 1.939\microns \ion{Ni}{II} feature (see Section~\ref{sec:average_spectra_LMIR} for a discussion of a MIR \ion{Ni}{III} feature) which is consistent with other double-detention model investigations \citep{Blondin2023}. 

Another intriguing result in the NIR is the emergence of a \ion{S}{I} feature at $\sim$1\microns in the 3D treatments of the explosion models (Figures~\ref{fig:Kromer_3DOneExpl} and \ref{fig:Kromer_3DTwoExpl}). Examination of Figure~\ref{fig:1DOneExplionplot} shows that for the \oneO model, \ion{S}{I} populations occur predominantly in the least dense outer regions of the ejecta. In the \one model, however, \ion{S}{I} populations are approximately two orders of magnitude larger and concentrated in the innermost dense regions (see Figure~\ref{fig:3DOneExplionplot}). This naturally explains the substantially stronger spectral feature in 3D. This inner-velocity component of \ion{S}{I} and other IMEs in the \one model originates in the portion of the ejecta where IMEs are swept around the companion.
The \twoO model possesses only a minor contribution from \ion{S}{I} in Figure~\ref{fig:Kromer_1DTwoExpl}, which is subdominant to the surrounding IME and IGE features. However, when performing the 3D calculation, the exact stratification and size of the \ion{S}{I} populations vary significantly (see Figures~\ref{fig:1DTwoExplionplot} and \ref{fig:3DTwoExplionplot}).
The distribution of \ion{S}{I} and other IMEs in the \two model differs significantly from the \one model. In the \two model, the detonation of the primary WD produces a shell of IMEs, and the secondary WD's detonation produces a large amount of IMEs in the innermost regions of the ejecta. This stratification of the \ion{S}{I} populations in the \two models is not well approximated by the \twoO as it is inherently asymmetric and off-centre; as such, it explains the emergence of the feature in 3D. 
Although the strength of the feature in the \one model is insufficient to place it in significant tension with observations of SNe~Ia, the feature in the \two model is considerably stronger and inconsistent with SN~2021aefx. However, if the ionisation were higher, which may improve other singly ionised features in the NIR, it may produce more \ion{S}{II} and bring the model closer to matching the observations.

We note that neither model produces the \ion{Ca}{IV} 3.206\microns feature, as this ion is absent from our atomic dataset. However, \cite{Blondin2023} successfully reproduced this feature using updated atomic data, suggesting both scenarios would likely yield this feature with a similar atomic dataset. Beyond the \ion{Ca}{IV} feature (omitted from the middle panel of Figure~\ref{fig:combined_spectra} for clarity but visible in the second panels of Figures~\ref{fig:Kromer_1DOneExpl}--\ref{fig:Kromer_3DTwoExpl}), all models predict negligible flux and fail to match SN~2021aefx’s observed low continuum. Instead, they exhibit trace emission from \ion{Fe}{III}, \ion{Fe}{II}, \ion{Ni}{III}, and \ion{Si}{I} which aligns with \cite{Blondin2023}\footnote{We note \cite{Blondin2023} suggest this continuum emission may arise from molecules or dust \citep[see also][ for further discussion]{Jerkstrand2012}}.

\subsubsection{Lower MIR Comparison (5-14\microns)}
\label{sec:average_spectra_LMIR}

We show the 5--30\microns region in the bottom panel of Figure~\ref{fig:combined_spectra}. In this Section, we focus on the observed MIR region (5--14\microns), which is dominated by singly ionised Ne and Co, doubly ionised Ar, Ni, and Co, and triply ionised S. 
The models reproduce several MIR features with varied success\footnote{The \ion{Ni}{IV} 8.405\microns transition is absent from our atomic dataset \citep[see][]{Shingles2020}. As such, its absence cannot be used to constrain either scenario. However, it has been produced in nebular calculations \citep{Blondin2023} of double-detonation models \citep{Gronow2021}.}.
Both scenarios exhibit sensitivity to multidimensional effects in the \ion{Ni}{III} 7.349\microns, \ion{Ar}{III} 8.991\microns, \ion{Ni}{III} 11.002\microns, and \ion{Co}{III} 11.888\microns features. These effects are strongest in the \twoscenario scenario, where the integrated flux of the \ion{Ar}{III} and \ion{Co}{III} lines are reduced by $\sim$38\% and $\sim$54\% respectively in the corresponding 3D calculation. In contrast, the \onescenario scenario shows more modest sensitivity to multidimensional effects, with the integrated flux of the \ion{Ar}{III} and \ion{Co}{III} features decreasing by approximately 6\% and 15\%, respectively, in the 3D calculation.
Similarly to the optical and NIR, singly ionised features such as \ion{Ni}{II} 6.636\microns and \ion{Ar}{II} 6.985\microns are weak or absent in the \onescenario scenario, while the \twoscenario scenario maintains a more balanced ionisation structure.

The 12.8\microns feature of SN~2021aefx was initially attributed to a \ion{Ni}{II} 12.73\microns line by \citet{Kwok2023} \citep[see also][]{Gerardy2007,Telesco2015}, but was later identified as \ion{Ne}{II} based on radiative transfer calculations of a violent merger \citep{Pakmor2012b} and delayed detonation model \citep{Seitenzahl2013} by \citet{Blondin2023}. We note that the violent merger model calculation by \citet{Blondin2023} showed a sharply peaked \ion{Ne}{II} feature, which was supported by observations of the 03fg-like SN~2022pul \citep{Kwok2024}. Our calculations show that this feature is composed entirely of \ion{Ne}{II}, supporting that identification. The total Ne mass is 0.006\mass\xspace and 0.019\mass\xspace in the \onescenario and  \twoscenario scenarios respectively, with \ion{Ne}{II} populations approximately four times higher in the latter, explaining its greater strength. We also find that the stratification of Ne differs between the scenarios. In the \one model, \ion{Ne}{II} is centrally concentrated and streams outward in one direction, whereas in the \two model, \ion{Ne}{II} is primarily distributed between 5,000 and 10,000$\,\mathrm{km}\,\mathrm{s}^{-1}$, resulting in a flatter-topped profile.

\subsubsection{Upper MIR Comparison (14-30\microns)}
\label{sec:average_spectra_UMIR}

Across both scenarios the upper MIR region (third panel of Figure~\ref{fig:combined_spectra}; bottom panels of Figures~\ref{fig:Kromer_1DOneExpl}--\ref{fig:Kromer_3DTwoExpl}) is dominated by several strong emission lines, including \ion{Co}{III} 16.391\microns, \ion{S}{III} 18.708\microns, and \ion{Fe}{III} 22.925\microns, which is consistent with predictions for other model classes \citep{Blondin2023}. Our calculations also show contributions from \ion{Ne}{III} 15.550\microns, \ion{Fe}{II} 17.936\microns, \ion{S}{III} 18.713\microns, \ion{Ar}{III} 21.832\microns, \ion{Co}{III} 24.067\microns, \ion{Fe}{II} 24.519\microns, and \ion{S}{I} 25.249\microns.

In all models, the \ion{Fe}{III} 22.925\microns line is blended on its blue wing with \ion{Ar}{III} and on its red wing with \ion{Co}{III}. In the \twoscenario scenario, it is further contaminated by \ion{Fe}{II}. Nevertheless, even with this blending, the feature remains well-defined relative to others in the upper MIR. All models in this spectral region exhibit a \ion{S}{III} 18.713\microns feature, which is blended with an \ion{Fe}{II} feature. This is weakest in the \one and \oneO models and strongest in the \twoO and \two models, reflecting the underlying ionisation balance of the different explosion models. 

The \ion{Co}{III} 16.391\microns feature is the least blended and appears centrally peaked, although its blue wing overlaps a \ion{Ne}{III} feature. This \ion{Ne}{III} feature primarily emerges in the \twoscenario scenario, similarly to the emergence of the \ion{Ne}{II} 12.815\microns feature. Moreover, in the \two model, \ion{Ne}{II} and \ion{Ne}{III} are mostly co-spatial (i.e., residing in an asymmetric shell), with the geometry not adequately captured in 1D. This results in differences in the line profiles between the 1D and 3D treatments.
Intriguingly, the 3D treatment of the explosion models results in a \ion{S}{I} 25.249\microns feature redward of the \ion{Co}{III} 24.067\microns feature. This feature is absent in the \oneO model and negligible in the \one and \twoO models. As such, its emergence in the \two model mirrors that seen in the NIR.

Although no upper MIR observations are available at 270 days post-explosion, additional spectra of SN~2021aefx were obtained at +323 days \citep{DerKacy2023} and +415 days \citep{Ashall2024} past rest-frame $B$-band maximum. As our calculations do not extend to such late epochs, we comment only briefly and qualitatively on how this realisation of the \dsix scenario performs. Comparing our models to the +415 day spectrum \citep[see figure~2 of][]{Ashall2024}, we find that the \ion{Co}{III} 16.391\microns feature in the \onescenario scenario is too sharply peaked relative to the flatter observed profile, whereas the \twoscenario scenario provides a better match in this respect. The neighbouring \ion{Fe}{II} 17.936\microns and \ion{S}{III} 18.713\microns features are not well reproduced in either scenario, as both models predict flatter, more top-hat-like profiles rather than the sharply peaked structures seen in the observation. Finally, the \ion{Fe}{III} 22.925\microns feature in the \onescenario scenario is more sharply peaked and therefore more consistent with the observations than in the \twoscenario scenario. Note that the exact strength of features will evolve over the $\sim165$ day difference between the epochs of our synthetic spectra and those presented by \cite{Ashall2024}. Therefore, to make more robust statements regarding the exact strength and line shapes of species, further theoretical modelling at these later epochs would be required, which should also include longer-lived decay chains that become significant at late times \citep[see][]{Seitenzahl2009}.

\begin{figure*}
\centering
\begin{subfigure}{\textwidth}
    \centering
    \includegraphics[width=0.99\textwidth,height=6.5cm]{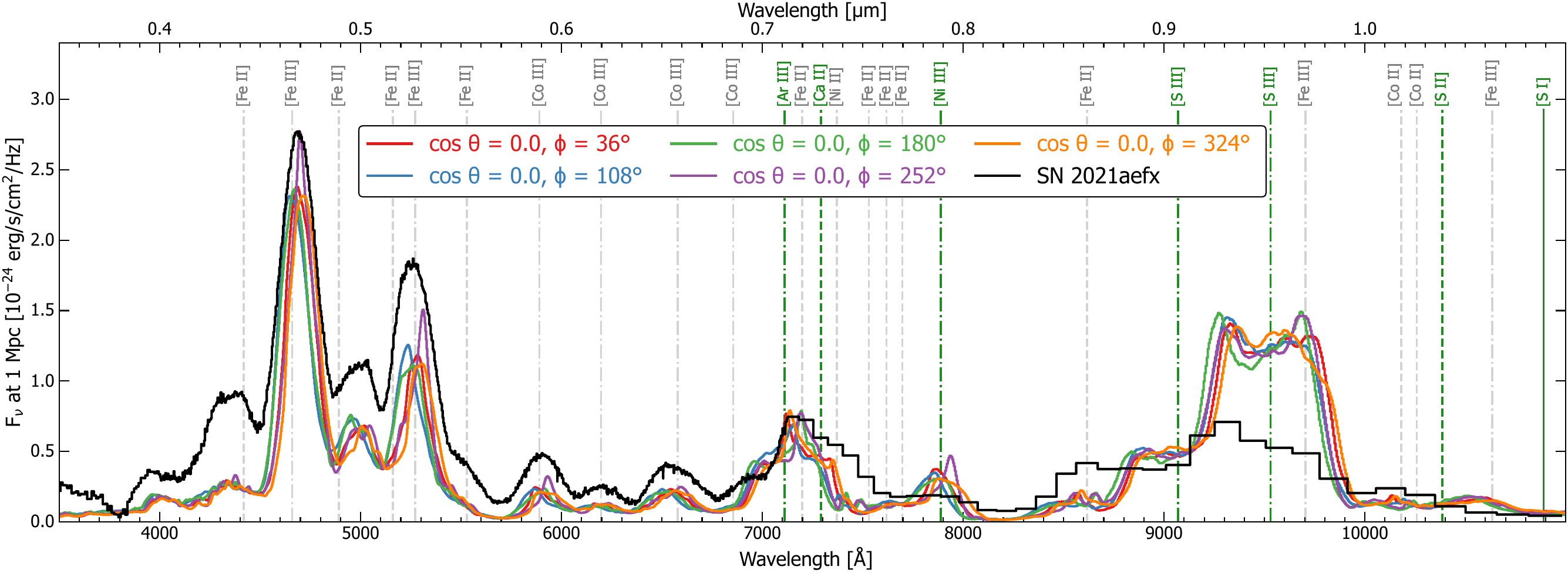}
\end{subfigure}

\vspace{1em}

\begin{subfigure}{\textwidth}
    \centering
    \includegraphics[width=0.99\textwidth,height=6.5cm]{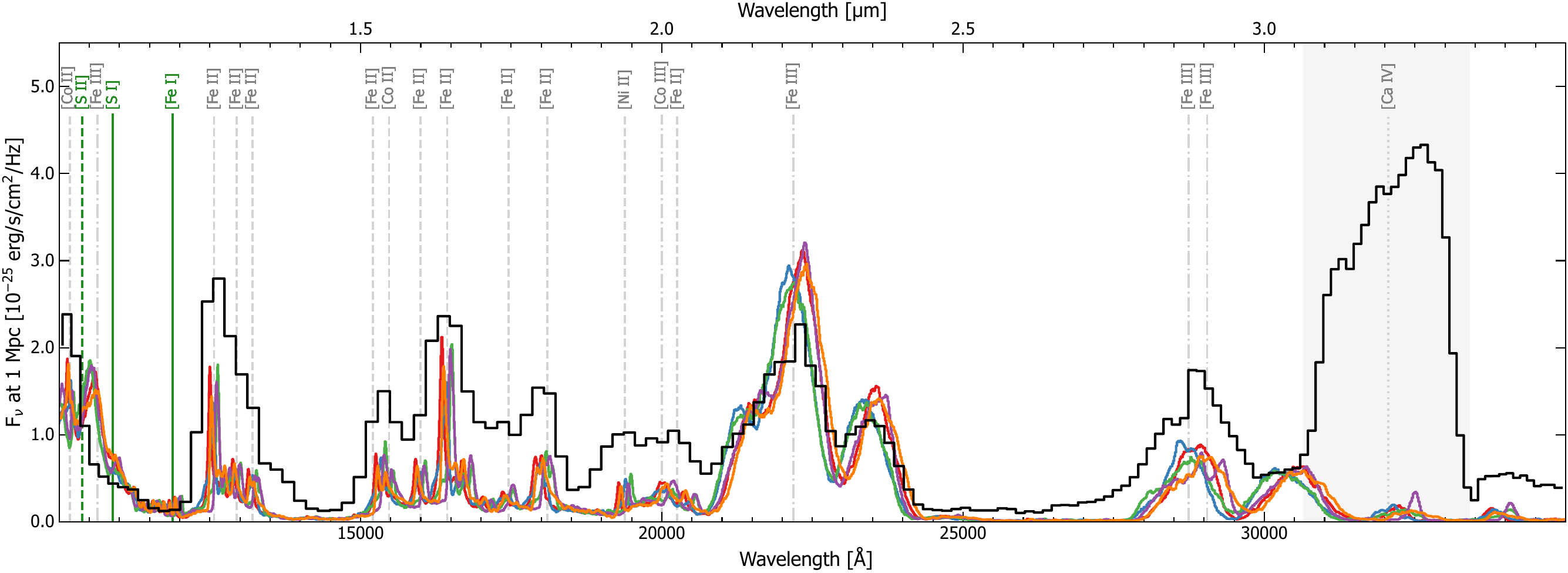}
    
\end{subfigure}

\vspace{1em}

\begin{subfigure}{\textwidth}
    \centering
    \includegraphics[width=0.99\textwidth,height=6.5cm]{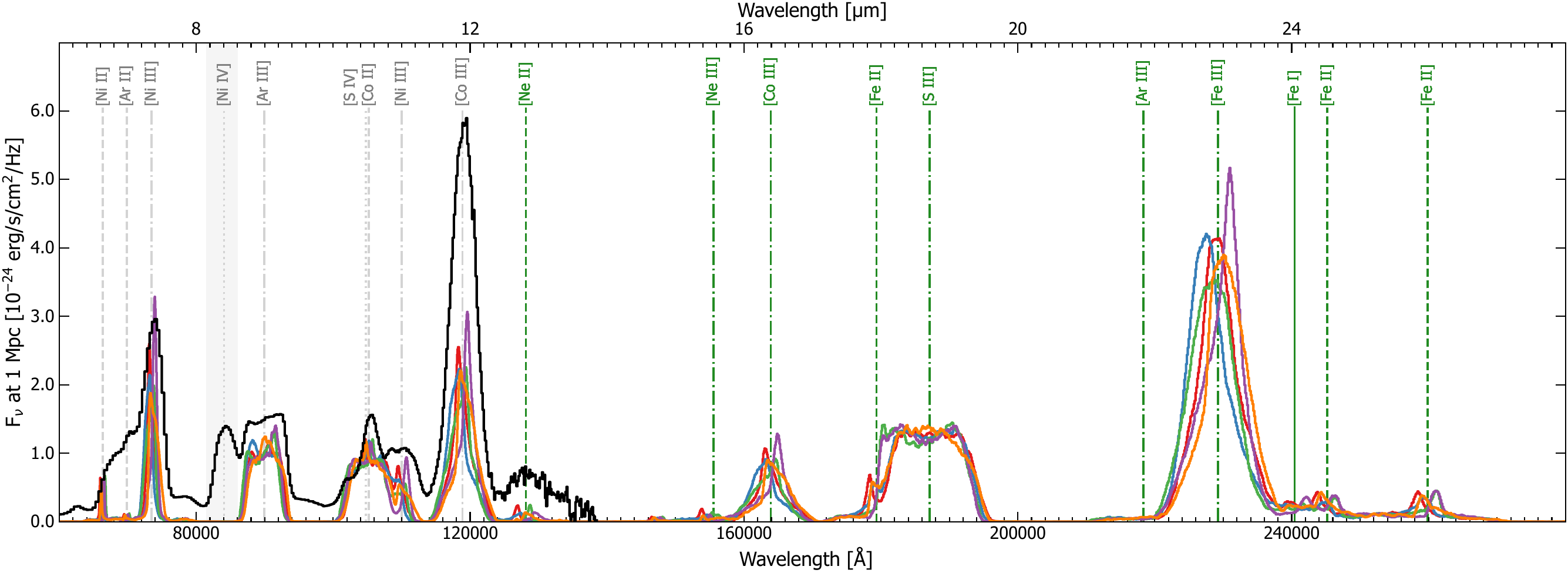}
\end{subfigure}

    \caption{Spectra of the \one model for different viewing angles at 270 days post-explosion for the optical (top), NIR (middle), and MIR (bottom). The lines of sight shown are oriented around the merger plane (i.e., $\cos(\theta)=0.0$), where the most significant variation in synthetic observables occurs. As in Figure~\ref{fig:combined_spectra}, observed spectra are corrected for redshift and extinction \protect\citep{Hosseinzadeh2022}, and all spectra are scaled to a distance of 1 Mpc. Vertical grey lines indicate the rest wavelengths of many prominent features identified by \protect\cite{Flores2020} and \protect\cite{Kwok2023}. In contrast, green lines highlight significant model features that diverge from observations and lie outside the spectral range of SN~2021aefx. The linestyles of the vertical lines indicate ionisation stages: solid for neutral species, dashed for singly ionised, dash-dotted for doubly ionised, and dotted for triply ionised species. Rest wavelengths identified for SN~2021aefx by \protect\cite{Kwok2023} are listed in Table~\ref{table:line_identifcation}.     
    Note that a Savitzky–Golay filter has been applied and the shaded grey regions highlight prominent features that we do not reproduce due to their absence in our atomic data.
    }    
    \label{fig:3DOneExpl_viewing_angle}
\end{figure*}

\begin{figure*} 
    \centering
    \includegraphics[height=9cm,width=3.5cm,trim=4 0 2 0, clip]{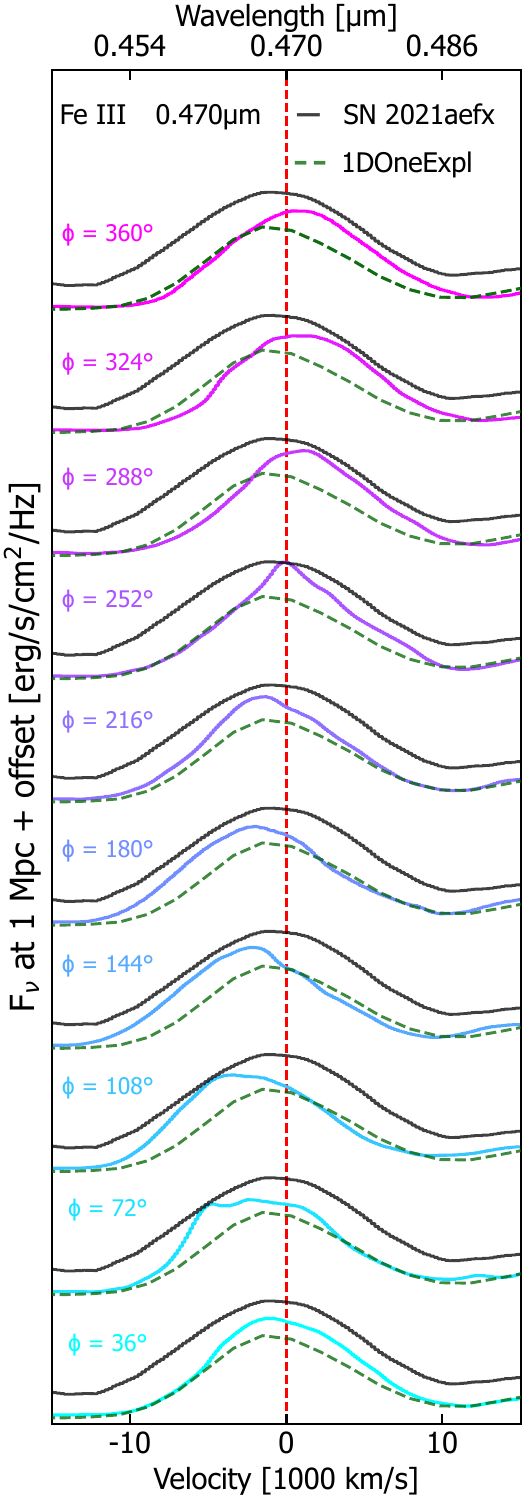}
    \includegraphics[height=9cm,width=3.5cm,trim=24 0 0 0, clip]{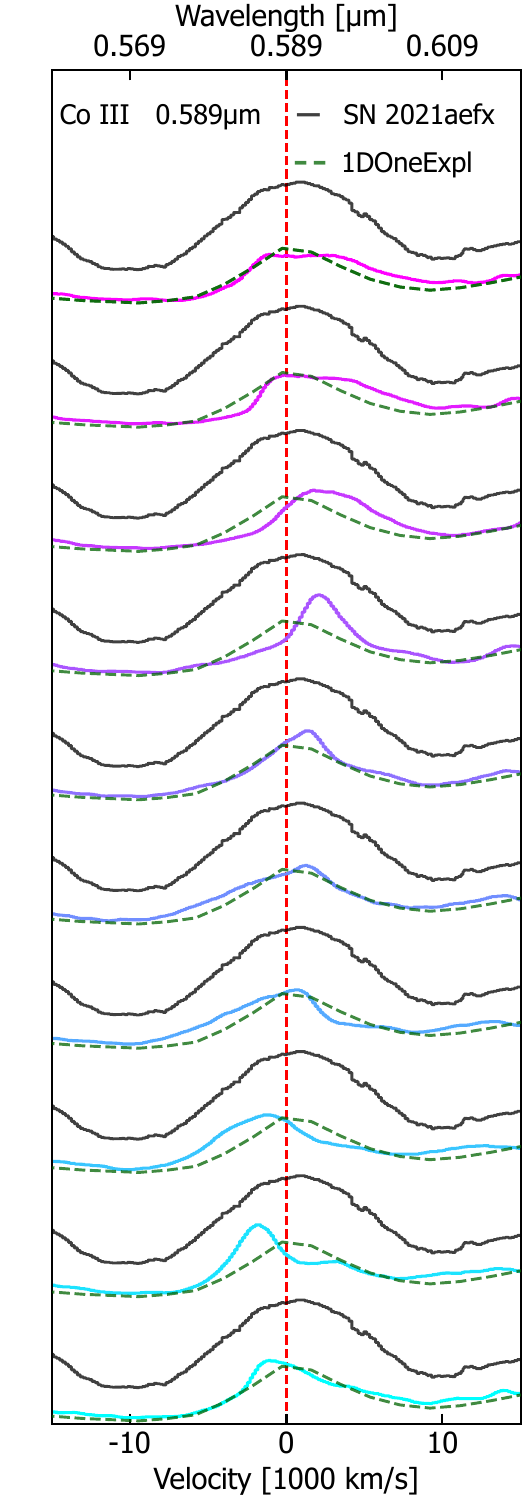}
    \includegraphics[height=9cm,width=3.5cm,trim=24 0 0 0, clip]{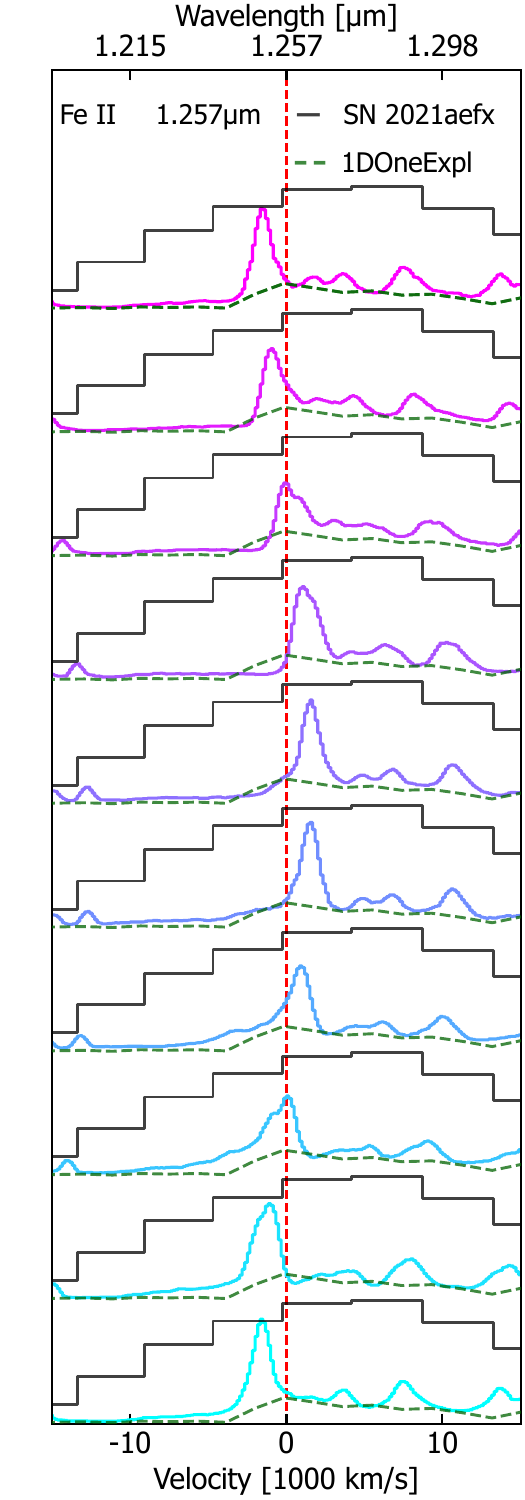}
    \includegraphics[height=9cm,width=3.5cm,trim=24 0 0 0, clip]{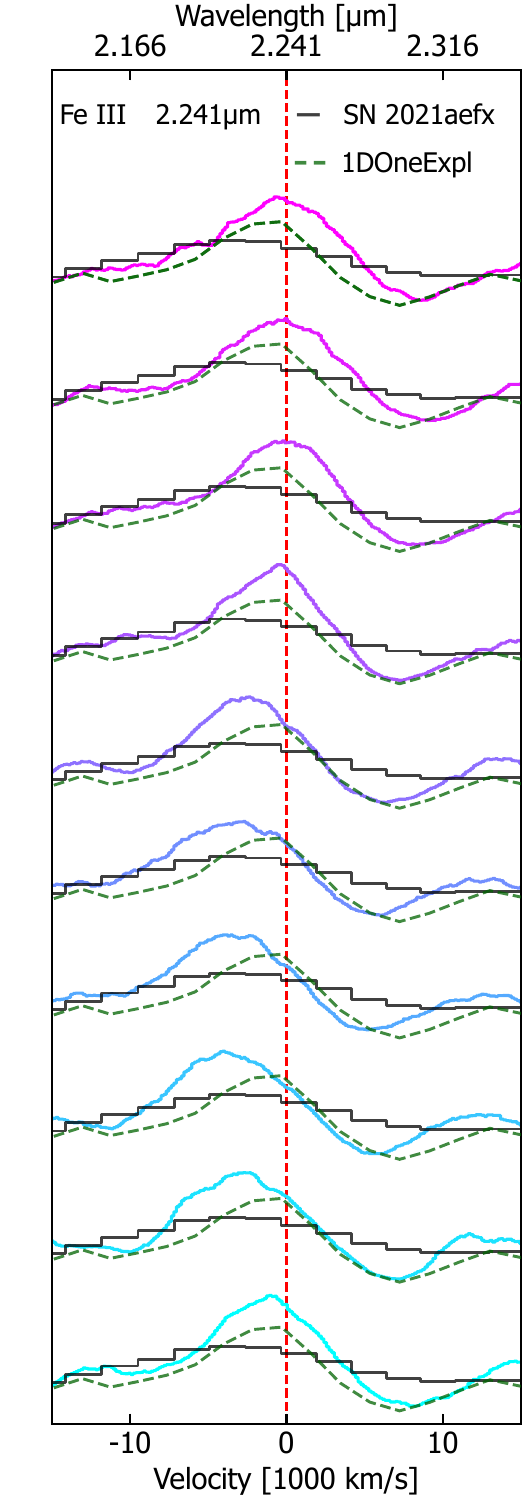}
    \includegraphics[height=9cm,width=3.5cm,trim=24 0 0 0, clip]{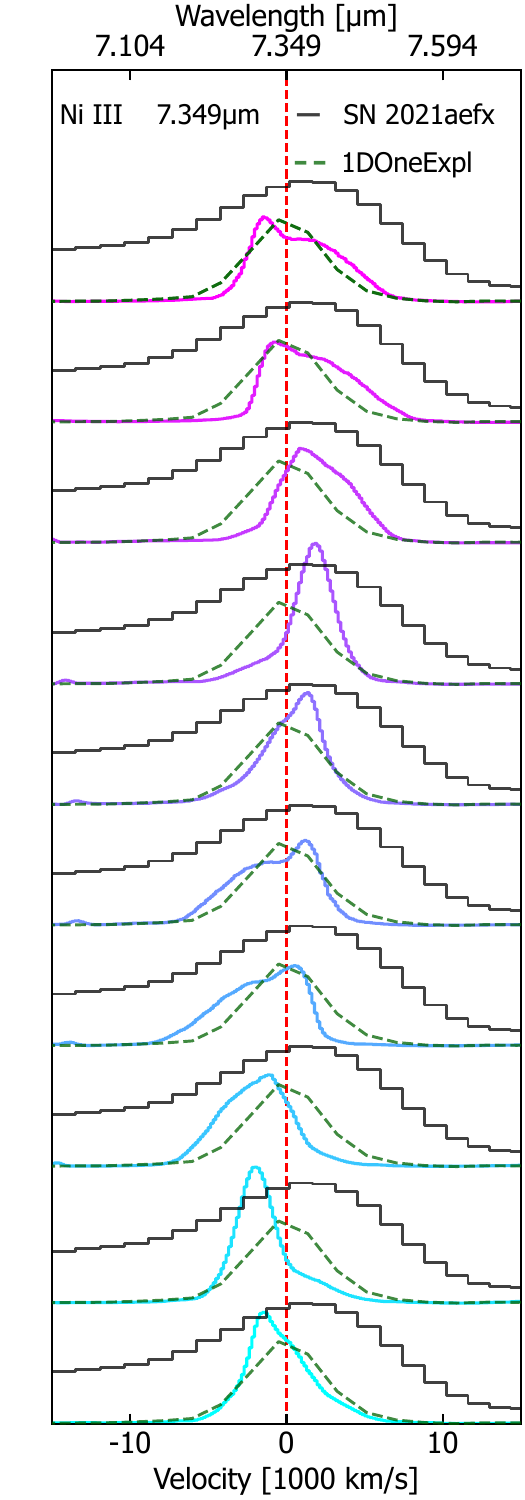}
    \includegraphics[height=9cm,width=3.5cm,trim=4 0 2 0, clip]{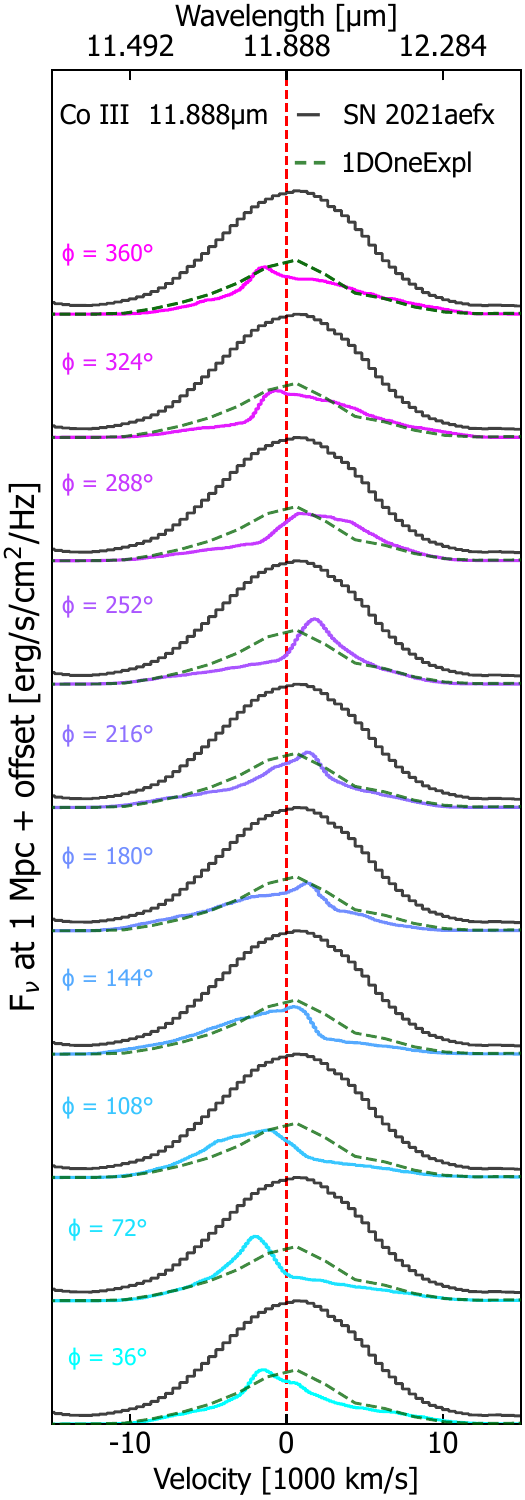}
    \includegraphics[height=9cm,width=3.5cm,trim=24 0 0 0, clip]{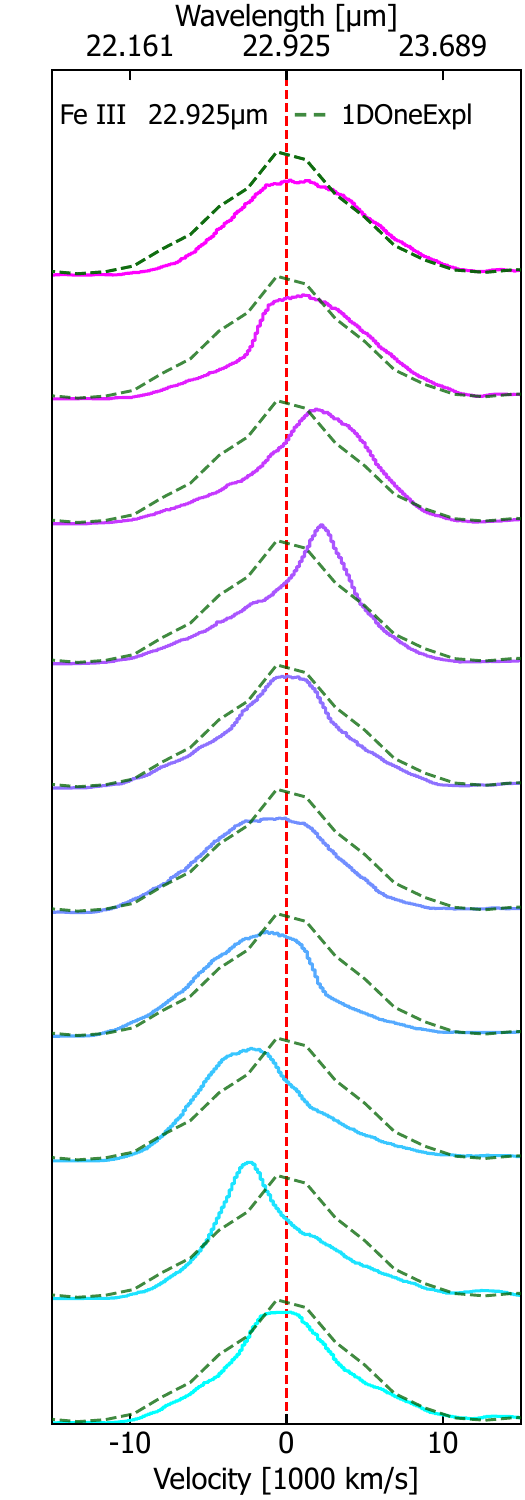}
    \includegraphics[height=9cm,width=3.5cm,trim=24 0 0 0, clip]{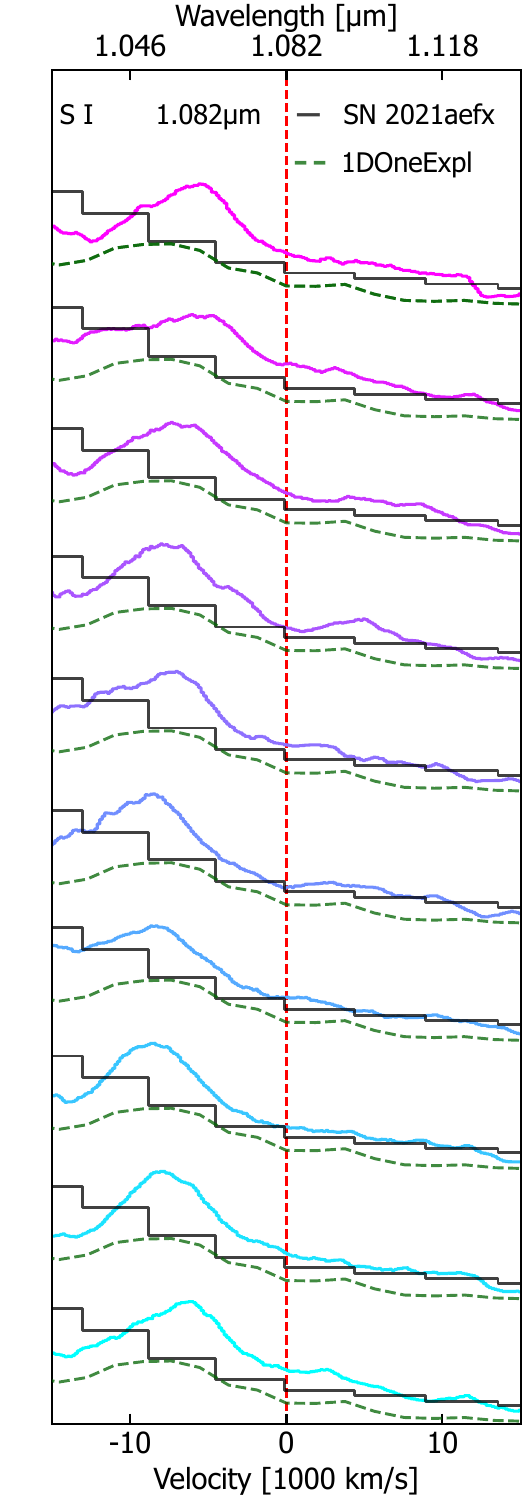}
    \includegraphics[height=9cm,width=3.5cm,trim=24 0 0 0, clip]{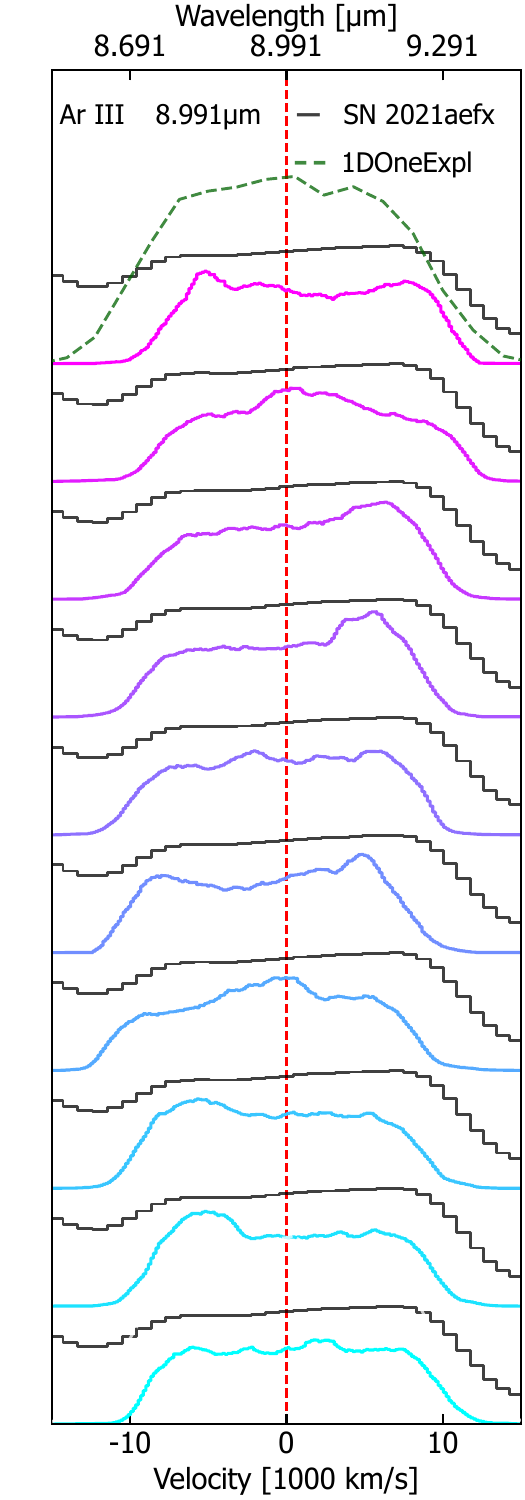}
    \includegraphics[height=9cm,width=3.5cm,trim=24 0 0 0, clip]{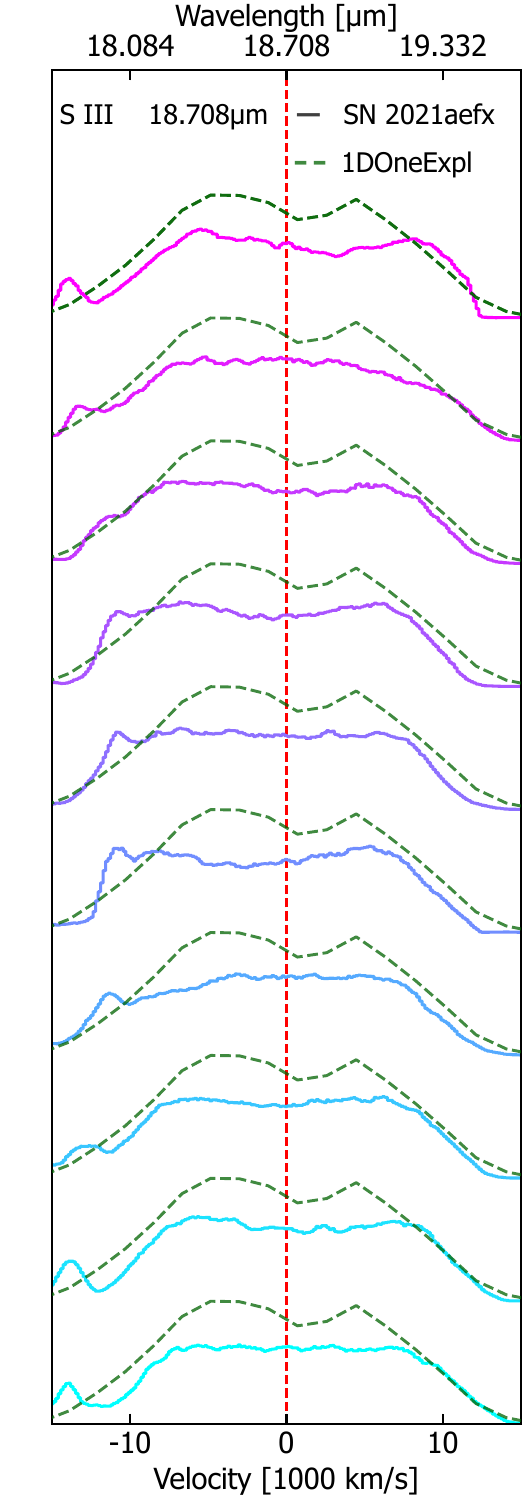}
    
    \caption{Viewing angle spectra for the \one model at $\cos(\theta) = 0$ (i.e., the merger plane), with the respective orientations (i.e., $\phi = 0$--$360\degree$) indicated in the first panel of each row, alongside the corresponding 1D profile from the \oneO model. Each spectrum is consistently offset, and where possible, compared to the observed spectra of SN~2021aefx (black). We show a set of prominent IGEs and IMEs: \ion{Fe}{III} 0.470\microns, \ion{Co}{III} 0.589\microns, \ion{Fe}{II} 1.257\microns, \ion{Fe}{III} 2.241\microns, \ion{Ni}{III} 7.349\microns, \ion{Co}{III} 11.888\microns, \ion{Fe}{III} 22.925\microns, \ion{S}{I} 1.082\microns, \ion{Ar}{III} 8.991\microns, and \ion{S}{III} 18.708\microns, with the red dashed line indicating the rest wavelength for reference. See Table~\ref{tab:feature_fits} for a detailed breakdown of the corresponding transitions.}
    \label{fig:velocity_stacked_plot_oneexpl_scenario}
\end{figure*}

\begin{figure} 
    \centering
    \includegraphics[width=0.47\textwidth]{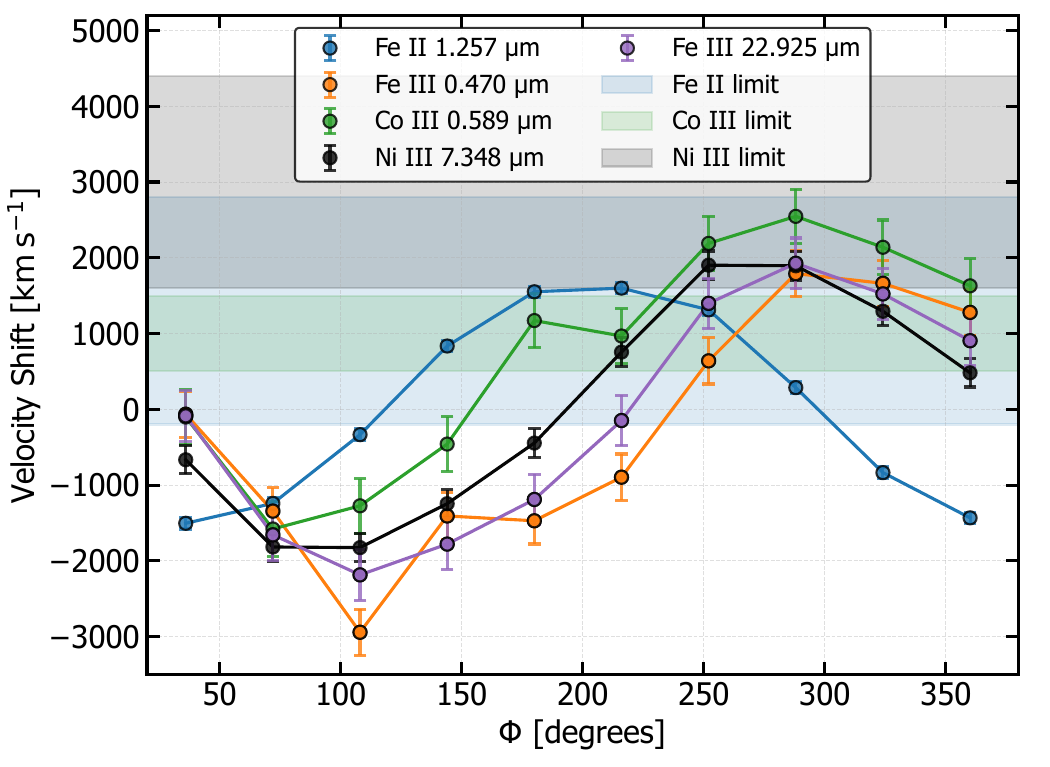}    
    \includegraphics[width=0.47\textwidth]{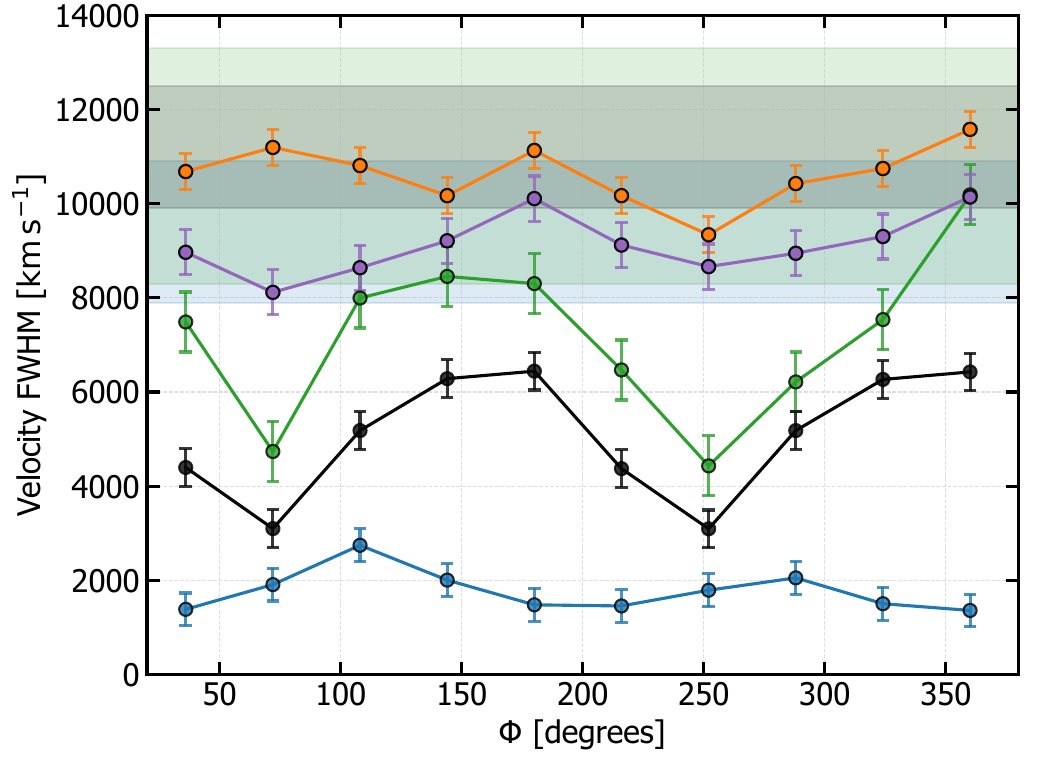}     
    \caption{Velocity shifts (top) of the \ion{Fe}{II} 1.257\microns, \ion{Fe}{III} 0.470\microns, \ion{Co}{III} 0.5890\microns, \ion{Ni}{III} 7.348\microns and \ion{Fe}{III} 22.925\microns features for different observer orientations at $\cos(\theta) = 0$, $\phi = 0$--$360\degree$ (i.e., the merger plane), and the corresponding FWHM (bottom) for the \one model. The shaded regions indicate the velocity limits determined by \citet{Kwok2023} for the corresponding features in SN~2021aefx. These velocity limits are shown for the \ion{Co}{III}, \ion{Ni}{III}, and \ion{Fe}{II} features. Note that no limit is shown for \ion{Fe}{III}, as no velocity measurement was determined.
    }
    \label{fig:3DOneExpl_Velocity_fits}
\end{figure}

\subsection{\one Model Orientation Effects}
\label{sec:3DOneExpl_Modle_Orientation_Effects}

In this section, we present synthetic spectra for the \one model viewed from different observer orientations within the merger plane (i.e., $\cos(\theta) = 0$, $\phi = 0$--$360\degree$), as shown in Figure~\ref{fig:3DOneExpl_viewing_angle}. These spectra allow us to directly assess how the 3D structure influences observable features. We focus on a selection of prominent IGE and IME features across the optical, NIR, and MIR ranges. These features are chosen for their diagnostic potential in probing the stratification of the ejecta, as illustrated in Figure~\ref{fig:velocity_stacked_plot_oneexpl_scenario}, which shows how their luminosity, Doppler velocity, and morphology vary with orientation. Crucially, we compare the synthetic spectra from the 3D calculation to those from the corresponding 1D model to highlight the impact of the multidimensional approach (a detailed breakdown of the corresponding transitions for each feature is provided in Table~\ref{tab:feature_fits}).

To quantitatively investigate the variations with rotation of the different IGEs shown in Figure~\ref{fig:velocity_stacked_plot_oneexpl_scenario}, we fit Gaussian(s) to features and extract the velocity and full width at half maximum (FWHM) for each species, as shown in Figure~\ref{fig:3DOneExpl_Velocity_fits}. Although Gaussians are an imperfect approximation for nebular-phase spectral features -- since line profiles are shaped by the underlying distribution of the emitting ejecta and often by the blending of multiple transitions from several species -- we adopt Gaussian fitting throughout this work as an empirical method to estimate velocities and understand how the geometry influences the spectra. The \onescenario scenario generally yields features reasonably well described as Gaussian-like (i.e., singly peaked with a symmetric profile). This description is commonly adopted for IGE features, as they are typically expected to be centrally located.
We follow a method similar to that used in observational studies: we fit a Gaussian to each emission feature by initially selecting continuum points at the edges of the feature interactively. In cases where a feature consists of a multiplet, such as the \ion{Fe}{III} 0.470\microns feature, we fit the multiplet with multiple Gaussians. As an alternative, we tested fitting a single Gaussian centred on the gf-weighted mean wavelength of the transitions. This method yielded a comparable velocity trend, although it required a correction to reproduce the expected net blueshift and redshift.
To estimate the uncertainties in the velocities and FWHMs, we refit the synthetic spectra three times, varying the initial continuum points, and determined the corresponding standard deviation at each angle. We subsequently adopted the largest standard deviation as the representative error for all points.

\begin{table*}
\centering
\caption{Summary of transition(s) and rest wavelengths for emission features displayed in Figure~\ref{fig:velocity_stacked_plot_oneexpl_scenario} and Figure~\ref{fig:Velocity_3DTwoExpl_stacked12}. Where features are due to a multiplet we list the corresponding individual transitions.}
\begin{tabular}{l l l}
\hline
Feature &  line(s) (\microns) &  Rest Wavelength (\microns)\\ \hline
\ion{Fe}{III} & 0.461, 0.466, 0.467, 0.470, 0.473, 0.476, 0.477, 0.478 & 0.470 \\
\ion{Co}{III} & 0.589 & 0.589 \\
\ion{Fe}{II} & 1.257, 1.270, 1.279, 1.294, 1.298, 1.321, 1.328 & 1.270 \\
\ion{Fe}{III} & 2.150, 2.218, 2.242, 2.348 & 2.241 \\
\ion{Ni}{III} & 7.349 & 7.349 \\
\ion{Co}{III} & 11.888 & 11.888 \\
\ion{Fe}{III} & 22.925 & 22.925 \\
\ion{S}{I} & 1.082 & 1.082 \\
\ion{Ar}{III} & 8.991 & 8.991 \\
\ion{S}{III} & 18.708 & 18.708 \\
\hline
\end{tabular}
\label{tab:feature_fits}
\end{table*}

\subsubsection{Iron Group Element Variations}

From Figure~\ref{fig:3DOneExpl_viewing_angle} it can be seen that the \one model exhibits distinct variations in line shifts depending on the observer orientation. This behaviour is particularly pronounced in the 0.470\microns \ion{Fe}{III} multiplet. Clear Doppler shifts are seen approximately $180\degree$ apart (i.e., at $\phi = 108\degree$ and $\phi = 252\degree$ in Figure~\ref{fig:3DOneExpl_viewing_angle}), which corresponds to orientations strongly shaped by the secondary WD's survival. This trend becomes even more apparent in Figure~\ref{fig:velocity_stacked_plot_oneexpl_scenario}, where the feature transitions smoothly from minimal Doppler blueshift  at $\phi = 36\degree$ to a maximum at approximately $\phi = 180\degree$, and then becomes redshifted with further rotation. A similar trend is observed for the \ion{Fe}{III} 2.241\microns multiplet, and it is also evident in the MIR \ion{Fe}{III} 22.925\microns feature.

The profiles of features also vary significantly with observer orientation. For example, in Figures~\ref{fig:3DOneExpl_viewing_angle} and \ref{fig:velocity_stacked_plot_oneexpl_scenario}, the \ion{Fe}{III} 0.470\microns feature at $\phi = 252\degree$ appears sharply peaked, whereas at $\phi = 360\degree$ the same feature has a relatively smooth peak. 
While many orientations produce smoothly varying profiles (i.e., from sharply peaked to more rounded peaks), certain angles reveal more complex structure. For instance, at $\phi = 72\degree$, the feature appears flatter, with subtle peaks emerging, particularly on the blue wing and near the centre of the profile, which aligns well with specific components of the multiplet. Similar structure can also be observed in the 2.241\microns feature, but it is noticeably absent in the 22.925\microns feature, which corresponds to a single transition and remains sharply peaked at the $\phi = 72\degree$ viewing angle.
These angle-dependent changes in the 0.470\microns \ion{Fe}{III} feature differ substantially from the profile predicted by the 1D calculation. As shown in Figure~\ref{fig:velocity_stacked_plot_oneexpl_scenario}, some orientations (e.g., $\phi = 36\degree$) closely match the 1D velocity and show a similar corresponding width, other orientations show significant deviations (e.g., $\phi = 252\degree$).

The spectra exhibit several \ion{Co}{III} features across the Optical, NIR and MIR we focus on the \ion{Co}{III} 0.589\microns feature as it is widely observed. We find that the viewing-angle dependence is broadly consistent with that of the \ion{Fe}{III} features (see Figure~\ref{fig:velocity_stacked_plot_oneexpl_scenario}). For example, at $\phi = 36\degree$ the feature shows negligible Doppler shift, then slowly increasing in its Doppler shift until it reaches a maximum blueshift, then becoming progressively redshifted. However, the viewing angles which correspond to maximum blueshift and redshift differ slightly from those of the 0.470\microns \ion{Fe}{III} feature. The 0.589\microns \ion{Co}{III} feature exhibits more rapid angular evolution than the \ion{Fe}{III} features, reaching a maximum Doppler shift at smaller angles and returning to a near-zero velocity shift by $\phi = 144\degree$ rather than $\phi = 252\degree$ which reflects that the stratification of the wake region of the secondary alters the distribution of the ion populations. This behaviour is mirrored in the MIR \ion{Co}{III} 11.888\microns and \ion{Fe}{III} 22.925\microns features, both of which arise from single transitions. 
We also find that the 0.589\microns \ion{Co}{III} feature shows clear orientation-dependent variations in the line wings, central wavelength, and overall profile width that the 1D model does not capture. We stress that these differences between the 3D and 1D profiles exist for all spectral features, as the 3D geometry introduces orientation-dependent variation.

The \one model exhibits a higher ionisation than that observed, and thus does not possess a strong \ion{Ar}{II} 6.983\microns feature. This leaves the nearby 7.349\microns \ion{Ni}{III} feature uncontaminated, enabling a direct probe of the Ni distribution. 
Across all viewing angles, the \one model predicts a \ion{Ni}{III} feature narrower than observed, though its luminosity remains comparable.
Notably, the \ion{Ni}{III} features are also narrower than the \ion{Fe}{III} features in both 1D and 3D calculations, which is consistent with the underlying distribution of populations shown in Figure~\ref{fig:1DOneExplionplot} and~\ref{fig:3DOneExplionplot}.
We find that the most blueshifted and redshifted profiles occur approximately $180\degree$ apart, which is consistent with other IGE features. 
However, the detailed morphology of the \ion{Ni}{III} feature does not always mirror that of the \ion{Fe}{III} lines across all viewing angles. 
For instance, at $\phi = 180\degree$, the \ion{Ni}{III} profile contains distinct velocity components, in contrast to the single blueshifted peak of the \ion{Fe}{III} features at 0.470\microns and 22.925\microns.
We attribute this difference in the profiles morphology to the \ion{Ni}{III} and \ion{Fe}{III} populations not being entirely co-spatial. As shown in Figure~\ref{fig:3DOneExplionplot}, the \ion{Ni}{III} distribution is slightly off-centre relative to that of \ion{Fe}{III}, with an extended bubble of \ion{Ni}{III} ($\sim$$0\,\mathrm{km}\,\mathrm{s}^{-1}$ to $-10,000$$\,\mathrm{km}\,\mathrm{s}^{-1}$) while the \ion{Fe}{III} is more evenly distributed between positive and negative velocities. Finally, we note that the 1D calculation does not accurately reproduce any line-of-sight profile from the 3D model in terms of width, luminosity, or profile morphology. 

Both \ion{Ni}{III} and \ion{Fe}{III} features shown in Figure~\ref{fig:3DOneExpl_Velocity_fits} follow broadly similar velocity trends, with both showing net maximum  blueshifts and redshifts of $\sim$2,500$\,\mathrm{km}\,\mathrm{s}^{-1}$. The \ion{Co}{III} feature closely follows the evolution of \ion{Ni}{III}, with the most significant deviation occurring at $\phi = 144\degree$. In contrast to \ion{Co}{III} and \ion{Ni}{III}, the \ion{Fe}{III} 0.470\microns feature does not follow the exact same trend. Instead, the \ion{Fe}{III} shows a more gradual evolution. The \ion{Fe}{II} 1.257\microns feature also shows a distinctly different velocity evolution, displaying a phase offset of $\sim$90$\degree$ compared to the \ion{Fe}{III} feature.
Figure~\ref{fig:3DOneExpl_Velocity_fits} also demonstrates that the FWHMs of \ion{Fe}{II} and \ion{Fe}{III} differ significantly from one another.
These differences in velocities and FWHMs arise from the distributions of their populations within the ejecta not being the same, as shown in Figure~\ref{fig:3DOneExplionplot}. The \ion{Fe}{II} populations are primarily concentrated in the innermost ejecta and the region influenced by the wake of the secondary WD, with only minor contributions from the outermost edges, which is significantly different from that of \ion{Fe}{III} as those are somewhat more uniform and extended. 
Figure~\ref{fig:3DOneExpl_Velocity_fits} also reveals that the FWHMs of the \ion{Co}{III} and \ion{Ni}{III} clearly show a viewing angle dependent double-peaked pattern. This arises from an elongated emitting region, which appears broader when aligned with the larger emitting region and narrower when viewed perpendicularly. As a result, two distinct maxima and minima appear. 
Moreover, the difference between the FWHMs of \ion{Co}{III} and \ion{Ni}{III} profiles is due to \ion{Co}{III} originating from a larger emitting region within the ejecta, reflecting the underlying ion distribution.
When comparing the \ion{Fe}{III} 0.470\microns and 22.925\microns features, we find that they exhibit slightly different velocities and FWHMs. 
Much of these differences are due to the 0.470\microns feature blending with other lines and ionisation stages. We note that single transitions are more representative of the models underlying velocity structure, while multiplets involve overlapping profiles.

The velocities and FWHMs of the \ion{Fe}{II}, \ion{Co}{III}, and \ion{Ni}{III} features for SN~2021aefx \cite{Kwok2023} are plotted as shaded regions in Figure~\ref{fig:3DOneExpl_Velocity_fits}. We find that the predicted velocities for \ion{Fe}{II}, \ion{Co}{III}, and \ion{Ni}{III} from the \one model lie within the observed ranges. However, no single orientation reproduces all observed velocities across all species simultaneously. We note that the $\phi=216-288\degree$ orientations comes closest, although they yield \ion{Ni}{III} or \ion{Co}{III} velocities slightly outside those observed. We also find that only the \ion{Co}{III} FWHM is consistent with the lower limits reported by \cite{Kwok2023}, and the \ion{Ni}{III} and \ion{Fe}{II} lines are too narrow to be consistent with observations of SN~2021aefx. 

Several large samples of optical and NIR SNe~Ia observations already exist \citep[e.g.,][]{Maeda2010,Maeda2011,Maguire2018,Flores2018}, but measurements at these wavelengths are affected by line blending, which can introduce substantial uncertainties in inferred velocities and FWHM values. As blending can be severe, any meaningful comparison between explosion models and observations in these regions requires a fully self-consistent fitting procedure applied to both synthetic and observed data. We have not implemented such an approach here, as to fully reconcile the explosion model properties with observed populations, this would require examining significantly more observer orientations and exploring many more explosion models, and as such lies beyond the scope of this work. Furthermore, our current models suffer from overionisation of \ion{Fe}{II} and \ion{Ni}{II}, rendering several key features too weak to provide a satisfactory match to the data; we therefore restrict ourselves to a brief comparison with the overall observational trends seen in SNe~Ia populations. Observationally, singly ionised species such as \ion{Fe}{II} typically exhibit velocities of $\sim$2,000--2,500$\,\mathrm{km}\,\mathrm{s}^{-1}$ and FWHM values of $\sim$6,000--9,000$\,\mathrm{km}\,\mathrm{s}^{-1}$ \citep[see figure 7 of][]{Maguire2018}.
In the \one model, the velocity spread of singly ionised features is too small, and the predicted FWHM values are lower by roughly a factor of four. For doubly ionised species such as \ion{Co}{III}, observed maximum velocity shifts are at most $\sim$1,000$\,\mathrm{km}\,\mathrm{s}^{-1}$ with FWHM values of $\sim$8,500--11,500$\,\mathrm{km}\,\mathrm{s}^{-1}$; in this case, the model \ion{Co}{III} maximum velocity shifts are approximately double those observed, while the corresponding FWHM values are, on average, only slightly too small to overlap with the observed sample.

\subsubsection{Intermediate Mass Element Variation}

In the \one model, IMEs such as S and Ar are predominantly located in the outer layers of the ejecta (see Figure~\ref{Fig: Model abundances plot}), which is also reflected in the stratification of the ion populations shown in Figure~\ref{fig:3DOneExplionplot}. As IMEs occupy a distinct region of the ejecta compared to IGEs, they exhibit different morphologies and velocity shifts.  
Examining the optical region in Figure~\ref{fig:3DOneExpl_viewing_angle}, the \ion{S}{III} multiplet feature $\sim$0.93\microns (0.907\microns and 0.953\microns) shows its strongest blueshift at $\phi = 180\degree$ and its strongest redshift at $\phi = 36\degree$. This behaviour differs from the IGEs such as the \ion{Fe}{III} 0.470\microns feature, which shows the strongest redshift at $\phi = 252\degree$.  
However, the difference in the morphology of the \ion{S}{III} feature is more striking than its net redshift or blueshift. Its profile is considerably more flat-topped than the IGE features, as the underlying distribution resembles a hollow shell (see \citealt{Jerkstrand2017} for a review). 
As illustrated in Figure~\ref{fig:velocity_stacked_plot_oneexpl_scenario}, the MIR 18.708\microns \ion{S}{III} feature, arising from a single transition with little blending, exhibits this hollow shell distribution more clearly as its profile is distinctly flat-topped across several orientations.
Moreover, the 1D profile of this MIR feature is not a good approximation for any 3D observer orientation as no profile exhibits a similar morphology, and all exhibit a lower luminosity than the 1D case

One of the most powerful IME features for diagnostic purposes is the \ion{Ar}{III} 8.991\microns line. In our calculation, this feature does not blend with the nearby \ion{Ni}{I} line and is well isolated, exhibiting a distinct flat-topped profile that evolves with rotation. As shown in Figure~\ref{fig:velocity_stacked_plot_oneexpl_scenario}, the feature is nearly entirely flat-topped at $\phi = 36\degree$, with a small bump emerging on the blue edge at $\phi = 72\degree$, gradually shifting redward by $\phi = 252\degree$.  
As seen in both Figure~\ref{Fig: Model abundances plot} and~\ref{fig:3DOneExplionplot}, most of the \ion{Ar}{III} is located in an outer shell; however, some IMEs are distributed from the outer ejecta to the inner ejecta, shaped by the wake of the secondary WD. This wake region produces the bump, which naturally explains the corresponding blueshift if viewed from $\phi = 72$--$108\degree$, and a redshift for $\phi = 252$--$288\degree$.  
The 1D model, in comparison to the 3D model, shows a broader profile that does not resemble any of the observer orientations and is notably over-luminous. As such, it has been omitted from several line-of-sight comparisons. 
As noted by \cite{Kwok2023}, the \ion{Ar}{III} profile in SN~2021aefx is only marginally sloped. In contrast, our \one model exhibits considerably more structure across many orientations, with $\phi = 36\degree$ and $\phi = 216\degree$ (separated by 180$\degree$) being the most comparable to the morphology of the feature in SN~2021aefx and would not be affected significantly by the wake of the secondary WD.  
When compared to previously published MIR nebular spectra of the normal SN~2003hv, the peculiar SN~2005df \citep{Gerardy2007}, and the normal SN~2014J \citep{Telesco2015}, \cite{Kwok2023} note that SN~2021aefx exhibits a more symmetric \ion{Ar}{III} profile. Therefore, other orientations of the \one model may be more consistent with these MIR spectra.

\subsection{\two Model Orientation Effects}
\label{sec:3DTwoExpl_Modle_Orientation_Effects}

\begin{figure*}
\centering
\begin{subfigure}{\textwidth}
    \centering
    \includegraphics[width=0.99\textwidth,height=6.5cm]{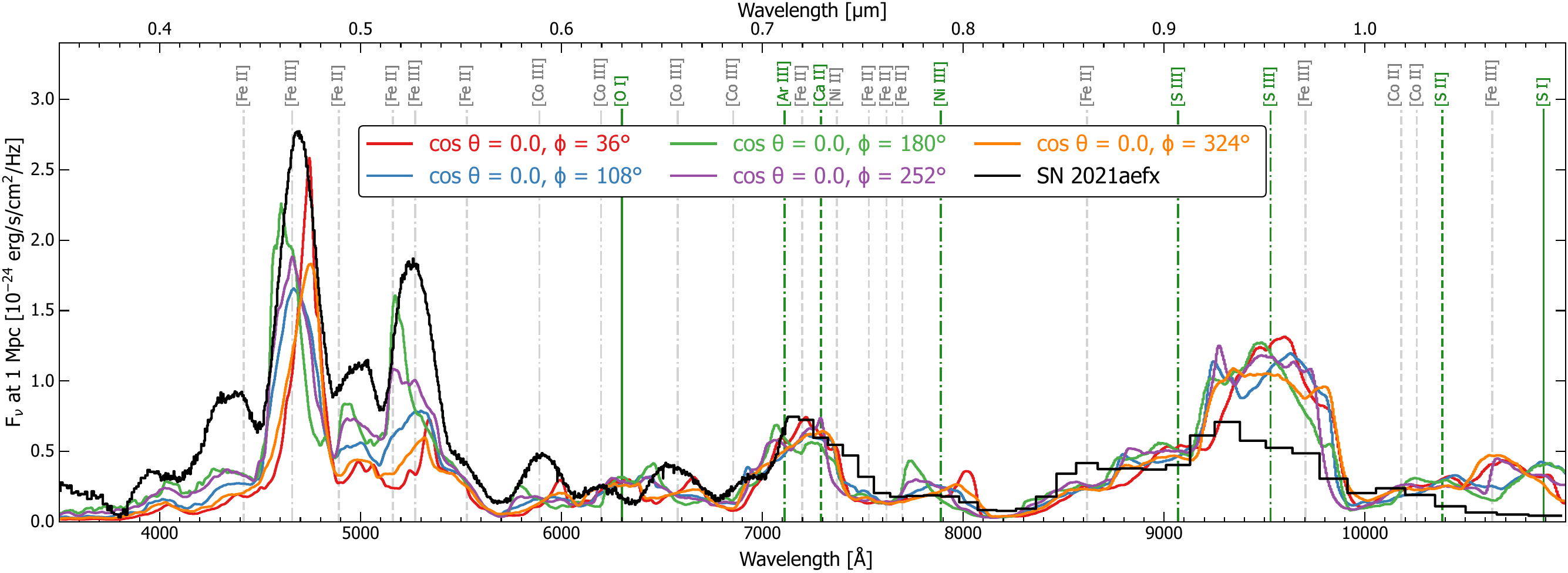}
\end{subfigure}

\vspace{1em}

\begin{subfigure}{\textwidth}
    \centering
    \includegraphics[width=0.99\textwidth,height=6.5cm]{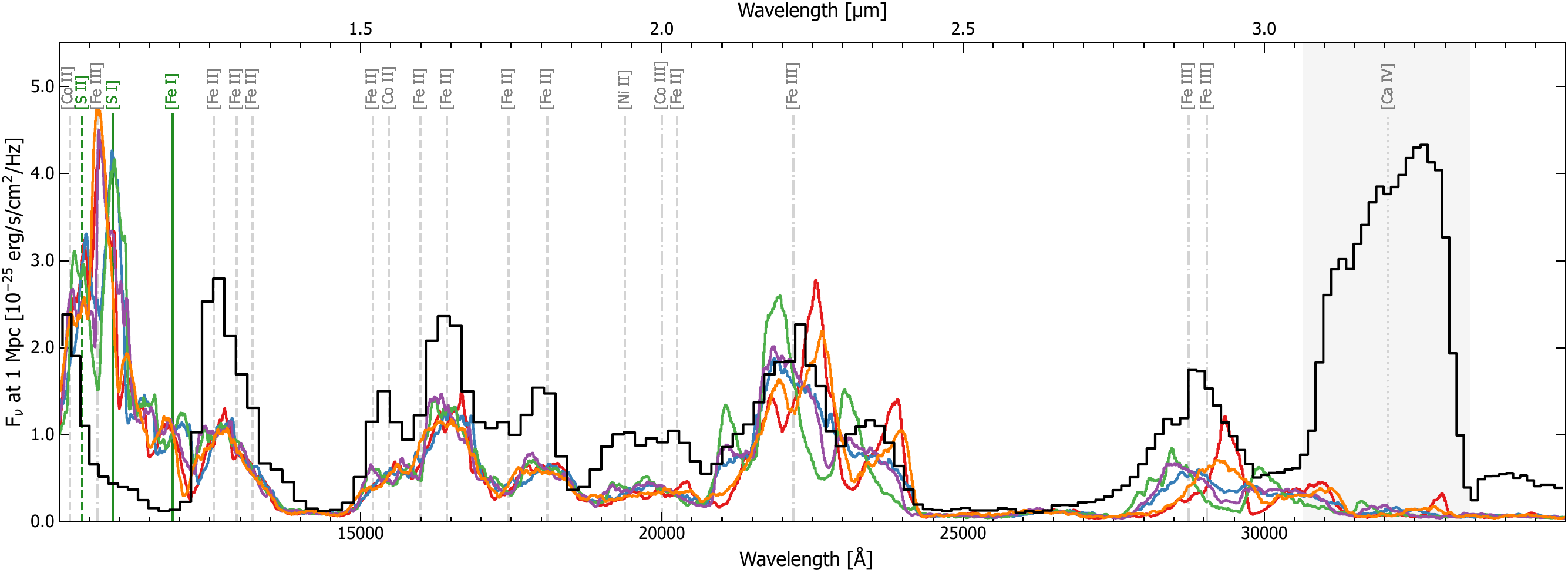}
\end{subfigure}

\vspace{1em}

\begin{subfigure}{\textwidth}
    \centering
    \includegraphics[width=0.99\textwidth,height=6.5cm]{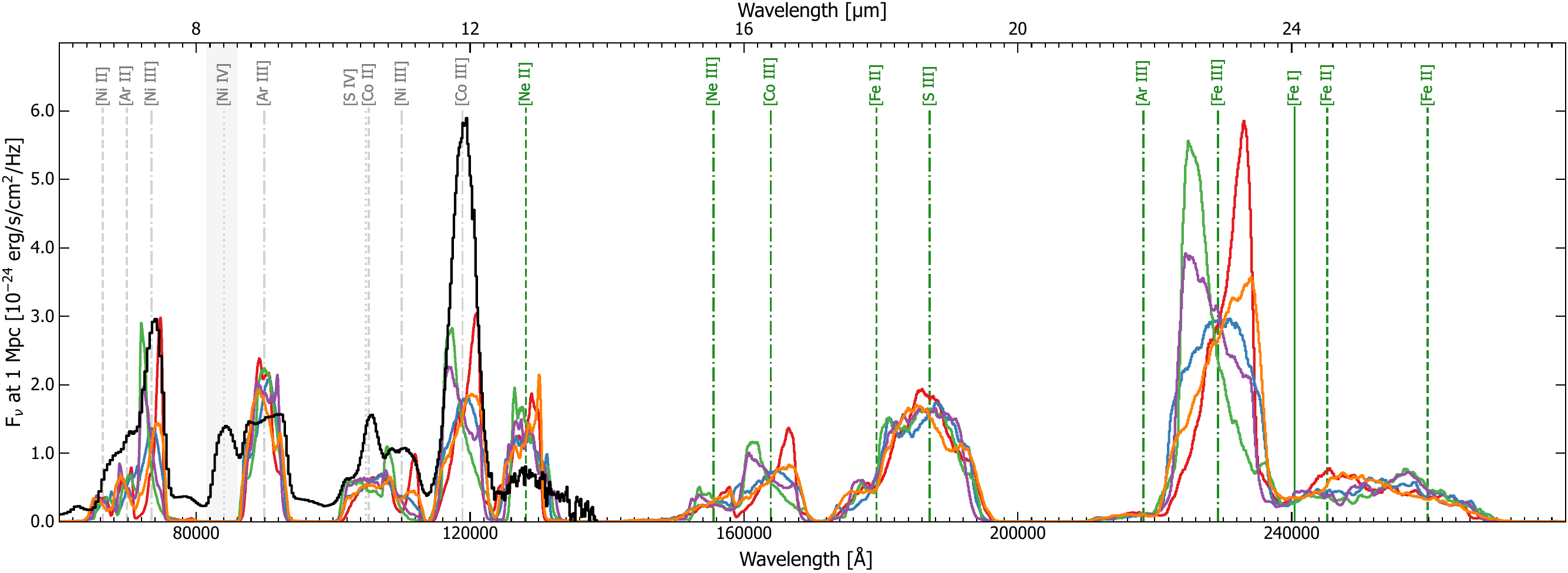}
\end{subfigure}

    \caption{Same as Figure~\ref{fig:3DOneExpl_viewing_angle} but for the \two model.}
    \label{fig:3DTwoExpl_viewing_angle}
\end{figure*}

\begin{figure*} 
    \centering
    \includegraphics[height=9cm,width=3.5cm,trim=4 0 2 0, clip]{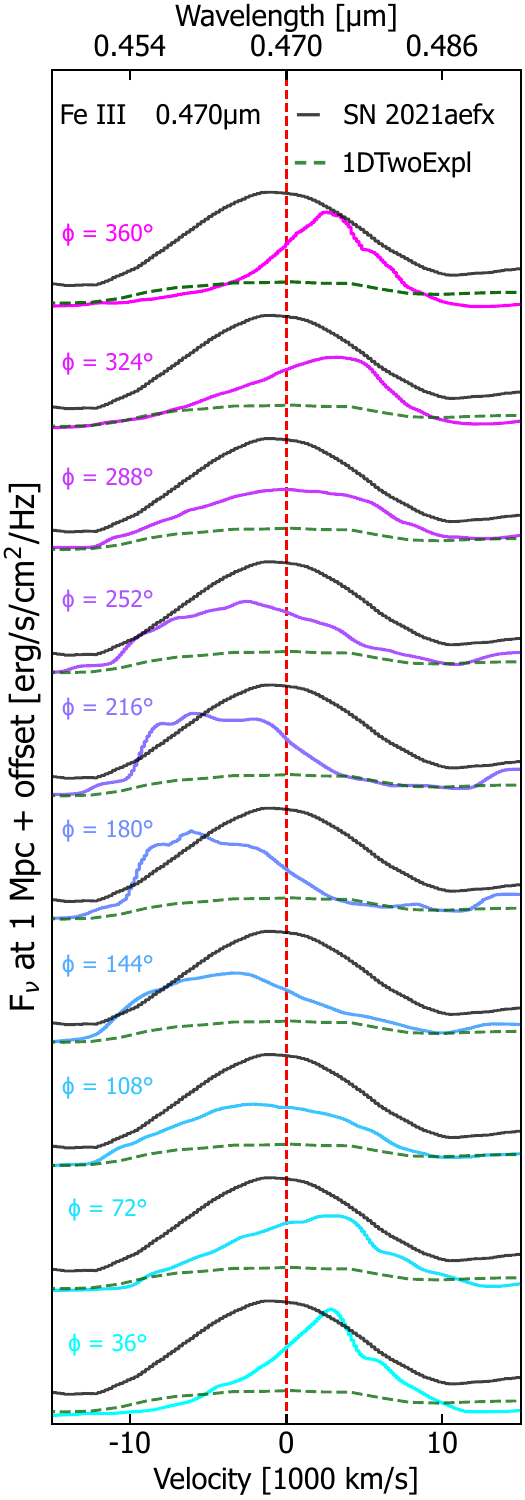}
    \includegraphics[height=9cm,width=3.5cm,trim=24 0 0 0, clip]{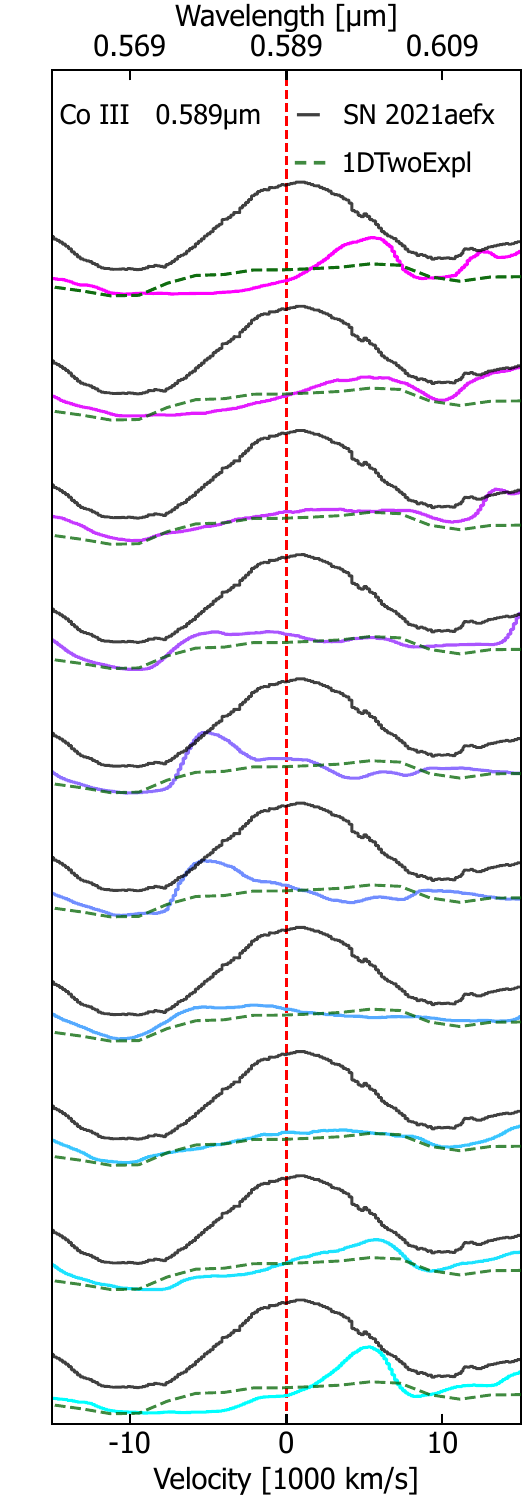}
    \includegraphics[height=9cm,width=3.5cm,trim=24 0 0 0, clip]{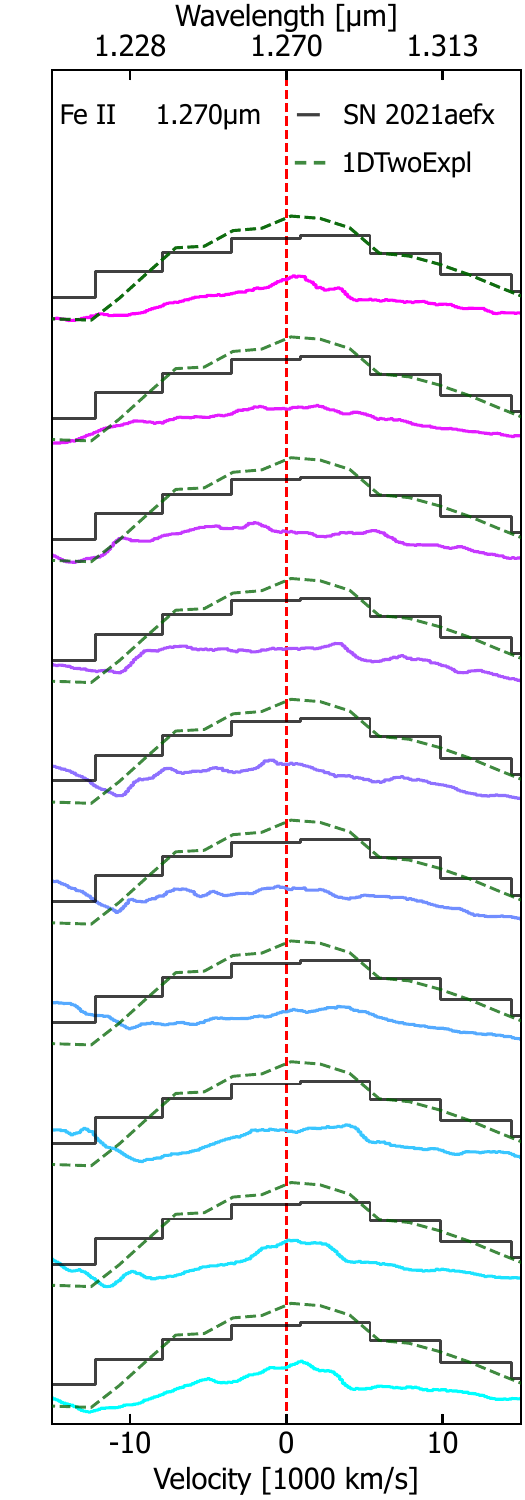}
    \includegraphics[height=9cm,width=3.5cm,trim=24 0 0 0, clip]{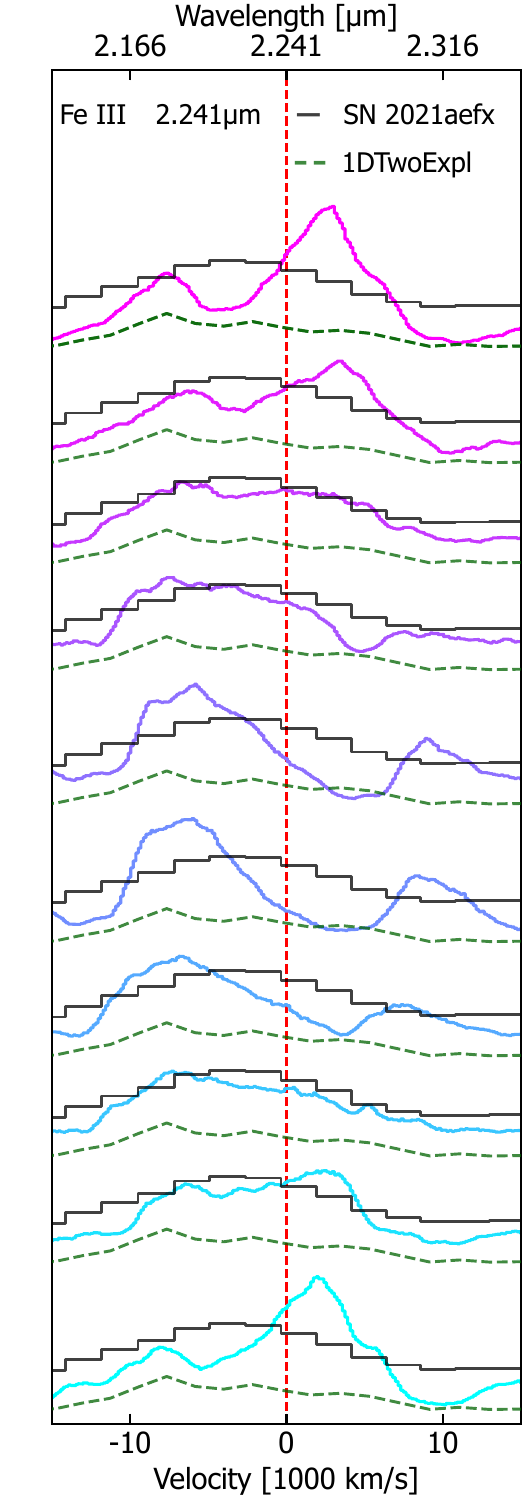}
    \includegraphics[height=9cm,width=3.5cm,trim=24 0 0 0, clip]{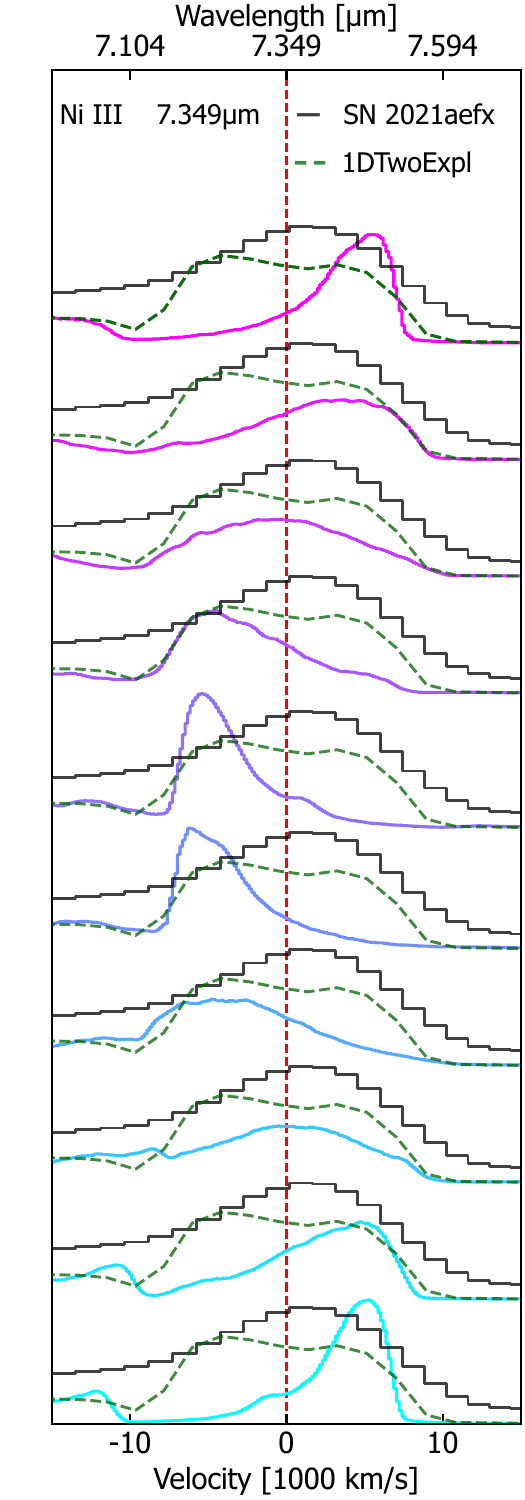}
    \includegraphics[height=9cm,width=3.5cm,width=3.5cm,trim=4 0 2 0, clip]{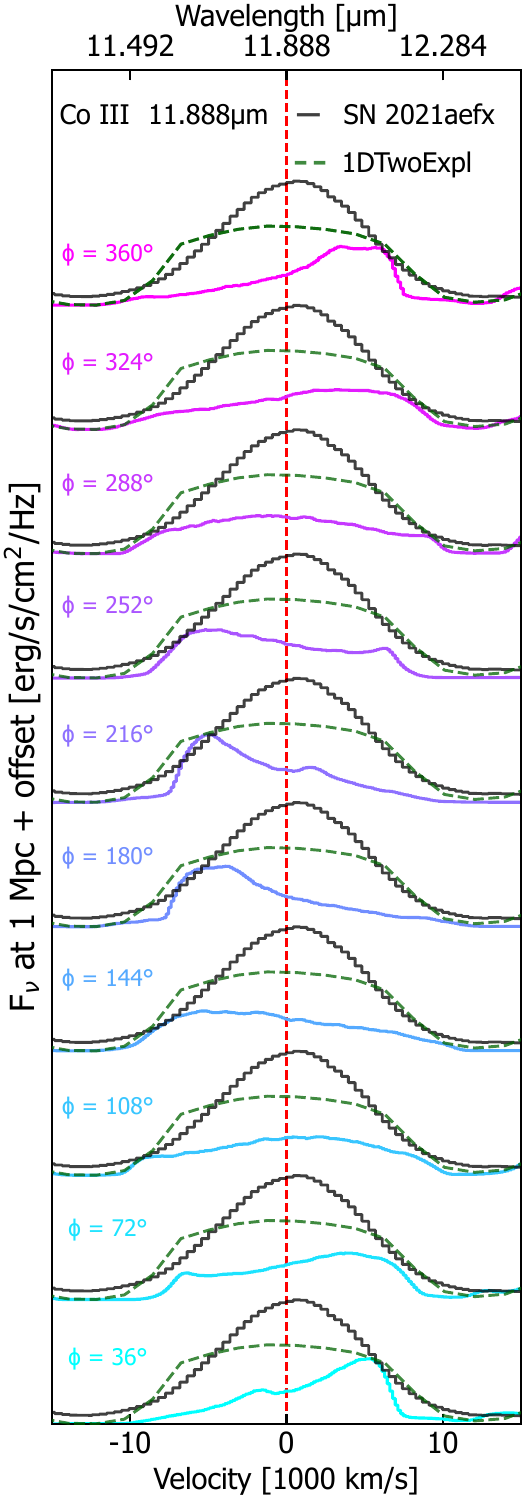}    
    \includegraphics[height=9cm,width=3.5cm,width=3.5cm,trim=24 0 0 0, clip]{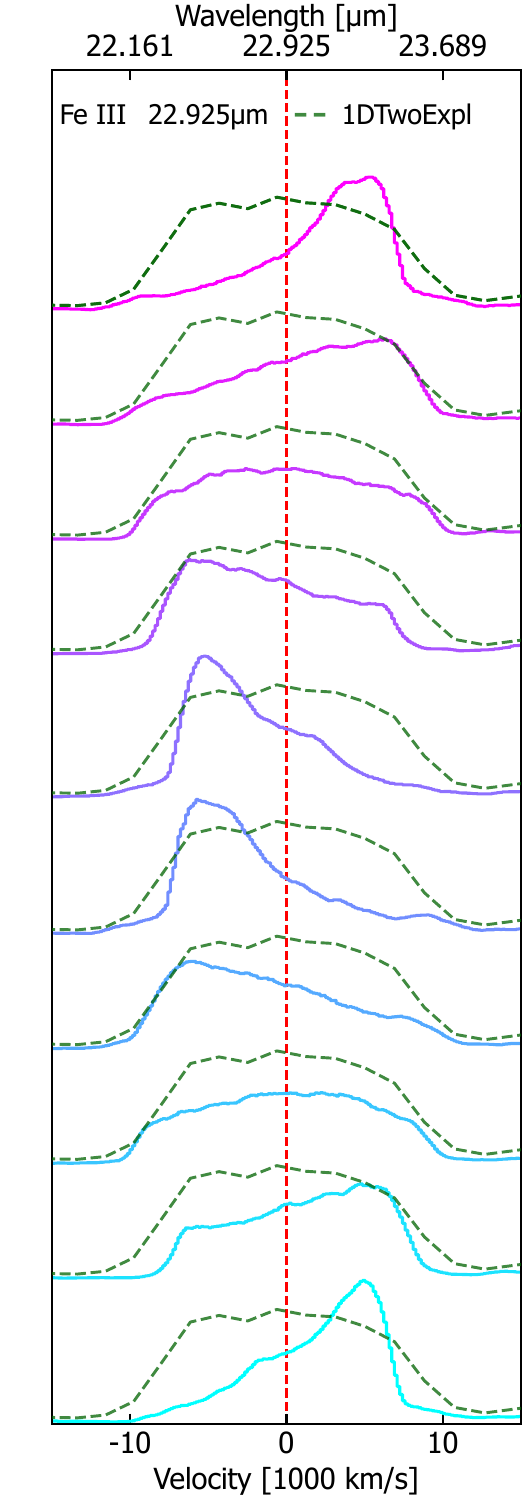}
    \includegraphics[height=9cm,width=3.5cm,width=3.5cm,trim=24 0 0 0, clip]{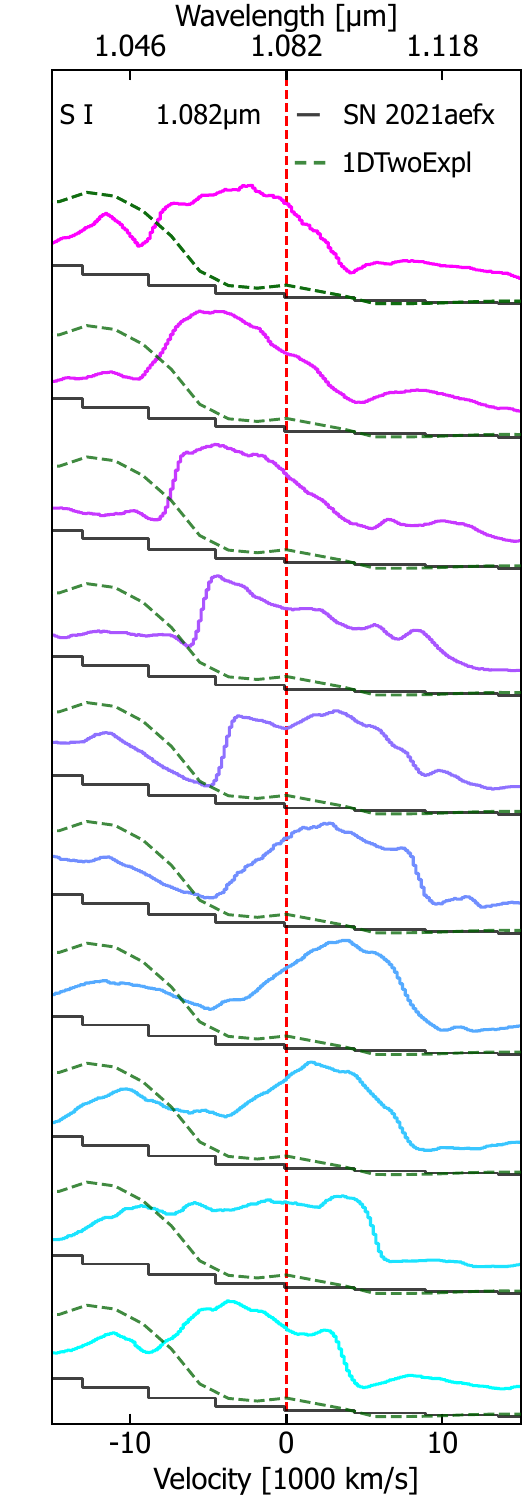}    \includegraphics[height=9cm,width=3.5cm,width=3.5cm,trim=24 0 0 0, clip]{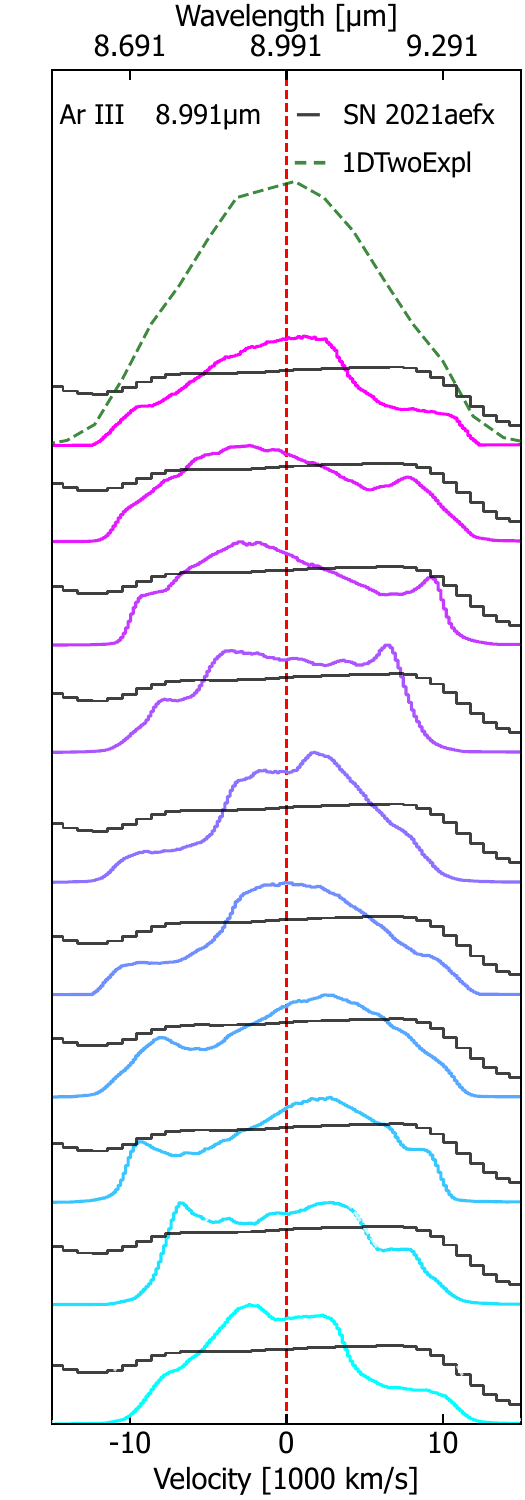}    \includegraphics[height=9cm,width=3.5cm,width=3.5cm,trim=24 0 0 0, clip]{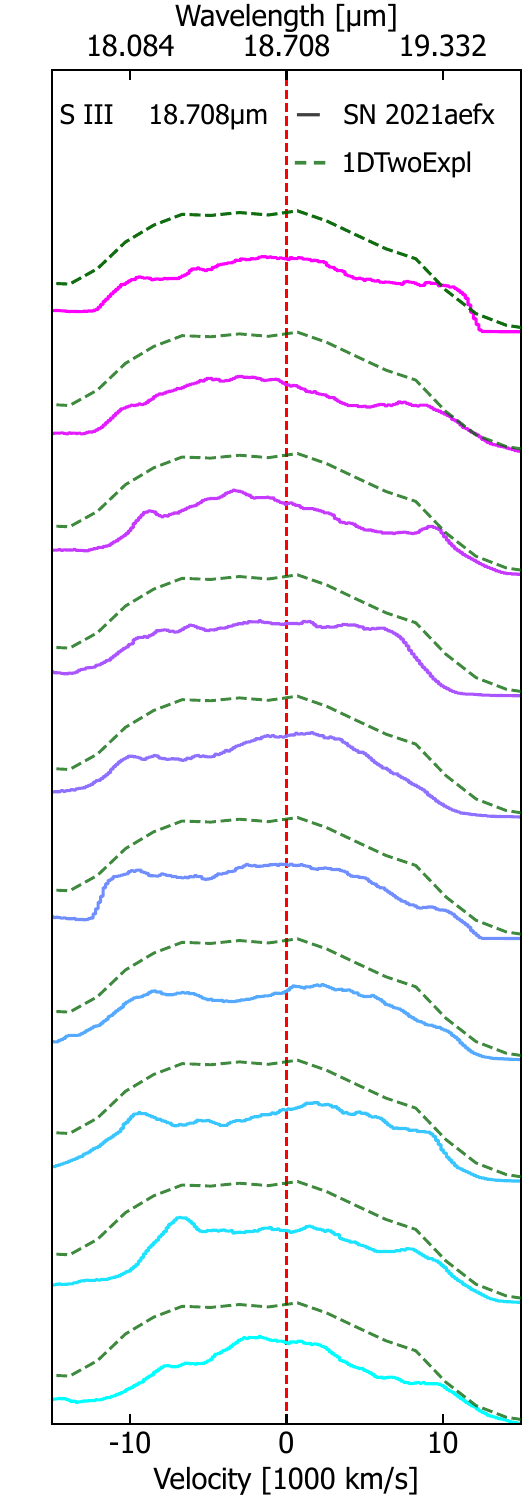}     
    \caption{Same as Figure~\ref{fig:velocity_stacked_plot_oneexpl_scenario} but for the \two models viewing angles and \twoO model spectra.}
    \label{fig:Velocity_3DTwoExpl_stacked12}
\end{figure*}

\begin{figure} 
    \centering
    \includegraphics[width=0.49\textwidth]{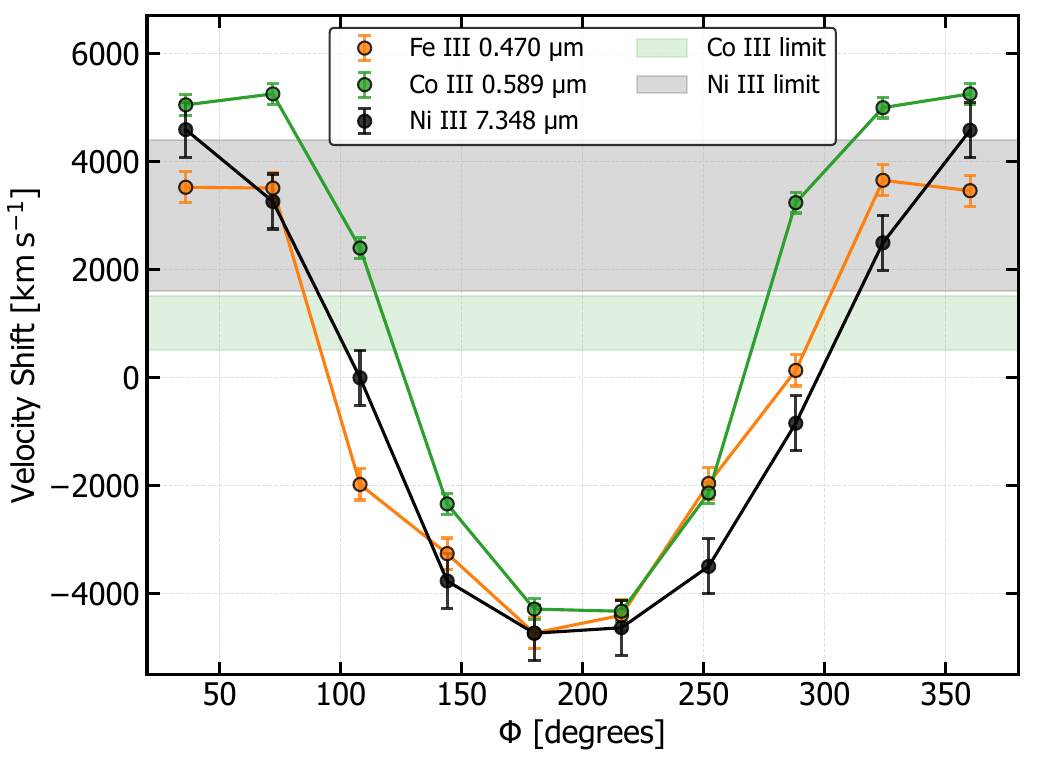}    
    \includegraphics[width=0.49\textwidth]{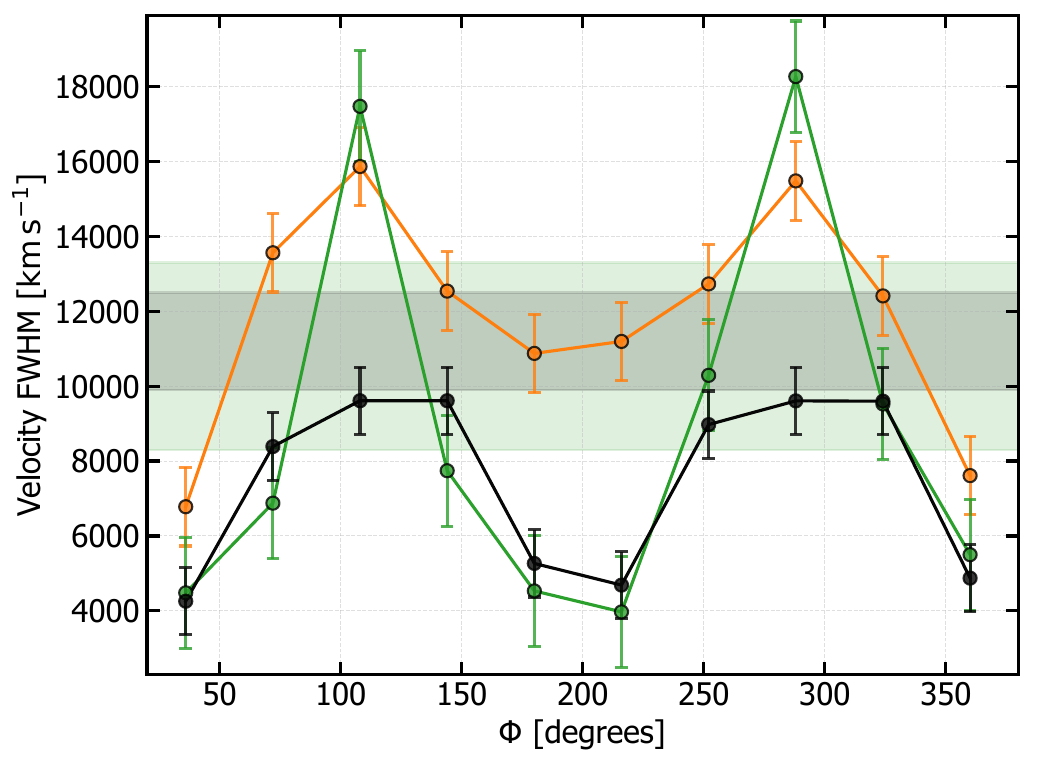}
        \caption{Similar to Figure~\ref{fig:3DOneExpl_Velocity_fits} but for the \ion{Fe}{III} 0.470\microns, \ion{Co}{III} 0.5890\microns, \ion{Ni}{III} 7.348\microns features of the \two model. Note the different velocity scales compared to Figure~\ref{fig:3DOneExpl_Velocity_fits}.}
    \label{fig:3DTwoExpl_Velocity_fits}
\end{figure}

In this section, we focus on the viewing-angle variation of the \two model. Figure~\ref{fig:3DTwoExpl_viewing_angle} shows the synthetic spectra of the \two model as viewed from five different orientations within the merger plane (i.e., $\cos(\theta) = 0$, $\phi = 0$–$360\degree$). We also investigate a selected set of IGE and IME features across all ten observer orientations in Figure~\ref{fig:Velocity_3DTwoExpl_stacked12}. Finally, we quantitatively compare the velocities and FWHMs of a subset of these IGE features in Figure~\ref{fig:3DTwoExpl_Velocity_fits}.

\subsubsection{Iron Group Element Variation}

It can be seen in Figure~\ref{fig:3DTwoExpl_viewing_angle} that the \two model produces significant variations in both feature velocities and profile morphology depending on the observer’s orientation. This variation is considerable across different viewing angles, with velocity shifts approximately double that of the \one model. The optical spectrum displays complex behaviour even among features of the same ionisation stage. For example, the central 0.470\microns \ion{Fe}{III} feature possesses a narrow peak at its most extreme redshift ($\phi=36\degree$), while the \ion{Fe}{III} 0.527\microns feature appears faintest and narrowest at the same orientation. This results from significant absorption and scattering by \ion{Fe}{I} (see Section~\ref{sec:average_spectra_optical}), indicating that the ejecta is not yet fully optically thin and that opacity still influences certain spectral features.

Focusing on the \ion{Fe}{III} 0.470\microns and \ion{Co}{III} 0.589\microns features in Figure~\ref{fig:Velocity_3DTwoExpl_stacked12}, we find that orientations showing a redshift in the \ion{Fe}{III} line (e.g., $\phi=36\degree$) also exhibit a redshift in the \ion{Co}{III} line. Similarly, orientations with a flatter \ion{Fe}{III} peak (e.g., $\phi=144\degree$) display a comparable profile in the \ion{Co}{III} feature. 
At other angles (e.g., $\phi=180\degree$), the 0.470\microns \ion{Fe}{III} multiplet shows a more complex structure which is also seen in the 2.241\microns \ion{Fe}{III} feature. 
However, in the MIR, the \ion{Fe}{III} 22.925\microns line more closely resembles the \ion{Co}{III} 11.888\microns feature. Additionally, the \ion{Fe}{III} 22.925\microns profile is not centrally peaked at orientations such as $\phi=252\degree$. While the \one model showed some differences between MIR and optical \ion{Fe}{III} profiles, variations are more pronounced in the \two model.
This change in profile morphology can be attributed to a non-negligible fraction of \ion{Fe}{III} residing outside the central ejecta combined with the reduced blending in the MIR.

We find that the more asymmetric 3D structure in the \twoscenario scenario results in more significant variation between lines of sight when compared to those in the \onescenario scenario.
A substantial difference between the 1D and 3D treatments appears in the 0.470\microns feature, which is more luminous and more centrally peaked across \two model orientations which is also true for the NIR \ion{Fe}{III} feature. However, the 1D model can match some features reasonably well, such as the \ion{Co}{III} 0.589\microns feature at flatter-peaked orientations (e.g., $\phi=288\degree$). As discussed in Section~\ref{sec:angleaveraged}, the \two model reproduces doubly ionised features better than the \twoO model but performs worse for singly ionised features. This ionisation balance, combined with flatter features and blending with \ion{Fe}{I} (see Figure~\ref{fig:Kromer_3DTwoExpl}), prevented the extraction of \ion{Fe}{II} velocities and FWHMs. Still, the overall larger width features in the \two model more closely match observations than in the \one scenario.

We also investigate the MIR \ion{Ni}{III} 7.349\microns feature, which in the \twoscenario scenario is slightly contaminated on the blue wing by \ion{Ar}{II}. The velocity shifts and morphologies of the \ion{Ni}{III} feature closely follow those of the other doubly ionised IGEs (i.e., the \ion{Co}{III} and \ion{Fe}{III} features), reflecting the similar spatial distributions of their ion populations in the ejecta. 
We find that the \ion{Ni}{III} feature in the \twoO model differs significantly from all viewing angles in the 3D calculation, with the 1D model generally producing a broad flatter-topped profile.
This behaviour reflects the underlying \ion{Ni}{III} distribution in \twoO, which is primarily concentrated in a thick outer shell between 5,000$\,\mathrm{km}\,\mathrm{s}^{-1}$ and 10,000$\,\mathrm{km}\,\mathrm{s}^{-1}$ (see Figure~\ref{fig:1DTwoExplionplot}). By comparison, the emitting region in the \two model is not confined to a symmetric outer shell. Consequently, some viewing angles produce more sharply peaked profiles, while others yield broader, relatively flatter-topped features.

It can be seen in Figure~\ref{fig:3DTwoExpl_Velocity_fits} that the \ion{Fe}{III}, \ion{Co}{III}, and \ion{Ni}{III} features show similar maximum blueshifts and redshifts of around 5,000$\,\mathrm{km}\,\mathrm{s}^{-1}$, approximately double those in the \one model. We also find larger FWHMs in these IGEs compared to the \one model. The \ion{Ni}{III} line generally shows the narrowest FWHMs, followed by \ion{Co}{III}, while \ion{Fe}{III} typically has the broadest FWHMs across most viewing angles, though there are separate angles where \ion{Co}{III} overlaps with \ion{Ni}{III} and \ion{Fe}{III} features. The secondary WD detonation compresses parts of the ejecta on the edge of the bubble into a half-crescent distribution on one side, with the opposite side being dominated by the secondary detonation ash. With this context, the velocity evolution of IGE features can be understood by interpreting Figure~\ref{fig:3DTwoExplionplot}. At $\phi=36$--$72\degree$, most of the material is moving away from this observer orientation, which results in a redshift, and the lack of material on this side results in a narrower FWHM. Rotating further results in an orientation where much of the material is still moving away, but the orientation is perpendicular to the elongated compression region, leading to a broader FWHM.
Overall, the emitting regions and Doppler velocity shifts in the \two model are more consistent between species, although their dynamic range is generally larger than in the \one model.

When compared to SN~2021aefx, we find several orientations in which the velocities and emitting regions of \ion{Co}{III} and \ion{Ni}{III} overlap with those observed. Similar to the \one model, no single viewing angle produces velocity shifts in both \ion{Co}{III} and \ion{Ni}{III} that align entirely with SN~2021aefx. Moreover, no individual orientation shows complete agreement in the corresponding FWHMs of the IGEs. These discrepancies arise primarily because the \ion{Ni}{III} FWHMs are too narrow.
Additionally, on average, there is a smaller net velocity offset between the \ion{Co}{III} and \ion{Ni}{III} in our calculation than that observed in SN~2021aefx, by approximately 500$\,\mathrm{km}\,\mathrm{s}^{-1}$. In general, we find that for several observer orientations, the \ion{Ni}{III} emitting region is significantly closer to matching that of SN~2021aefx than in the \one model. However, we stress the need for a broader set of theoretical models and MIR observations.

We now compare to the same observational samples discussed in Section~\ref{sec:3DOneExpl_Modle_Orientation_Effects}, in which doubly ionised features such as \ion{Co}{III} typically exhibit velocities of $\sim$1,000$\,\mathrm{km}\,\mathrm{s}^{-1}$ and FWHM values of $\sim$8,500--11,500$\,\mathrm{km}\,\mathrm{s}^{-1}$. In the \two model, the \ion{Co}{III} velocities span from $\sim$5,000$\,\mathrm{km}\,\mathrm{s}^{-1}$ to $\sim$-5,000$\,\mathrm{km}\,\mathrm{s}^{-1}$, and therefore results in velocity shifts that are approximately five times larger than those observed. However, the average FWHM of the \ion{Co}{III} features are $\sim$10,000$\,\mathrm{km}\,\mathrm{s}^{-1}$, which is broadly consistent with the observed range of $\sim$8,500--11,500$\,\mathrm{km}\,\mathrm{s}^{-1}$. This suggests that, while the extent of the emitting material is comparable to that inferred from observations, the degree of underlying asymmetry in the model is too pronounced. We also note that for a subset of angles ($\phi=108\degree$ and $\phi=288\degree$) the FWHM values are much larger than those found in observational samples such as \cite{Maguire2018}. A more comprehensive examination, incorporating a broader number of explosion models, a wider range of observer orientations, and a self-consistent fitting approach between both synthetic and observed spectra, would be required to robustly assess compatibility of the theoretical models with the observed populations of SNe~Ia, and as such lies beyond the scope of the present investigation.

\subsubsection{Intermediate Mass Element Variation}

In the \two model, IMEs are located in the outer and innermost regions of the ejecta (see Figures~\ref{Fig: Model abundances plot} and~\ref{fig:3DTwoExplionplot}) with the secondary detonation increasing the abundance of IMEs such as S and Ar by approximately 50\%. This distribution arises from the detonation of both the primary and secondary WDs and their detonation timings. During the primary detonation, the secondary remains intact, causing IMEs from the primary to be located in the outer ejecta. Subsequently, the secondary detonation produces a significant amount of IMEs in the innermost regions of the ejecta. As such, the resulting stratification of IMEs is markedly different from the \one model.

The \ion{Ar}{iii} 8.991\microns feature exhibits substantial variation across different orientations. For instance, at $\phi=72\degree$, the model predicts a profile much closer to being entirely flat-topped, while $\phi=216\degree$ yields a profile that diverges significantly from being entirely flat-topped. The complexity of the \ion{Ar}{III} feature arises from its distribution throughout the ejecta (see Figures~\ref{Fig: Model abundances plot} and~\ref{fig:3DTwoExplionplot}), which is neither entirely confined to the outer shell (which would produce a flat-topped profile) nor completely centralised (which would result in a sharply peaked profile). Moreover, compared to other IMEs such as \ion{S}{III}, a greater proportion of \ion{Ar}{III} is located in the innermost ejecta compared to the outer ejecta, resulting in a profile morphology less well described as a being entirely flat-topped profile.
We find that the \twoO model produces an \ion{Ar}{III} profile more than twice as luminous as the 3D calculation and particularly centrally peaked, exhibiting only a small flat-topped region at its centre. As such, we exclude it from comparison across many orientations in Figure~\ref{fig:Velocity_3DTwoExpl_stacked12}, as it hinders meaningful comparisons between viewing angles and SN~2021aefx. While some orientations in the \two model yield a somewhat flatter-topped profile, most orientations do not resemble the feature observed in SN~2021aefx. Instead, the \ion{Ar}{III} profile more closely resembles those seen in other MIR SNe~Ia (e.g., the normal SN 2003hv and peculiar SN 2005df \citealt{Gerardy2007}).

From Figure~\ref{fig:3DTwoExpl_viewing_angle}, it can be seen that the \ion{S}{III} multiplet at 0.93\microns deviates from the Doppler velocity shifts seen in the IGEs, exhibiting significantly less variation in the shifts between different observer orientations. More notably, the feature is more flat-topped than the IGE features across all orientations. However, not all orientations are perfectly flat-topped, as some show additional structure in their profile morphology.
We also find that the \ion{S}{III} 18.708\microns feature follows broadly similar behaviour to the 0.93\microns feature. As illustrated in Figure~\ref{fig:Velocity_3DTwoExpl_stacked12}, we find that the 1D calculation yields a more luminous \ion{S}{III} 18.708\microns line, and no 3D observer orientation reproduces the 1D case in both luminosity and width.  
Despite these differences, the agreement between the 1D and 3D calculations is generally better for \ion{S}{III} than for other doubly ionised IMEs such as \ion{Ar}{III}. 
However, the 1D approximation does not always provide a good representation of other ionisation stages of S. In particular, the NIR \ion{S}{I} feature emerges strongly only in the \two model (see Figures~\ref{fig:velocity_stacked_plot_oneexpl_scenario} and~\ref{fig:Velocity_3DTwoExpl_stacked12}) and is entirely absent in the 1D calculation. This feature is also highly sensitive to observer orientation, exhibiting a large variation in velocity shift.
The feature shows a distinct redshift at $\phi=180\degree$, which can be understood by examining the ion distribution (see Figure~\ref{fig:3DTwoExplionplot}).
This orientation probes a region with low \ion{S}{I} populations due to IGEs compressed by the secondary WD detonation filling that region, and some highly ionised S occupying the region surrounding those compressed IGEs. Contrasting this redshift, orientations such as $\phi=288\degree$ show a clear blueshift, caused by the compression region of the primary and secondary detonation ash creating a region with a large amount of \ion{S}{I} moving toward that direction. Moreover, the feature exhibits a greater width at angles such as $\phi=252\degree$, as this orientation views both the outer shell and the innermost \ion{S}{I} populations.
In the context of the ionisation distributions, the evolution of the velocity and widths of the \ion{S}{I} feature can be understood, and provides a clear example of how multidimensional calculations enable the investigation of geometry-dependent information for features not present in 1D calculations. Furthermore, as this feature is absent in observations of SN~2021aefx, it serves as a valuable diagnostic to constrain the model.

\section{Discussion and Conclusions}
\label{sec:Discussion_and_Conclusions}

We performed both 1D and 3D nebular radiative transfer calculations for the \dsix explosion models developed by \cite{pakmor_2021}, where the primary difference between scenarios lies in the fate of the secondary WD. We compared the spectra produced by the 1D and 3D models from 0.35--30\microns to assess the impact of multidirectional effects on the synthetic observables. We also analysed line-of-sight spectra for both explosion scenarios, extracting velocities and FWHMs for IGE features. These were also compared to SN~2021aefx \citep{Kwok2023}, with a particular focus on how different observer orientations in the 3D calculation compare to those observed. There are four key results from our investigation, as follows:

\begin{enumerate}  
    \item \textbf{Multidimensional Structure of Explosion Models}: We find that a multidimensional treatment of explosion models can significantly affect the luminosity and strength of spectral features, such as the 0.470\microns \ion{Fe}{III} feature in the \two model. Comparing 1D and 3D calculations, the extent of change in the synthetic observables closely 
    reflects the ejecta's underlying asphericity level. Both 1D and 3D models also suffer from overionisation; however, the \two model generally attains a better ionisation balance across \ion{Fe}{I--III}, which is reflected in the NIR spectral features. Ion populations can also be notably asymmetric and off-centre, especially in the \twoscenario scenario, where the 3D structure alters ionisation and leads to the emergence of \ion{S}{I} and \ion{Fe}{I} features in the NIR, which are suppressed in the 1D calculation. The \onescenario scenario also shows some sensitivity to multidimensional effects, notably altering fluxes of features such as the MIR 8.991\microns \ion{Ar}{III} feature. However, it is generally better approximated by spherically averaged ejecta. As emphasised by \cite{Pakmor2024}, modern explosion models are inherently multidimensional, and our results strengthen the necessity of performing 3D nebular phase calculations to capture the full diversity in synthetic observables.

    \item \textbf{Orientation-Dependent Observables}:
    We have demonstrated the ability to extract line-of-sight spectra from 3D calculations, allowing us to investigate velocity shifts, FWHM variations, and diversity in profile morphologies of key nebular-phase features. Across several features, the angle-averaged spectra fail to capture the luminosity, width, velocity, or morphology of any observer orientation. In both explosion scenarios, velocity shifts, FWHMs, and feature profiles exhibit distinctive patterns with rotation that can be understood by analysing the underlying distribution of ion populations. As expected, the \twoscenario scenario shows a heightened sensitivity to orientation effects, with velocity shifts approximately double those of the \onescenario scenario.
    We also find that the FWHMs of \ion{Co}{III} and \ion{Ni}{III} features are on average about 2,000$\,\mathrm{km}\,\mathrm{s}^{-1}$ larger in the \two model. However, the \one model still exhibits substantial and detectable variation in velocity and FWHM across different observer orientations, as is evident by the differing trends of \ion{Fe}{II} and \ion{Fe}{III} features. Importantly, our results show that the innermost geometry shapes the morphology of IME spectral features like the \ion{Ar}{III} 8.991\microns feature, with certain viewing angles in both the \one and \two models producing synthetic spectra that better match the observations of SN~2021aefx than others.

\item \textbf{Fate of the Secondary WD}:   
    The detonation of the secondary WD impacts nebular-phase synthetic observables more significantly than during the photospheric phase \citep{Pollin2024a}, as the average model spectra differ substantially, evident across the optical, NIR, and MIR regions
    We find other key differences between the scenarios emerge: the \one model yields over-ionised and excessively narrow NIR spectra, inconsistent with SN~2021aefx and typical nebular-phase SNe~Ia. In contrast, the \two model more accurately reproduces the observed NIR flux, but introduces discrepancies, including the absence of sharply peaked \ion{Co}{III} features, the presence of a strong NIR \ion{S}{I} feature, and a MIR \ion{Ar}{III} feature that lacks the expected flat-topped profile. While viewing-angle variations in the \two model can bring the \ion{Ar}{III} feature closer to being somewhat flat-topped, we tentatively favour the \one model as a more plausible candidate for normal SNe~Ia than this realisation of the \two model. Conversely, the weak optical \ion{O}{I}, and strong MIR \ion{Ne}{II} feature suggest that the \two model may be a better match for peculiar 02es-like (e.g., SN~2010lp; \citealt{Taubenberger2013}) or 03fg-like SNe~Ia (e.g., SN~2022pul; \citealt{Kwok2024}) classes. We, however, note that favouring the \onescenario scenario is in tension with the observed population of hypervelocity white dwarfs, which are expected products of the \dsix scenario when the secondary survives and have been regarded as key evidence for the scenario \citep{Shen2018b,El_Badry2023,Hollands2025}. These hypervelocity runaways can currently only account for 2\% of SNe~Ia \citep{Shen2025}, suggesting that while the \one model may be responsible for a small subset, it is unlikely to represent the dominant progenitor channel for SNe~Ia.

\item \textbf{Diagnostic Potential of the MIR}:  
    As discussed above, the contrasting spectral signatures of explosion models in the MIR provide a valuable means to distinguish between different explosion pathways. This contrasts with the optical and NIR, where interpreting variations is challenging due to blending, particularly in the main optical Fe complex and the 2.24\microns \ion{Fe}{III} feature. Consistent with previous investigations (e.g., \citealt{Gerardy2007,Blondin2023}), we confirm that no feature is entirely free from contamination, though the level of blending is significantly reduced in the MIR. Moreover, the MIR offers a valuable opportunity to test predictions of both IGE and IME features, such as the \ion{Co}{III} 11.888\microns, \ion{Ni}{III} 7.349\microns, \ion{Ne}{II} 12.815\microns features and, in particular, the 8.991\microns \ion{Ar}{III} feature. We find that the MIR viewing-angle spectra of the \one model produce an \ion{Ar}{III} feature more consistent with observations, whereas only specific orientations in the \two model produce a feature that somewhat resembles SN~2021aefx. This suggests that while centrally located IMEs cannot be entirely ruled out, their presence tends to result in more complex feature morphologies. These conclusions are only possible due to MIR observations, where reduced blending offers strong diagnostic power and allows for a clearer probe of IMEs than optical or NIR wavelengths.
 
\end{enumerate} 

Our multidimensional simulations reveal that a 3D treatment of explosion models can produce spectral features absent in 1D calculations, while also revealing systematic variations in these features depending on the viewing angle, including velocity shifts, FWHMs, and changes in feature morphology. These variations arise from asymmetries in the innermost ejecta and show how different observer orientations can give rise to distinct IGE and IME features. Together, these results show that 3D effects reshape which ions dominate nebular-phase spectra and allow for the determination of model-dependent orientation signatures. Hence, to identify which progenitor channel(s) are most likely to create SNe~Ia, it is paramount to investigate a suite of explosion models and compare their spectral variations with samples of observed SNe~Ia. We note that ground-based surveys have produced high-quality samples of optical and NIR spectra of SNe~Ia (e.g., \citealt{Maeda2010,Silverman2013,Childress2015,Maguire2016,Black2016,Maguire2018,Flores2020}), which provide opportunities to test 3D explosion models already. Combining these samples with future observations, particularly those from JWST, will be critical for thoroughly assessing the optical, NIR and MIR spectra produced by multidimensional models. Our investigation indicates that both \dsix models generated by \cite{pakmor_2021} may represent plausible pathways capable of producing normal SNe~Ia; however, as discussed, each model faces similar challenges, in particular the underproduction of singly ionised features.
Nevertheless, we stress that neither pathway can be definitively ruled out based on a single realisation alone, which currently possesses a somewhat artificial initial distribution of helium and in which the detonation of the secondary WD was not entirely self consistent, together with the limited number of JWST observations. Of particular interest for furthering our understanding of normal SNe~Ia and the \dsix scenario are:

\begin{enumerate}  
    \item \textbf{Parameter Space Exploration}: A comprehensive suite of nebular-phase calculations for the \dsix scenario are required to assess the role of WD mergers as the progenitors of normal and peculiar SNe~Ia. Multidimensional investigations should explore how variations in the following influence explosion geometry and the subsequent impact on nebular spectra (i.e., velocities, FWHMs, and line profiles): different mass pairs, helium shell masses, shell detonation mechanisms (e.g., converging shock, scissors mechanisms, edge-lit scenarios). Additionally, multidimensional investigation of all classes of explosion models will be essential for determining which nebular features can be reliably reproduced by all progenitor models and which features are truly unique to specific classes of models.
    
    \item \textbf{Evolution of Nebular Features}:  
    Future investigations should explore the evolution across multiple epochs in the nebular-phase to determine the impact of additional radioisotopes synthesised in explosion models \citep{Seitenzahl2009}.

    \item \textbf{Atomic Data \& Ionisation Challenges}: Updated atomic datasets should be incorporated into future 3D simulations to test model predictions against critical features such as the NIR $\sim$3.5\microns and MIR $\sim$8.5\microns features. Similar to other calculations (see \cite{Ruiz1996,Mazzali2015,Wilk2018,Shingles2020}), we also find that the NIR is over-ionised compared to observations. One possible solution to this discrepancy is clumping \citep{Wilk2020,Mazzali2020,Blondin2023} of the ejecta, which results in increased recombination and thus a reduced ionisation state bringing synthetic spectra more in line with observations.     
\end{enumerate}  

\section*{Acknowledgements}

JMP thanks Aysha Aamer for discussions related to the observational spectra and flux conversions.
SAS and FPC, acknowledge funding from STFC grant ST/X00094X/1. This work used the DiRAC Memory Intensive service Cosma8 at Durham University, managed by the Institute for Computational Cosmology on behalf of the STFC DiRAC HPC Facility (www.dirac.ac.uk). The DiRAC service at Durham was funded by BEIS, UKRI and STFC capital funding, Durham University and STFC operations grants. DiRAC is part of the UKRI Digital Research Infrastructure.
The authors gratefully acknowledge the Gauss Centre for Supercomputing e.V. (www.gauss-centre.eu) for funding this project by providing computing time on the GCS Supercomputer \cite{JUWELS} at Jülich Supercomputing Centre (JSC).
JMP acknowledges the support of the Department for Economy (DfE) and the use of Grammarly for proofreading and grammar checking. 
The work of F.K.R.\ and A.H. \ is supported by the Klaus Tschira Foundation and by the Deutsche Forschungsgemeinschaft (DFG, German Research Foundation) -- RO 3676/7-1, project number 537700965. F.K.R.\ acknowledges support by the European Union (ERC, ExCEED, project number 101096243). Views and opinions expressed are, however, those of the authors only and do not necessarily reflect those of the European Union or the European Research Council Executive Agency. Neither the European Union nor the granting authority can be held responsible for them. LJS acknowledges support by the European Research Council (ERC) under the European Union’s Horizon 2020 research and innovation program (ERC Advanced Grant KILONOVA No. 885281). 
A.H. is a fellow of the International Max Planck Research School for Astronomy and Cosmic Physics at the University of Heidelberg (IMPRS-HD) and acknowledges financial support from IMPRS-HD.
CEC is funded by the European Union’s Horizon Europe research and innovation programme under the Marie Skłodowska-Curie grant agreement No.~101152610.
We acknowledge Numpy \citep{harris2020array}, SciPy \citep{2020SciPy-NMeth}, Matpoltlib \citep{harris2020array} and
\href{https://zenodo.org/records/8302355} {\textsc{artistools}}\footnote{\href{https://github.com/artis-mcrt/artistools/}{https://github.com/artis-mcrt/artistools/}} \citep{artistools2024} for data processing and plotting. We also wish to thank the anonymous referees for their constructive comments and helpful suggestions.
\section*{Data Availability}

The spectra presented here will be made publicly available via the Heidelberg Supernova Model Archive (HESMA; \citealt{Kromer2017}; https://hesma.hits.org). They may also be obtained directly from the corresponding author upon request.
 
\bibliographystyle{mnras}
\bibliography{bibliography} 

\appendix
\clearpage
\section{Photoionisation Testing}
\label{apen:Photoionisation}

As noted in Section~\ref{sec:Radiative Transfer}, modifications have been made to the treatment of photoionisation in \textsc{ARTIS} to facilitate full 3D NLTE calculations.
\textsc{ARTIS} can employ two distinct methods to determine photoionisation coefficients. The first adopts a detailed approach in which specific photoionisation Monte Carlo estimators are recorded, following the methodology of \citet[eqn 44]{lucy2003}. The second method uses a radiation field model, which is based on fitting the parameters of dilute Planck function models to frequency-binned Monte Carlo estimators and then integrating photoionisation cross-sections over this model \citep[see ][]{Shingles2020}. The detailed approach is more accurate since every contribution to the estimator involves the photoionisation cross-section at the correct co-moving frequency, but this comes at the expense of involving a distinct estimator for every process considered. In contrast, obtaining values by integrating the radiation field model might sacrifice some accuracy\footnote{Relying on integration of the radiation field model will not resolve any effects of cross-section structure on frequency scales smaller than the frequency bin size of the model} but can reduce the total number of estimators that need to be recorded.

In our previous 1D nebular studies \citep{Shingles2020,Shingles2022a}, the detailed approach was applied to every photoionisation process. However, moving from a 1D model ($\sim$100 grid cells) to a 3D model ($\sim$100,000 grid cells) would result in prohibitively large memory requirements if only this approach is used.
We have therefore implemented a hybrid approach. Specifically, the detailed approach is used to obtain photoionisation rates for selected photoionisation processes (typically those associated with the low-lying states of each ion), while values obtained by integration of the model radiation field are adopted for higher energy levels.

\begin{figure} 
    \centering
    \includegraphics[width=0.47\textwidth]{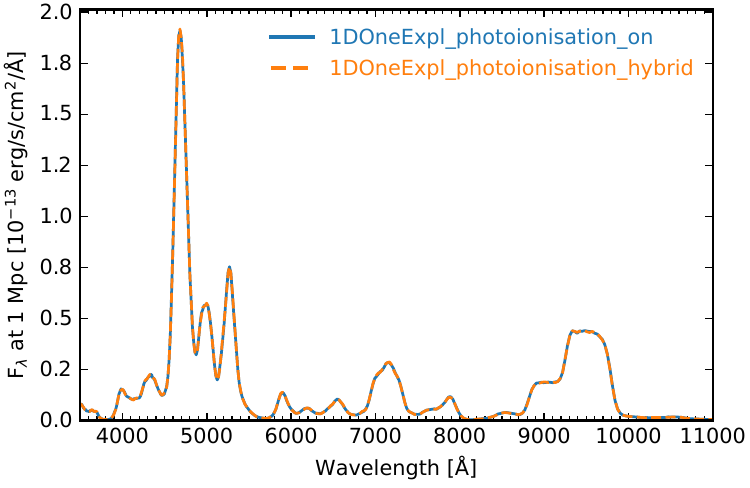}        
    \caption{Spectra of the \oneO model at 270 days post–explosion, shown with full detailed photoionisation and in a hybrid mode where detailed photoionisation is only applied to levels treated in NLTE.}
    \label{fig:Photoionisation_comparisons}
\end{figure}

Figure~\ref{fig:Photoionisation_comparisons} shows two calculations of the \oneO model, comparing the following scenario: (1) photoionisation rate estimators determined using the  detailed approach in all cases and (2) our hybrid scheme in which detailed estimators are only retained for  bound‐free transitions whose lower levels are included in the NLTE solution (and all other photoionisation processes are estimated using the binned radiation-field model). It can be seen that the hybrid scheme effectively reproduces the detailed approach and as such we employ this hybrid scheme throughout our investigation.

\section{Computational Cost and Computational Domain}
\label{apen:Resolution}

We use \textsc{ARTIS} v2024.5.1 \citep{artis_collaboration_2024_11230916}, which has been modified from previous versions to be capable of 3D NLTE nebular calculations for explosion models with up to $50^3$ grid cells. As discussed in Section~\ref{sec:Radiative Transfer}, our primary 3D nebular simulations employ a $40^3$ resolution due to the large computational runtime associated with increasing the number of grid cells. Both 1D and 3D calculations were run on 3072 cores (128 cores per node, 1 TB RAM), with 1D nebular calculations requiring $\sim$17,000 CPU core hours and the $40^3$ simulations averaging $\sim$473,000 CPU core hours. In total, the combined production cost for two 3D ($40^3$) and two 1D simulations was $\sim$980,000 CPU core hours. This cost is primarily dominated by the amount of time needed to update the plasma conditions for each grid cell. Given that we aim to carry out multiple 3D nebular simulations in a future parameter search, any savings in computational costs are extremely valuable. As such, we investigated removing outer grid cells, which are dominated by low-density cells. As such this would make the computational grid 24,000$\,\mathrm{km}\,\mathrm{s}^{-1}$ instead of 30,000$\,\mathrm{km}\,\mathrm{s}^{-1}$.

\begin{figure} 
    \centering
    \includegraphics[width=0.47\textwidth]{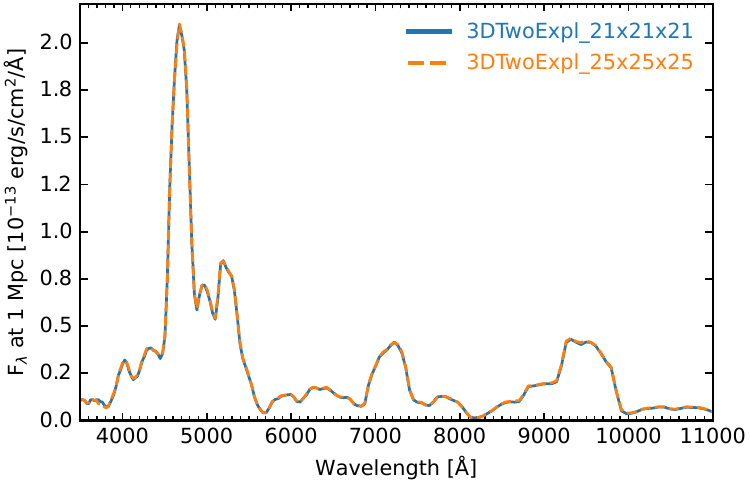}       
    \caption{3D angle-averaged spectra for the \two model at resolutions of $21^3$ and $25^3$ at 270 days post-explosion. In the former case, the outer cells were removed from the $25^3$ model.}
    \label{fig:downscale_test}
\end{figure}

To assess the impact of this approach, we conducted two lower-resolution simulations. The first is a 3D calculation of the \two models at a resolution of $25^3$ (114,000 CPU core hours). In the second case, we removed the outer grid cells of the $25^3$ model to create a $21^3$ resolution model (with the outer 2,400$\,\mathrm{km}\,\mathrm{s}^{-1}$ of the ejecta removed; 74,000 CPU core hours). As shown in Figure~\ref{fig:downscale_test}, we find that removing these outer cells has a negligible effect on the emergent spectra. Based on these results, we performed our main simulations at a resolution of $40^3$, which maintains the same inner resolution as our photospheric phase $50^3$ calculation \citep{Pollin2024a} but reduces the total wall-clock time significantly.

\section{Additional Viewing Angles}
\label{apen:additional_viewing_angles}

\begin{figure*}
\centering
\begin{subfigure}{\textwidth}
    \centering
    \includegraphics[width=0.99\textwidth]{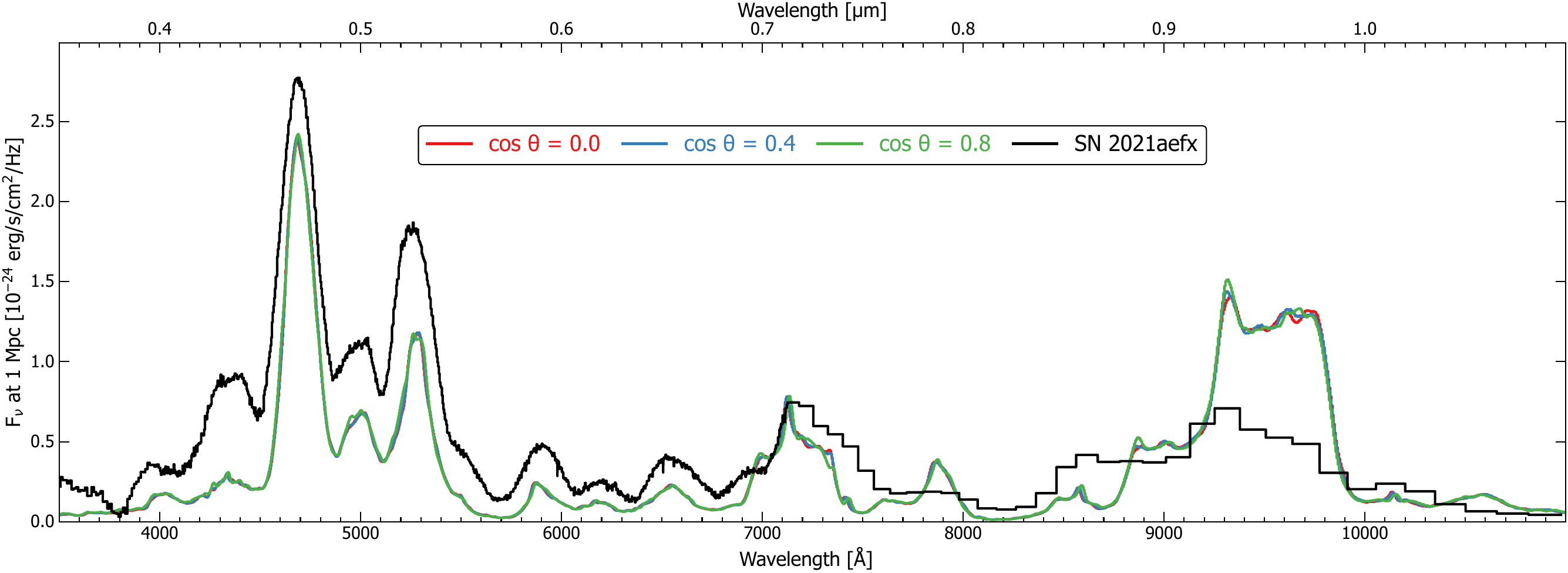}
\end{subfigure}

\vspace{1em}

\begin{subfigure}{\textwidth}
    \centering
    \includegraphics[width=0.99\textwidth]{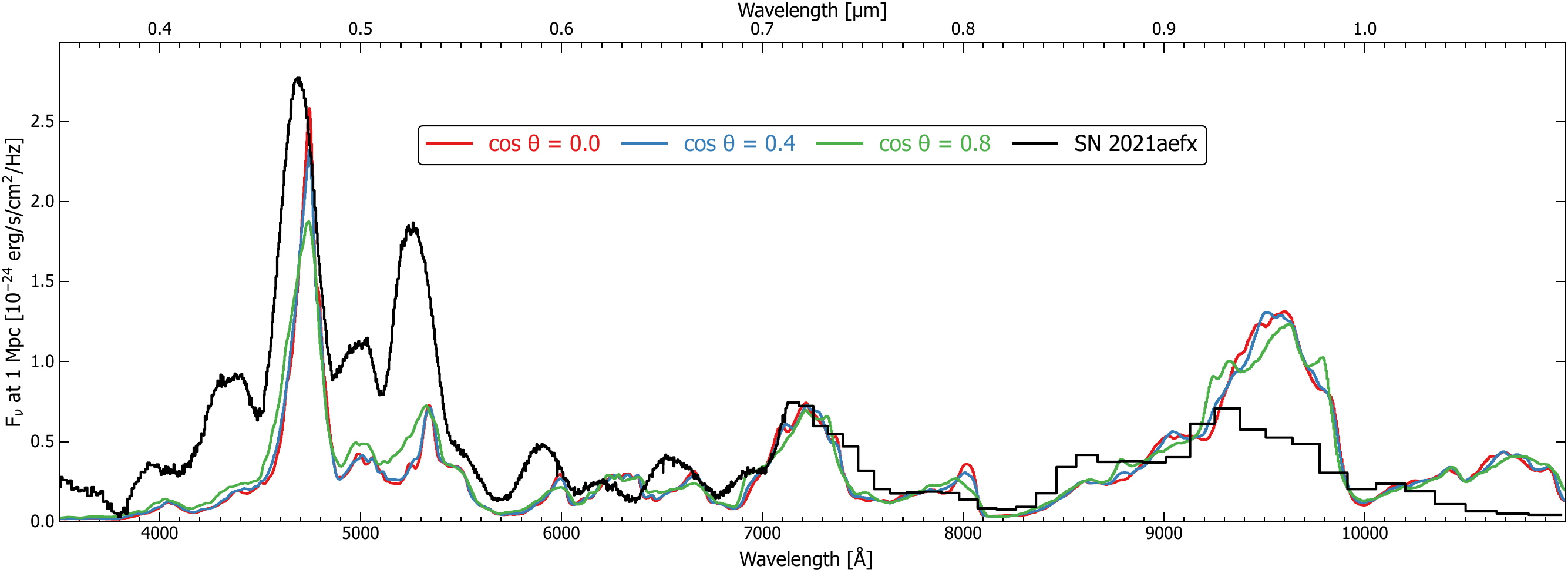}
\end{subfigure}
    \caption{Spectra of the \one (top) and \two (bottom) models at 270 days post-explosion for different viewing angles. All lines of sight are oriented at $\phi=36\degree$ with varying azimuthal angles ($\cos\theta=0.0$, $0.4$ and $0.8$). For comparison, we include SN~2021aefx \protect\citep{Kwok2023}.}
    \label{fig:apen_va}
\end{figure*}

Figure~\ref{fig:apen_va} displays spectra for different viewing angles at constant $\phi$ with varying azimuthal angles ($\cos(\theta)=$ 0.0, 0.4 and 0.8). The \one model shows only minor variations in spectral features. The \two model, however, displays more pronounced changes, particularly in the luminosity of central optical IGE features, though most features remain relatively similar. 
While a broader comparison of velocity offsets and profile evolution across viewing angles at these inclinations would be needed for complete observational reconciliation, the overall behaviour of the models can be understood by examining the rotation in $\phi$ at $\cos(\theta) = 0.0$, as this produces characteristic spectra that capture the key features of the models.

\section{Additional Density, Temperature, and Ion Distributions}

\begin{figure*}
\centering
\begin{subfigure}{\textwidth}
    \centering
    \includegraphics[width=0.91\textwidth]{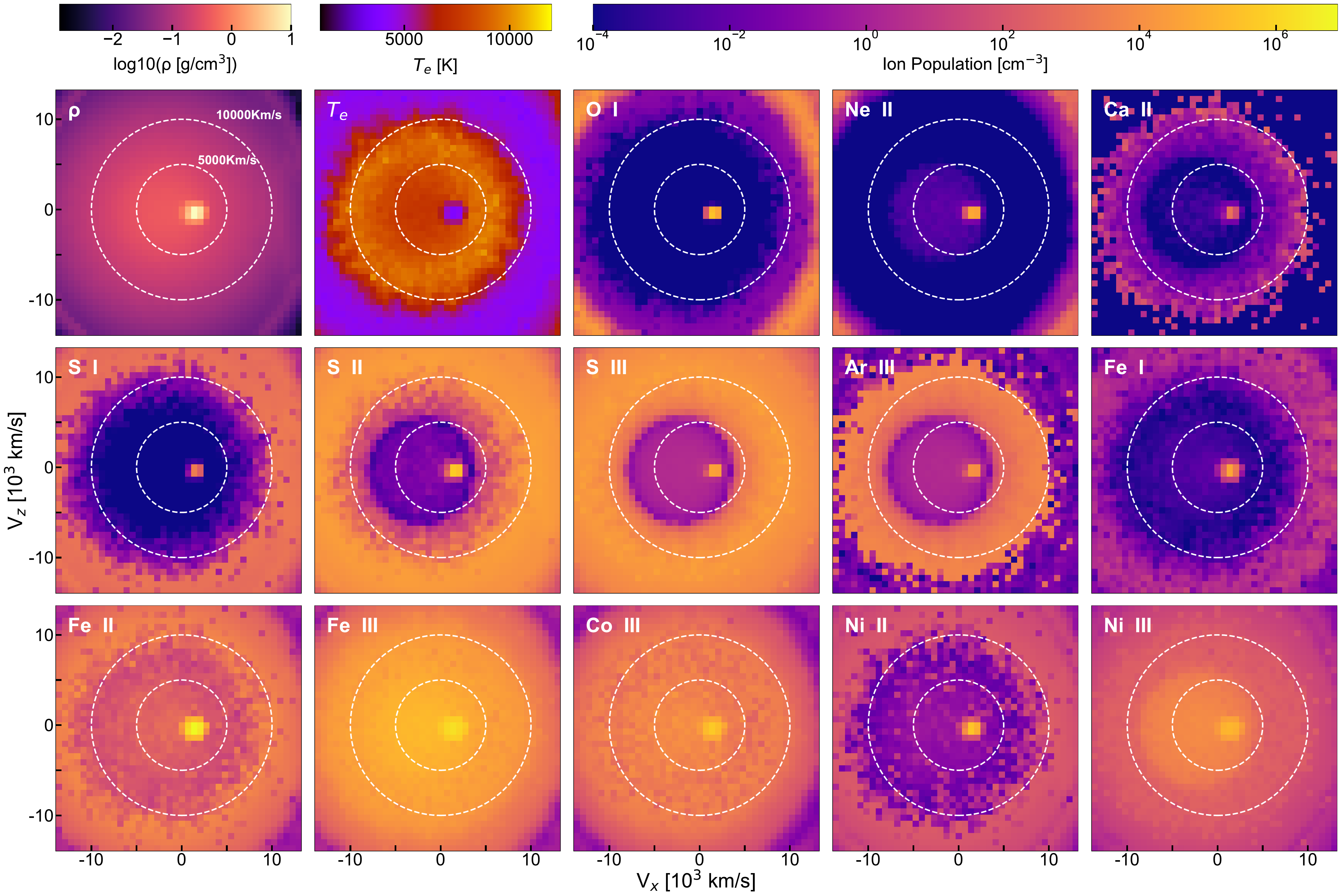}
\end{subfigure}

    \caption{Same as Figure~\ref{fig:3DOneExplionplot} but for the x-z plane}
    \label{fig:3DOneExpl_combined_ion_distribution_x_z}
\end{figure*}

\begin{figure*}
\centering
\begin{subfigure}{\textwidth}
    \centering
    \includegraphics[width=0.91\textwidth]{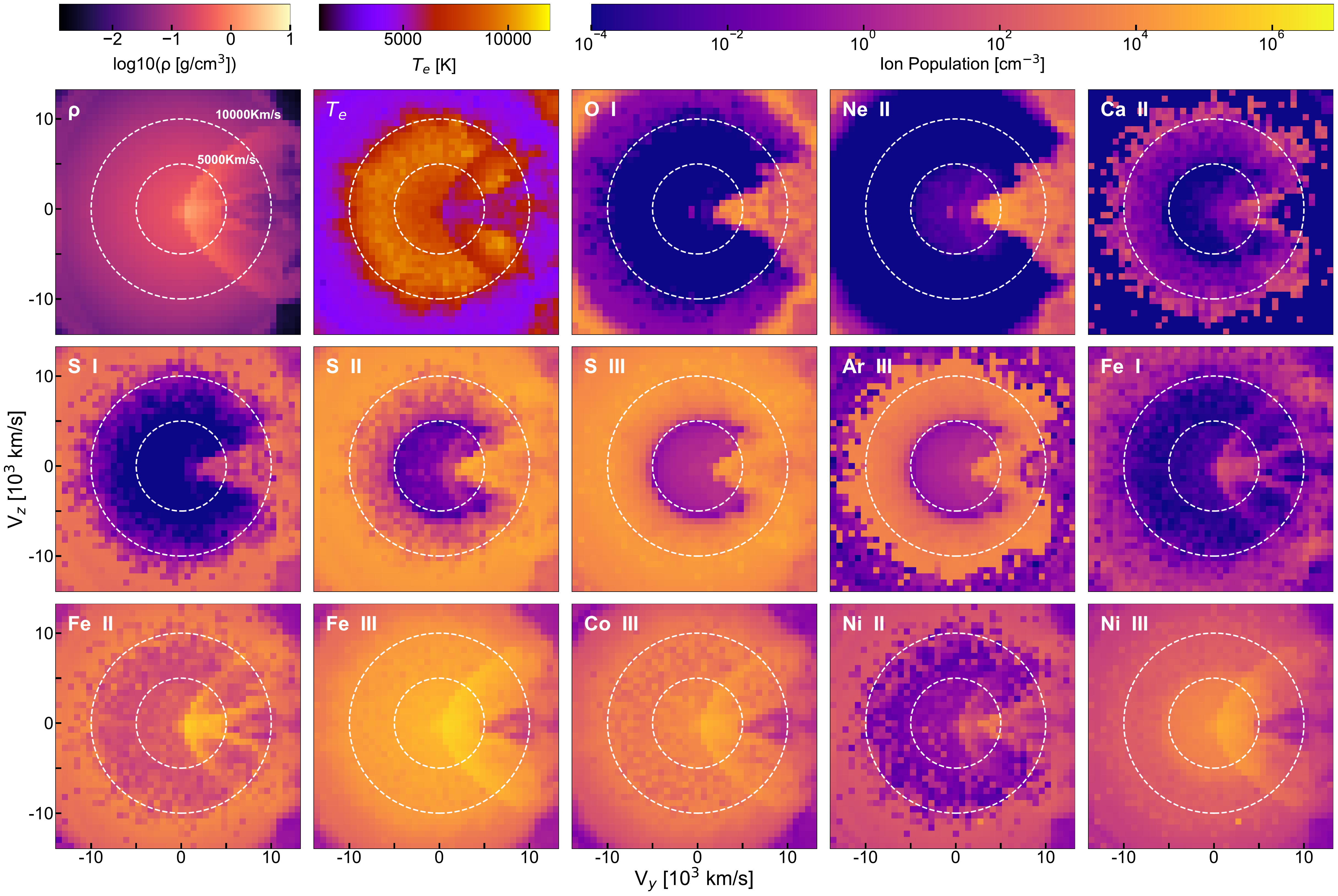}
\end{subfigure}

    \caption{Same as Figure~\ref{fig:3DOneExplionplot} but for the y-z plane}
    \label{fig:3DOneExpl_combined_ion_distribution_y_z}
\end{figure*}

\begin{figure*}
\centering
\begin{subfigure}{\textwidth}
    \centering
    \includegraphics[width=0.91\textwidth]{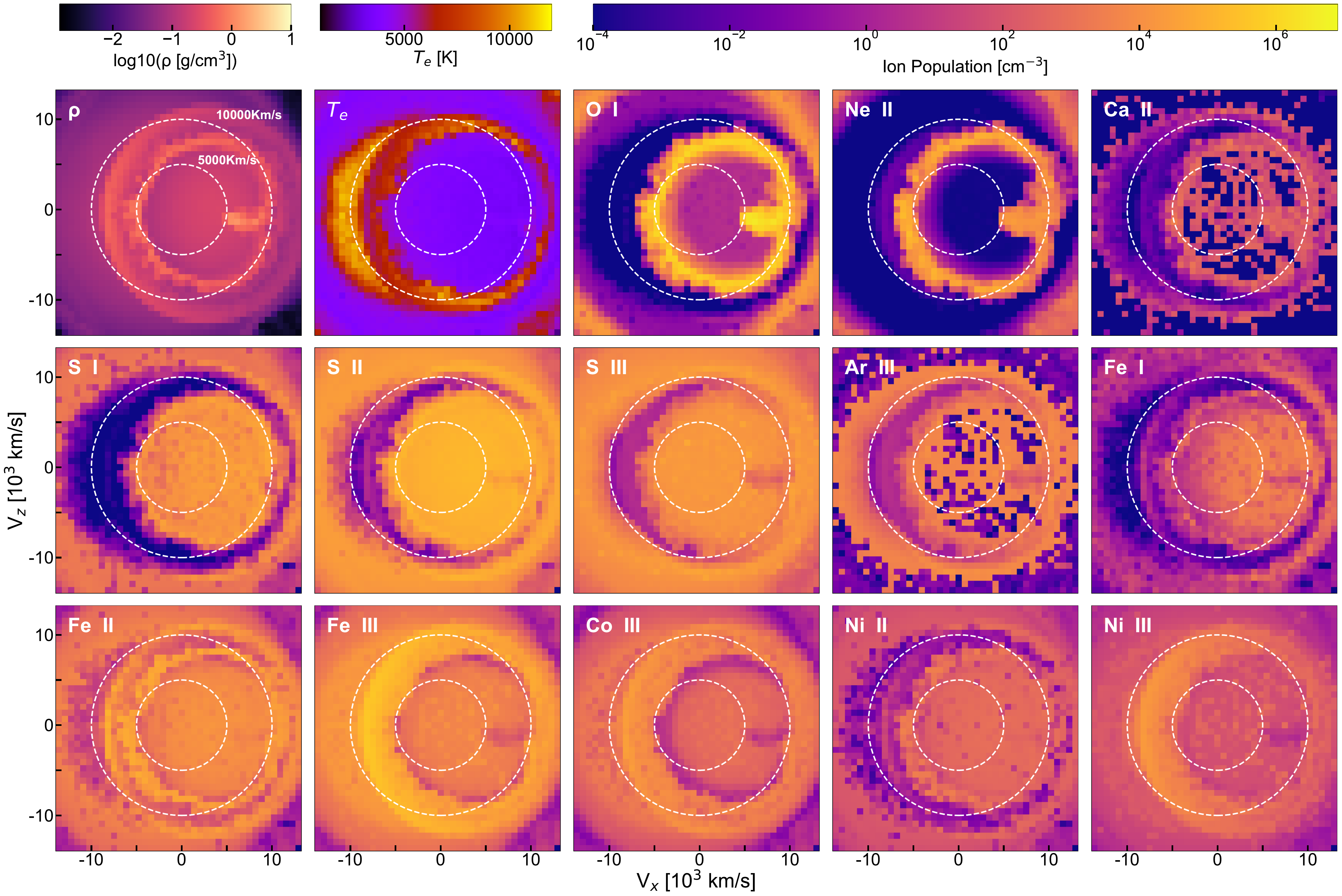}
\end{subfigure}

    \caption{Same as Figure~\ref{fig:3DTwoExplionplot} but for the x-z plane}
    \label{fig:3DTwoExpl_combined_ion_distribution_x_z}
\end{figure*}

\begin{figure*}
\centering
\begin{subfigure}{\textwidth}
    \centering
    \includegraphics[width=0.91\textwidth]{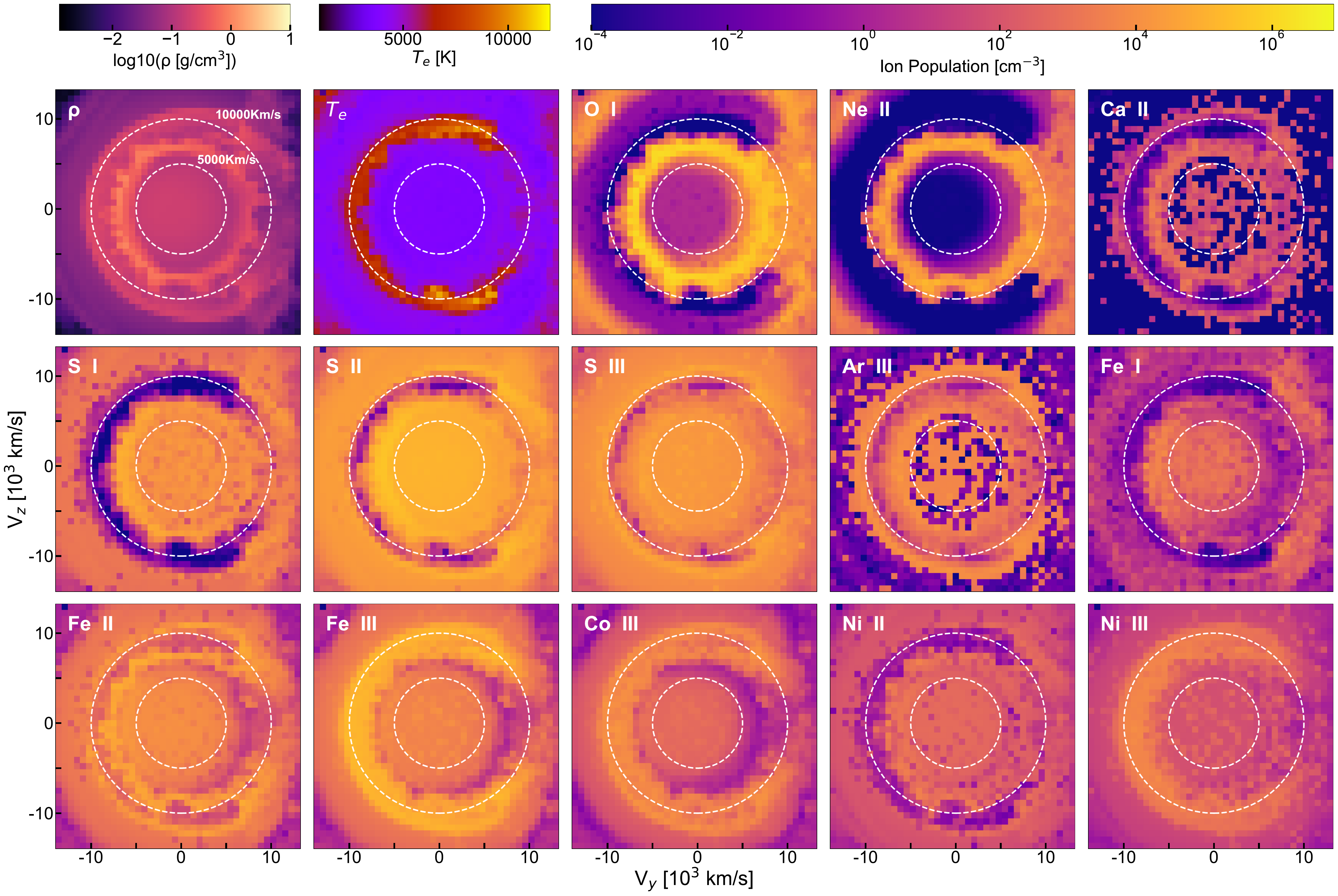}
\end{subfigure}

    \caption{Same as Figure~\ref{fig:3DTwoExplionplot} but for the y-z plane}
    \label{fig:3DTwoExpl_combined_ion_distribution_y_z}
\end{figure*}

As noted in Section~\ref{sec:Results}, we primarily discuss the x–y plane when interpreting spectral features, as the merger plane synthetic observables are the primary focus of our investigation. However, both explosion models are inherently multidimensional and, as such, the 1D temperatures, densities and ion distributions do not capture the dynamic range of the 3D models. Moreover, the 1D models do not uniquely correspond to the x–y plane of the corresponding 3D calculation. Instead, the 1D radially averaged composition describes the generic properties of the entire ejecta. Therefore, for completeness, we also provide the x-z and y-z slices for the \one model in Figures~\ref{fig:3DOneExpl_combined_ion_distribution_x_z} and~\ref{fig:3DOneExpl_combined_ion_distribution_y_z}, and for the \two models x-z and y-z slices in Figures~\ref{fig:3DTwoExpl_combined_ion_distribution_x_z} and~\ref{fig:3DTwoExpl_combined_ion_distribution_y_z}.

\bsp	
\label{lastpage}
\end{document}